\documentclass[smallextended]{svjour3}       % onecolumn (second format)
\smartqed  % flush right qed marks, e.g. at end of proof
\usepackage{lineno,hyperref}
\usepackage{amssymb,amsmath}
\usepackage{bm}
\usepackage{lipsum}
\usepackage{amsfonts}
\usepackage{graphicx,subfigure}
\usepackage{epstopdf}
\usepackage{algorithmic}
\usepackage{color}
\usepackage{tikz}

\begin{document}

\title{A Three-Dimensional Continuum Simulation Method for Grain Boundary Motion Incorporating Dislocation Structure}%\tnoteref{mytitlenote}}
%\tnotetext[mytitlenote]{Fully documented templates are available in the elsarticle package on %\href{http://www.ctan.org/tex-archive/macros/latex/contrib/elsarticle}{CTAN}.}

\titlerunning{A Three-Dimensional Continuum Simulation Method for Grain Boundary Motion}

\author{Xiaoxue Qin \and Luchan Zhang \and Yang Xiang}

\authorrunning{X. X. Qin, L. C. Zhang, Y. Xiang}

\institute{X. X. Qin \at Department of Mathematics, Hong Kong University of Science and Technology, Clear Water Bay, Kowloon, Hong Kong. \email{maxqin@ust.hk}
\and L. C. Zhang \at College of Mathematics and Statistics, Shenzhen University, Shenzhen, 518060,  China. Corresponding author. \email{zhanglc@szu.edu.cn}
\and Y. Xiang \at Department of Mathematics, Hong Kong University of Science and Technology, Clear Water Bay, Kowloon, Hong Kong. Corresponding author. \email{maxiang@ust.hk} }
%\author[mymainaddress,mysecondaryaddress]{Department of Mathematics, Hong Kong University of Science and Technology}

%\ead{xqinac@connect.ust.hk, lzhangas@connect.ust.hk, maxiang@ust.hk}
%
%\address[mymainaddress]{Department of Mathematics, Hong Kong University of Science and Technology}
%\address[mysecondaryaddress]{Clearwater Bay, Kowloon, Hong Kong}

%%\ead{malczhang@ust.hk,maxiang@ust.hk}
%
%\corref{mycorrespondingauthor}
%\ead{malczhang@ust.hk}

%\author{First Author         \and
%        Second Author %etc.
%}
%
%%\authorrunning{Short form of author list} % if too long for running head
%
%\institute{F. Author \at
%              first address \\
%              Tel.: +123-45-678910\\
%              Fax: +123-45-678910\\
%              \email{fauthor@example.com}           %  \\
%%             \emph{Present address:} of F. Author  %  if needed

\date{}

\maketitle

\begin{abstract}
We develop a continuum model for the dynamics of grain boundaries in three dimensions that incorporates the motion and reaction of the constituent dislocations.  The continuum model is based on a simple representation of densities of curved dislocations on the grain boundary. Illposedness due to nonconvexity of the total energy is fixed by a numerical treatment based on a projection method that maintains the connectivity of the constituent dislocations.
 An efficient simulation method is developed, in which the critical but computationally expensive long-range interaction of dislocations is replaced by another projection formulation that maintains the constraint of equilibrium of the dislocation structure described by the Frank's formula.
This continuum model is able to describe the grain boundary motion and grain rotation due to both coupling and sliding effects, to which the classical motion by mean curvature model does not apply. Comparisons with atomistic simulation results show that our continuum model is able to give excellent predictions of evolutions of low angle grain boundaries and their dislocation structures.
\keywords{Grain boundary dynamics \and Coupling and sliding motions \and
Dislocation dynamics \and Frank's formula \and Projection methods}
\end{abstract}

%\linenumbers

\section{Introduction}
Grain boundaries are indispensable components in polycrystalline materials. The energy and dynamics of grain boundaries play essential roles in the mechanical and plastic behaviors of the materials \cite{Sutton1995}. %There are extensive studies in the literature on this motion of grain boundaries by using molecular dynamics or continuum simulations, e.g. \cite{chen1994computer,upmanyu1998atomistic,kazaryan2000generalized,liuchun2001,upmanyu2002boundary,feng2003numerical,zhang2005curvature,kinderlehrer2006variational,zhang2009numerical,Selim2009,lazar2010more,elsey2013simulations,dai2018convergence}.
Most of the available continuum models for the dynamics of grain boundaries are  based on the motion driven by the capillary force which is proportional to the local mean curvature of the grain boundary \cite{Sutton1995,Herring1951,mullins1956two}. This motion is a process to reduce the interfacial energy $\int_S\gamma dS$, where $S$ is the grain boundary and $\gamma$ is the grain boundary energy density. If the energy density $\gamma$ is fixed, the driving force given by variation of the total energy  is in the normal direction of the grain boundary and is proportional to its mean curvature. There are many
atomistic simulations and continuum models in the literature for the grain boundary motion driven by mean curvature, e.g., \cite{chen1994computer,kazaryan2000generalized,liuchun2001,upmanyu2002boundary,ChenLQ2002,feng2003numerical,zhang2005curvature,Kirch2006,SrolovitzNature2007,DuQ2009,zhang2009numerical,Selim2009,lazar2010more,dai2018convergence,Du-Feng2020}.

%For low angle grain boundaries, the grain boundary energy density $\gamma(\theta)$ is an increasing function of the misorientation angle $\theta$.
The decreasing of grain boundary energy density $\gamma(\theta)$ can also reduce the total energy. For a low angle grain boundary,
this implies the decreasing the misorientation angle $\theta$.
In this case, the two grains on different sides of the grain boundary will rotate and cause a relatively rigid-body translation of the two grains along the boundary. This process is called sliding motion of grain boundaries \cite{li1962possibility,shewmon1966energy,harris1998grain,Kobayashi2000,upmanyu2006simultaneous,esedoglu2016grain,epshteyn2019motion}.

There is a different type of grain boundary motion which is called coupling motion \cite{li1953stress,srinivasan2002challenging,cahn2004unified}, in which
 the normal motion of the grain boundary induces a tangential motion proportionally. In the coupling motion, the energy density $\gamma(\theta)$ can increase although the total energy $\int_S\gamma dS $  is decreasing. Cahn and Taylor \cite{cahn2004unified} proposed a unified theory for the coupling and sliding motions of the grain boundary and demonstrated the theory based on dislocation mechanisms for a circular low angle grain boundary in two dimensions. Especially, the coupling motion of the grain boundary is associated with dislocation conservation during the motion of the grain boundary.  The  Cahn-Taylor theory and mechanisms of motion and reaction of the constituent dislocations have been examined by atomistic simulations and experiments \cite{srinivasan2002challenging,cahn2006coupling,molodov2007low,Molodov2009,trautt2012grain,wu2012phase,mcreynolds2016grain,yamanaka2017phase,salvalaglio2018defects}. It has been shown in Ref.~\cite{Rath2007} by a dislocation model and experimental observations that conservation and annihilation of the constituent dislocations may lead to cancelation of the coupling and sliding motions of the grain boundary, leading to the classical motion by curvature.    A continuum model  has been developed based on the motion and reaction of the constituent dislocations  for the dynamics of low angle grain boundaries in two dimensions \cite{zhang2018motion}. Their model can describe both the coupling  and sliding motions of low angle grain boundaries. Recently, they have proposed a more efficient numerical formulation \cite{zhang2019new}.
A continuum model that generalizes the Cahn-Taylor theory based on mass transfer by diffusion confined on the grain boundary has been proposed \cite{Taylor2007}, and numerical simulations based on this generalization were performed using the level set method \cite{Gupta2014}. Crystal plasticity models that include shear-coupled grain boundary motion in the phase field framework of Ref.~\cite{Kobayashi2000} have been developed \cite{AdmalIJP2018,AskJMPS2018}, in which the geometric necessary dislocation (GND) tensor/lattice curvature tensor was used to approximate the actual dislocation distributions on the grain boundaries. All these continuum models are for grain boundaries in two dimensions.

There are only limited studies in the literature for the three-dimensional coupling and sliding motions of grain boundaries.
Grain boundary motion and grain rotation in bcc and fcc bicrystals composed of
a spherical grain embedded in a single crystal matrix were studied by using three-dimensional phase field crystal model~\cite{yamanaka2017phase} and amplitude expansion phase field crystal model~\cite{salvalaglio2018defects}, and  properties of grain boundaries and their dislocation structures as the grain boundary evolves have been examined.
Although these atomistic-level phase field crystal simulations are able to provide detailed information associated with the coupling and sliding motions of grain boundaries in three dimensions, three-dimensional continuum models of the dynamics of grain boundaries incorporating their dislocation structures are still desired for larger scale simulations.

In this paper, we generalize the two-dimensional continuum model for grain boundary dynamics in Ref.~\cite{zhang2018motion,zhang2019new} to three dimensions, where grain boundaries and their constituent dislocations are curved in general. The three-dimensional continuum model for the dynamics of grain boundaries incorporates the motion and reaction of the constituent dislocations, and is able to describe  both coupling and sliding motions of the grain boundaries, to which the classical motion by mean curvature model does not apply. The continuum model includes evolution equations for both the motion of the grain boundary and the evolution of dislocation structure on the grain boundary. The evolution of orientation-dependent continuous distributions of dislocation lines on the grain boundary is based on the simple representation using dislocation density potential functions \cite{zhu2014continuum}.
This simple representation method also guarantees the continuity of the dislocation lines on the grain boundaries during the evolution. This continuum simulation framework for the distribution and dynamics of curves on  curved surfaces can be applied more generally beyond  the dynamics of dislocations and grain boundaries.

In a straightforward formulation of the continuum model, the variational force for the evolution of dislocations comes from a non-convex total energy, which leads to illposedness of the model. This problem is fixed by an alternative  formulation with constraints, whose geometric meaning is to  maintain the connectivity of dislocation lines.
A numerical treatment based on a projection method is developed to solved the constrained evolution equations.
The continuum model contains a long-range force in the form of singular integrals, whose evaluation is time-consuming especially in the three dimensional case. We generalize the projection method developed in two dimensional case~\cite{zhang2019new} that replaces the long-range force by a constraint of the Frank’s formula~\cite{Frank1950,Bilby1955,zhu2014continuum} describing equilibrium of the long-range force. The projection procedure in three dimensional case can be solved, by generalizing the ideas in two dimensional case~\cite{zhang2019new} with extra treatments to handle the Frank's formula in three dimensions and the connectivity of dislocations.

%With the fact that the long-range force is so strong that an equilibrium state described by the Frank’s formula~\cite{Frank1950,Bilby1955,zhu2014continuum} is quickly reached during the evolution, we replace the long-range force by a constraint of the Frank’s formula. We further solve the constraint evolution problem using projection methods, leading to a new, efficient, and well-posed continuum formulation.

Using the obtained continuum model, we perform numerical simulations for the evolution of low angle grain boundaries by coupling and sliding motions, and compare the results with those of atomistic simulations using phase field crystal model~\cite{yamanaka2017phase} and amplitude expansion phase field crystal model~\cite{salvalaglio2018defects} for validation of our continuum model. We also explain the anisotropic motion observed in these atomistic simulations based on our continuum model.

This paper is organized as follows. In Sec.~\ref{sec:illposed}, we present a three dimensional continuum model for the evolution of grain boundaries with dislocation structures that is directly based on the simple representation of curved dislocation lines on curved grain boundaries and the associated energies and driving forces~\cite{zhu2014continuum}. Illposedness of this formulation is discussed.
In Sec.~\ref{sec:cm}, in order to fix the illposedness problem, we present an alternative continuum formulation with constraints for the dynamics of grain boundaries in three dimensions, and propose
a numerical treatment based on a projection method  to solved the constrained evolution equations.
In Sec.~\ref{sec:projection}, we develop a more efficient formulation in which the computationally time-consuming long-range force is replaced by the constraint of the Frank’s formula, and obtain an explicit solution formula of the projection procedure.  Numerical simulations using our continuum model for the evolution of low angle grain boundaries by coupling and sliding motions are performed, and comparisons with the results of atomistic simulations using phase field crystal model~\cite{yamanaka2017phase} and amplitude expansion phase field crystal model~\cite{salvalaglio2018defects}
are made in Sec.~\ref{sec:nr}.

\section{Straightforward generalization to three dimensional model: Illposedness}\label{sec:illposed}

We have already  developed two dimensional continuum model for the evolution of grain boundaries with dislocation structures that is able to describe the coupling and sliding motions of grain boundaries~\cite{zhang2018motion,zhang2019new}.  Recall that in two dimensions where the grain boundary is a curve and the dislocations are points, dislocation densities on the grain boundary can be described directly by scalar functions. However, in three dimensions where the grain boundary is a surface and dislocations are lines on the surface, scalar densities are not able to describe the distributions of orientation-dependent, connected dislocation lines.
A simple representation  using scalar functions (dislocation density potential functions) for the densities of connected, curved dislocation lines on curved grain boundaries and the associated energies and driving forces have been proposed in Ref.~\cite{zhu2014continuum}.
   Using this representation, the orientation dependent dislocation densities are described based on surface gradient of the scalar dislocation density potential functions, instead of the scalar dislocation densities themselves in two dimension. This leads to illuposedness in the straightforward generalization of the continuum dynamics model to three dimensions;
see the discussion at the end of this section and more details in Theorem 1 in Sec.~\ref{sec:analysis}. In this section, we  present this straightforward generalization of the continuum model to three dimensions. An alternative form of this formulation that fixes the illposedness will be presented in the next section.

Using the dislocation representation and dynamics formulation in  Ref.~\cite{zhu2014continuum}, we have the following evolution equations of a grain boundary $S$ and its dislocation structure:
\begin{eqnarray}
&&v_n= M_\mathrm{d}\sum_{j=1}^J\frac{\|\nabla_S \eta_j\|}{\sum_{k=1}^J\|\nabla_S \eta_k\|}(\mathbf{f}^{(j)}_{\mathrm{long}}+\mathbf{f}^{(j)}_{\mathrm{local}})\cdot\mathbf{n}, \label{eqn:mod1v}\\
&& \frac{\partial \eta_j}{\partial t}=-M_{\rm d} \, \mathbf{f}^{(j)}_{\mathrm{long}}\cdot\nabla_S\eta_j
-M_\eta \mathbf{f}^{(j)}_{\mathrm{local}}\cdot\frac{\nabla_S \eta_j}{\|\nabla_S \eta_j\|}, \ \ j=1,2,\cdots,J.\label{eqn:mod1e}
\end{eqnarray}
Eq.~\eqref{eqn:mod1v} governs the evolution of the grain boundary, and Eq.~\eqref{eqn:mod1e} describes evolution of the constituent dislocations on the grain boundary. The first term on the right-hand side of Eq.~\eqref{eqn:mod1e} describes the motion of the constituent dislocations on the grain boundaries, and the second term models the change of dislocations due to dislocation reaction. Here it is assumed that  there are $J$ arrays of dislocations with Burgers vectors $\mathbf b^{(j)}$, $j=1,2,\cdots,J$, respectively, on the grain boundary, and they are described by the dislocation density potential functions $\eta_j$,  $j=1,2,\cdots,J$, respectively. In these evolution equations, $M_{\rm d}>0$ is the mobility of the constituent dislocations, and $M_\eta>0$ is the mobility associated with dislocation reaction.

%{\bf (1) Representation of curved dislocations.}

\begin{figure}[htbp]
	\centering
	\includegraphics[width=0.6\textwidth]{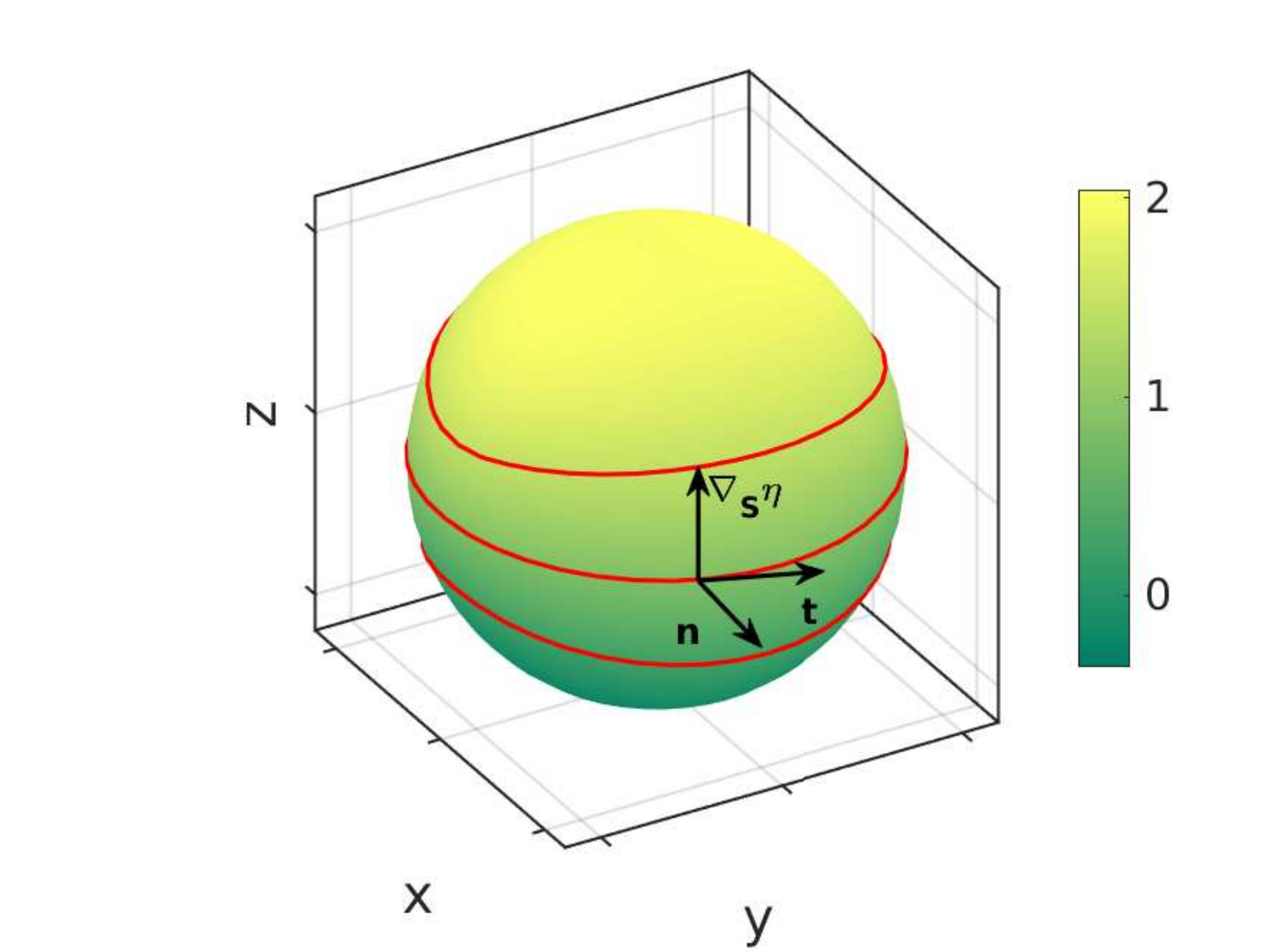}
	\caption{A dislocation density potential function $\eta$ defined on a grain boundary $S$. Its integer-value contour lines  represent the array of dislocations with the same Burgers vector. $\mathbf n$ is the unit normal vector of the grain boundary, and $\mathbf t$ is the local dislocation line direction. }
	\label{fig:dislocation}
\end{figure}

For a dislocation density potential function $\eta$ defined on a grain boundary $S$, the constituent dislocations of Burgers vector $\mathbf b$ are given by the contour lines of $\eta$: $\eta=i$, for integer $i$. See Fig.~\ref{fig:dislocation} for an example of dislocation structure on a spherical grain boundary and $\eta$ defined on it.  From the dislocation density potential function $\eta$, the inter-dislocation distance $D$ can be calculated by
$D=\dfrac{1}{ \| \nabla_S\eta\|}$,
and the dislocation direction is given by
$\mathbf t=\dfrac{\nabla_S\eta \times \mathbf n }{\| \nabla_S\eta\| }$,
where $\nabla_S\eta$ is the surface gradient of $\eta$ on $S$:
%\begin{equation}
$\nabla_S\eta=\left(\nabla-\mathbf n (\mathbf n \cdot \nabla)\right)\eta$, and $\mathbf n$ is the unit normal vector of the grain boundary.
%\end{equation}
Multiple dislocation density potential functions are used for dislocations with different Burgers vectors. From the physical meaning, we always have $\|\nabla_S \eta\|\leq
\dfrac{1}{b}$, meaning that the inter-dislocation distance $D\geq b$.

%{\bf (2) Energies and forces formulations.}

 The continuum formulation of the total  energy is
\begin{flalign}
E_{\rm tot}=&E_{\rm long}+E_{\rm local},\label{eqn:energy} \\
E_{\rm long} = &\frac{1}{2}\sum_{i=1}^J\sum_{j=1}^J \int_S  \mathrm{d}S_i \int_S  \mathrm{d}S_j \left[ \right. \frac{\mu}{4\pi}\frac{(\nabla_S \eta_i\! \times \!\mathbf{n}_i\cdot\mathbf b^{(i)})(\nabla_S \eta_j\! \times \!\mathbf{n}_j\cdot\mathbf b^{(j)}) }{r_{ij}}\nonumber\\
&  -\frac{\mu}{2\pi}\frac{(\nabla_S \eta_i\! \times \!\mathbf{n}_i )\times (\nabla_S \eta_j\! \times \!\mathbf{n}_j ) \cdot(\mathbf b^{(i)}\times \mathbf b^{(j)} ) }{r_{ij}}\nonumber\\
&\left.+ \frac{\mu}{4\pi(1-\nu)} (\nabla_S \eta_i\! \times \!\mathbf{n}_i\cdot\mathbf b^{(i)})\cdot (\nabla\otimes \nabla r_{ij})\cdot (\nabla_S \eta_j\! \times \!\mathbf{n}_j\cdot\mathbf b^{(j)})\right], \label{eqn:longenergy}  \\
 E_{\rm local } = &\int_S  \gamma_{\rm gb} \ \mathrm{d}S, \label{eqn:localenergy}\\
 \gamma_{\rm gb}=&\sum_{j=1}^J\frac{\mu(b^{(j)})^2}{4\pi(1-\nu)}\!\left(1-\nu\frac{(\nabla_S \eta_j\! \times \!\mathbf{n} \!\cdot\! \mathbf{b}^{(j)})^2}{(b^{(j)})^2 {\|\nabla_S \eta_j\|}^2}\right)\|\nabla_S \eta_j\| \log\! \frac{1}{r_g \|\nabla_S \eta_j\|}.\label{eqn:gamma_gb}
\end{flalign}
Here  $E_{\rm long}$ is the long-range interaction energy of dislocations, and $E_{\rm local}$ is the local dislocation line energy with energy density $\gamma_{\rm gb}$. In the formulation of $E_{\rm long}$ in Eq.~\eqref{eqn:longenergy},
$r_{ij}=\|\mathbf X_i-\mathbf X_j\|$, where $\mathbf X_i$ and $\mathbf X_j$ are the points varying on the grain boundary $S$ and are associated with the surface integral $dS_i$ and $dS_j$, respectively. $\mathbf n_j$ is the normal direction of the surface S associated with the surface integral $dS_j$, notation $\otimes$ is the tensor product operator, the gradient in the term $\nabla \otimes \nabla r_{ij}$ is taken with respect to $\mathbf X_i$, and $b^{(j)}=\|\mathbf b^{(j)}\|$. The elastic constants $\mu$ is the shear modulus and $\nu$ is the Poisson's ratio. The parameter $r_g$  in  $\gamma_{\rm gb}$ in Eq.~\eqref{eqn:gamma_gb}
depends on the size and energy of the dislocation core, and is of the order of $b$.

The driving forces for the dynamics of the grain boundary and the dislocation structure are associated with the variations of the total energy. In the grain boundary dynamics equations \eqref{eqn:mod1v} and \eqref{eqn:mod1e},
%\begin{eqnarray}
%&&\frac{\delta E_{\rm tot}}{\delta r}=-\sum_{j=1}^{J}\| \nabla_S\eta_j\|(\mathbf f_{\rm long} ^{(j)}+\mathbf f_{\rm local} ^{(j)} )\cdot \mathbf n,\label{eqn:var_r}\\
%&&\frac{\delta E_{\rm tot}}{\delta \eta_j}=(\mathbf f_{\rm long} ^{(j)}+\mathbf f_{\rm local} ^{(j)} )\cdot \frac{\nabla_S\eta_j }{\| \nabla_S\eta_j\|},
%\end{eqnarray}
 $\mathbf f_{\rm long} ^{(j)} $ is the continuum long-range force, and $\mathbf f_{\rm local} ^{(j)}$ the local force on the $\mathbf b ^{(j)}$-dislocations.
% In fact, the total force $\mathbf f_{\rm long} ^{(j)}+\mathbf f_{\rm local} ^{(j)} $ is the continuum limit \cite{zhu2014continuum} from the Peach-Koehler force on dislocations in discrete dislocation dynamics~\cite{HirthLothe1982}.
These  forces  have the following formulations:
\begin{flalign}
\mathbf f^{(j)}_{\rm long}=&(\pmb \sigma^{\rm tot}\cdot \mathbf b^{(j)})\times \left(\frac{\nabla_S \eta_j}{\|\nabla_S \eta_j\|}\times \mathbf n\right), \label{eqn:c3flong}\\
\pmb \sigma^{\rm tot}=&\sum_{j=1}^J\frac{\mu}{4\pi}\int_S\left[ \left(\nabla\frac{1}{r}\times\mathbf b^{(j)}\right)\otimes (\nabla_S\eta_j\times\mathbf n)+(\nabla_S\eta_j\times\mathbf n)\times\left(\nabla\frac{1}{r}\times\mathbf b^{(j)} \right) \right. \nonumber\\
&\left.+\frac{1}{1-\nu}\left(\mathbf b^{(j)}\times (\nabla_S\eta_j\times\mathbf n)\cdot \nabla\right)(\nabla\otimes\nabla-I\Delta)r   \right]\mathrm{d}S+ \pmb \sigma^{\rm app}, \label{eqn:lsig}\\
\mathbf f_{\rm local} ^{(j)}=&\frac{\mu}{4\pi(1-\nu)}\kappa_d \left[(1+\nu)(b^{(j)}_t)^2+(1-2\nu)(b^{(j)}_N)^2+(b^{(j)}_B)^2\right]\log \frac{1}{r_g \| \nabla_S\eta_j\|}\mathbf n_d^{(j)}  \nonumber \\
&-\frac{\mu\nu}{2\pi(1-\nu) }\kappa_db^{(j)}_Nb^{(j)}_B\mathbf t^{(j)}\times \mathbf n_d^{(j)}
 +\frac{\mu}{4\pi(1-\nu)}\kappa_p^{(j)}\left[(b^{(j)})^2-\nu(b_t^{(j)})^2 \right]\mathbf n_p^{(j)}\nonumber \\
 &+\frac{\mu}{4\pi(1-\nu)}\left[(b^{(j)})^2-\nu(b_t^{(j)})^2\right]\frac{(\nabla_S \nabla_S \eta_j) \cdot \nabla_S \eta_j}{\|\nabla_S \eta_j\|^2}.\label{eqn:flt}
\end{flalign}
Here $\pmb \sigma^{\rm tot}$ is the total stress field, which includes the
long-range stress field generated by the dislocation arrays on $S$, i.e., the first integral term, and the other stress fields $\pmb \sigma^{\rm app}$, and in this formulation of  $\pmb \sigma^{\rm tot}$,
 $r=\|\mathbf X-\mathbf X_S \| $ with points $ \mathbf X_S$ varying on the grain boundary S and $\nabla_S\eta_j$ and $\mathbf n$ being evaluated at $\mathbf X_S$. In the local force $\mathbf f_{\rm local} ^{(j)}$ in Eq.~\eqref{eqn:flt},   $\kappa_d$ is the curvature of dislocation line, $\mathbf n_d$ is the normal direction of dislocation, $\kappa_p$ and $\mathbf n_p$ are the curvature and normal direction of the curve on $S$ that is normal to the location dislocation, respectively, and $\kappa_d \mathbf n_d=(\nabla_S\mathbf t )\cdot \mathbf t=\nabla_S\left(\frac{\nabla_S \eta_j}{\|\nabla_S \eta_j\|}\times \mathbf n \right)\cdot \left(\frac{\nabla_S \eta_j}{\|\nabla_S \eta_j\|}\times \mathbf n\right)$, $\kappa_p \mathbf n_p=\left(\nabla_S \frac{\nabla_S \eta_j}{\|\nabla_S \eta_j\|} \right)\cdot \frac{\nabla_S \eta_j}{\|\nabla_S \eta_j\|} $, $b_t=\mathbf b\cdot \mathbf t$, $b_N=\mathbf b\cdot \mathbf n_d$, $b_B=\mathbf b\cdot (\mathbf t \times \mathbf n_d ) $.

\vspace{0.1in}
\noindent
\underline{\bf Illposedness of this formulation}

Unfortunately, Eqs.~\eqref{eqn:mod1v} and \eqref{eqn:mod1e} do not form a wellposed formulation. Especially,
the evolution equation of dislocation structure in \eqref{eqn:mod1e} is illposed.
In fact, Eq.~\eqref{eqn:mod1e} is a second order evolution equation of $\eta_j$, which is determined by the second term $-M_\eta \mathbf{f}^{(j)}_{\mathrm{local}}\cdot\dfrac{\nabla_S \eta_j}{\|\nabla_S \eta_j\|}=-M_\eta \dfrac{\delta E_{\rm local}}{\delta \eta_j}$,  where $\mathbf{f}^{(j)}_{\mathrm{local}}$ is given in Eq.~\eqref{eqn:flt}. Note that $\kappa_d$, $\mathbf n_d$ and $\kappa_p$, $\mathbf n_p$ in $\mathbf{f}^{(j)}_{\mathrm{local}}$ are all expressed in terms of second partial derivatives of $\eta_j$.
However,
the  energy density $\gamma_{\rm gb}$ of the local energy  $E_{\rm local}$ in \eqref{eqn:localenergy} and \eqref{eqn:gamma_gb} is not convex as a function of $\nabla_S \eta_j$.
This nonconvexity leads to an illposed formulation when using the gradient flow $\dfrac{\partial \eta_j}{\partial t}=-M_\eta \dfrac{\delta E_{\rm local}}{\delta \eta_j}=-M_\eta \mathbf{f}^{(j)}_{\mathrm{local}}\cdot\dfrac{\nabla_S \eta_j}{\|\nabla_S \eta_j\|}$ for the evolution of $\eta_j$.
This can be understood as follows. Neglecting the orientation dependence factor, the contribution of $\eta_j$ in the energy density $\gamma_{\rm gb}$ is essentially $-\|\nabla_S \eta_j\|\log\|\nabla_S \eta_j\|$, which is a concave function of $\|\nabla_S \eta_j\|$. As a result, the gradient flow gives a backward-diffusion like illposed evolution equation of $\eta_j$.  See Theorem 1 in Sec.~\ref{sec:analysis} for detail of the proof.
Therefore, the continuum model in Eqs.~\eqref{eqn:mod1v} and \eqref{eqn:mod1e} cannot be used directly to simulate the evolution of the grain boundary and its dislocation structure.

\section{Continuum model for grain boundary dynamics in three dimensions}\label{sec:cm}

In this section, we present a  continuum model for the dynamics of grain boundaries in three dimensions  incorporating the coupling and sliding motions, which fixes the illposedness problem in the formulation in  Eqs.~\eqref{eqn:mod1v} and \eqref{eqn:mod1e}.

 In order to obtain  a gradient flow formulation that avoids the above discussed illposedness for the evolution of the dislocation structure represented by dislocation density potential functions $\eta_j$, $j=1,2,\cdots,J$,
 we use the components of $\nabla_S \eta_j$  as independent variables instead of $\eta_j$ itself in the evolution equation of dislocation structure. That is, when the grain boundary $S$ is expressed by $\mathbf r(u,v)$, where $(u,v)$ is an orthogonal parametrization with $\|\mathbf r_u\|=\|\mathbf r_v\|=1$, we have
 \begin{equation}\label{eqn:local-grad0}
 \nabla_S \eta_j=\eta_{ju}\mathbf r_u+\eta_{jv}\mathbf r_v,
 \end{equation}
 where $\eta_{ju}$ and $\eta_{jv}$ are partial derivatives of $\eta_j$ with respect to $u$ and $v$, and $\mathbf r_u$ and $\mathbf r_v$ are partial derivatives of $\mathbf r$ with respect to $u$ and $v$.
 We use $\eta_{ju}$ and $\eta_{jv}$ as independent variables for the evolution  of dislocation structure.
  Gradient flow based on
 variations of the local energy  taken with respect to $\eta_{ju}$ and $\eta_{jv}$ gives:
 \begin{flalign}
 \dfrac{\partial \eta_{ju}}{\partial t}=&-M_r\dfrac{\delta E_{\rm local}}{\delta \eta_{ju}}=-M_r\dfrac{\partial \gamma_{\rm gb}}{\partial \eta_{ju}}, \label{eqn:ode1} \\
 \dfrac{\partial \eta_{jv}}{\partial t}=&-M_r\dfrac{\delta E_{\rm local}}{\delta \eta_{jv}}=-M_r\dfrac{\partial \gamma_{\rm gb}}{\partial \eta_{jv}}, \label{eqn:ode2}
 \end{flalign}
where $M_r>0$.  Noticing that the local energy density $\gamma_{\rm gb}$ is a function of $\nabla_S \eta_j=\eta_{ju}\mathbf r_u+\eta_{jv}\mathbf r_v$,
    the gradient flow equations in  \eqref{eqn:ode1} and \eqref{eqn:ode2} are ODEs of $\eta_{ju}$ and $\eta_{jv}$ with respect to time $t$. For this ODE system, we have the standard local existence for the solution \cite{ODE}. As a result,
 illposedness due to the backward-diffusion like PDEs of $\eta_j$ in the original formulation is avoided.

However,
 this alternative formulation leads to a new problem that as partial derivatives of the same function,
 $\eta_{ju}$ and $\eta_{jv}$ are not independent. In fact, they are related by
 $ \dfrac{\partial \eta_{ju}}{\partial v}-\dfrac{\partial \eta_{jv}}{\partial u}=0$.  Recalling that dislocations are contour lines of the functions $\{\eta_j\}$ on the grain boundary,
 the physical meaning of these relations is that the dislocations are connected lines on the grain boundary, i.e., there is no dislocation source/sink at any point on the grain boundary.\footnote{In fact, the net dislocation flux across the boundary of any region $\Omega$ on the grain boundary is $\int_{\partial \Omega}\nabla_S \eta_j\cdot d\mathbf r=\int_\Omega\left(\frac{\partial \eta_{jv}}{\partial u}-\frac{\partial \eta_{ju}}{\partial v}\right)dudv=0$ using this condition.}  In order to fix this new problem,
 we include these relations of $\eta_{ju}$ and $\eta_{jv}$, $j=1,2,\cdots,J$,
 as  constraints in the continuum model. Using these treatments and combining the contribution from the long-range energy, the continuum formulation can be rewritten as:

 \vspace{0.1in}
\noindent
\underline{\bf Continuum model with constraints}
\vspace{0.05in}
\begin{flalign}
&v_n= M_\mathrm{d}\sum_{j=1}^J\frac{\|\nabla_S \eta_j\|}{\sum_{k=1}^J\|\nabla_S \eta_k\|}(\mathbf{f}^{(j)}_{\mathrm{long}}+\mathbf{f}^{(j)}_{\mathrm{local}})\cdot\mathbf{n},  \label{eqn:mod2vn}\\
& \frac{\partial \eta_{ju}}{\partial t}=-M_{\rm d} \frac{\partial}{\partial u}\big( \mathbf{f}^{(j)}_{\mathrm{long}}\cdot\nabla_S\eta_j\big)-M_\mathrm{r}\frac{\partial \gamma_{\rm gb}}{\partial \eta_{ju}},  \label{eqn:mod2enu}\\
&\frac{\partial \eta_{jv}}{\partial t}=-M_{\rm d} \frac{\partial}{\partial v}\big( \mathbf{f}^{(j)}_{\mathrm{long}}\cdot\nabla_S\eta_j\big)-M_\mathrm{r}\frac{\partial \gamma_{\rm gb}}{\partial \eta_{jv}},\label{eqn:mod2env}\\
&\text{subject to} \ \
 \frac{\partial \eta_{ju}}{\partial v}-\frac{\partial \eta_{jv}}{\partial u}=0.\label{eqn:gcon}
\end{flalign}
Here  $M_\mathrm{r}>0$ is the mobility associated with dislocation reaction based on the energy variations with respect to $\eta_{ju}$ and $\eta_{jv}$.

Numerically, we implement the
 constraint in Eq.~\eqref{eqn:gcon}  using a projection method similar to that for  fluid dynamics problems \cite{Chorin1968}. Since evolution of $\eta_{ju}$ and $\eta_{jv}$ due to the first  term in Eqs.~\eqref{eqn:mod2enu} and \eqref{eqn:mod2env} satisfies the constraint, we only need to focus on the deviation from the constraint due to the second terms therein.

 Recall that the second terms in the evolution of  $\eta_{ju}$ and $\eta_{jv}$ in Eqs.~\eqref{eqn:mod2enu} and \eqref{eqn:mod2env} come from the gradient flow of the
local energy $E_{\rm local} = \int_S \gamma_{\rm gb} dS$. In order to implement the  constraint in Eq.~\eqref{eqn:gcon}, we introduce a Lagrangian function:
\begin{equation}
L=\int_S \left(\gamma_{\rm gb}+\sum_{j=1}^{J}\lambda_j\left(\frac{\partial \eta_{ju}}{\partial v}-\frac{\partial \eta_{jv}}{\partial u}\right)\right)  \mathrm{d}S,
\end{equation}
where $\lambda_j$, $j=1,2,\cdots,J$, are Lagrange multipliers associated with the constraints. Using the Lagrangian function $L$ instead of $E_{\rm local}$ in the gradient flow,
the evolution of dislocation structure in   Eqs.~\eqref{eqn:mod2enu} and \eqref{eqn:mod2env}  becomes
\begin{flalign}
&\frac{\partial\eta_{ju} }{\partial t}=-M_{\rm d} \frac{\partial}{\partial u}\big( \mathbf{f}^{(j)}_{\mathrm{long}}\cdot\nabla_S\eta_j\big) -M_\mathrm{r}\frac{\partial \gamma_{\rm gb}}{\partial \eta_{ju}}+\frac{\partial\lambda_{j}}{\partial v},\\
&\frac{\partial\eta_{jv} }{\partial t}=-M_{\rm d} \frac{\partial}{\partial v}\big( \mathbf{f}^{(j)}_{\mathrm{long}}\cdot\nabla_S\eta_j\big)-M_\mathrm{r}\frac{\partial \gamma_{\rm gb}}{\partial \eta_{jv}}-\frac{\partial\lambda_{j}}{\partial u}.
\end{flalign}
Here the coefficients of $\dfrac{\partial\lambda_{j}}{\partial v}$ and $\dfrac{\partial\lambda_{j}}{\partial u}$ in these equations are set to be $1$.

During the evolution in the time step from $t_n$ to $t_{n+1}=t_n+\delta t$, we separate the evolution of $\eta_{ju}$ and $\eta_{jv}$  into two steps:
\begin{flalign}
&\eta_{ju}^* = \eta_{ju}^n-\left[M_{\rm d} \frac{\partial}{\partial u}\big( \mathbf{f}^{(j)}_{\mathrm{long}}\cdot\nabla_S\eta_j\big)+M_{\rm r}\frac{\partial \gamma_{\rm gb}}{\partial \eta_{ju}}\right]_{t_n}\cdot \delta t,\label{eqn:dp01} \\
& \eta_{jv}^* = \eta_{jv}^n-\left[M_{\rm d} \frac{\partial}{\partial v}\big( \mathbf{f}^{(j)}_{\mathrm{long}}\cdot\nabla_S\eta_j\big)+M_{\rm r}\frac{\partial \gamma_{\rm gb}}{\partial \eta_{jv}}\right]_{t_n}\cdot \delta t, \vspace{1ex} \label{eqn:dp1}\\
&  \eta_{ju}^{n+1}=\eta_{ju}^*+\frac{\partial\lambda_{j}^{n+1}}{\partial v}\delta t, \ \ \eta_{jv}^{n+1}=\eta_{jv}^*-\frac{\partial\lambda_{j}^{n+1}}{\partial u}\delta t. \label{ddd}
\end{flalign}
In order to satisfy the constraint $\dfrac{\partial \eta_{ju}^{n+1}}{\partial v}-\dfrac{\partial \eta_{jv}^{n+1}}{\partial u}=0 $, using Eq.~\eqref{ddd}, we have the formula for updating $\lambda_j$:
\begin{equation}
\bigtriangleup \lambda_{j}^{n+1}=\frac{1}{\delta t}\left(\frac{\partial \eta_{jv}^{*}}{\partial u}- \frac{\partial \eta_{ju}^{*}}{\partial v}\right), \label{eqn:dp2}
\end{equation}
where $\bigtriangleup$ is the Laplace operator. This Poisson equation  for $\lambda_{j}^{n+1}$ can be solved using a finite difference method.  See Theorem 2 in Sec.~\ref{sec:analysis} for the property of this projection method.

The numerical algorithm for solving the continuum model with constraints is summarized as follows:

\newpage
\vspace{0.05in}
\noindent
\underline{\bf Numerical algorithm for solving constrained evolution}
\vspace{0.05in}

From $t_n$ to $t_{n+1}=t_n+\delta t$,
\begin{flalign*}
\mathbf r^{n+1}=&\mathbf r^n +v_n\mathbf n\big|_{t_n} \cdot \delta t,\\
\eta_{ju}^* = &\eta_{ju}^n-\left[M_{\rm d} \frac{\partial}{\partial u}\big( \mathbf{f}^{(j)}_{\mathrm{long}}\cdot\nabla_S\eta_j\big)+M_{\rm r}\frac{\partial \gamma_{\rm gb}}{\partial \eta_{ju}}\right]_{t_n}\cdot \delta t, \\
 \eta_{jv}^* = &\eta_{jv}^n-\left[M_{\rm d} \frac{\partial}{\partial v}\big( \mathbf{f}^{(j)}_{\mathrm{long}}\cdot\nabla_S\eta_j\big)+M_{\rm r}\frac{\partial \gamma_{\rm gb}}{\partial \eta_{jv}}\right]_{t_n}\cdot \delta t, \vspace{1ex} \\%\label{eqn:dp1}\\
\bigtriangleup \lambda_{j}^{n+1}=&\frac{1}{\delta t}\left(\frac{\partial \eta_{jv}^{*}}{\partial u}- \frac{\partial \eta_{ju}^{*}}{\partial v}\right),\\%\label{eqn:Poisson}\\
  \eta_{ju}^{n+1}=&\eta_{ju}^*+\frac{\partial\lambda_{j}^{n+1}}{\partial v}\delta t, \ \ \eta_{jv}^{n+1}=\eta_{jv}^*-\frac{\partial\lambda_{j}^{n+1}}{\partial u}\delta t.
\end{flalign*}

\section{Continuum model without long-range force}\label{sec:projection}

The continuum model given by Eqs.~\eqref{eqn:mod2vn}--\eqref{eqn:gcon} contains the long-range elastic force
$\mathbf{f}^{(j)}_{\mathrm{long}}$ (given in Eqs.~\eqref{eqn:c3flong} and \eqref{eqn:lsig}), which is a singular integral over the entire grain boundary surface. Numerically,
computation of such long-range force with reasonable accuracy is complicated and time-consuming
even in two-dimensional cases \cite{zhang2018motion,zhang2019new}. It has been shown in two-dimensional cases \cite{zhang2019new} by comparison with discrete dislocation dynamics simulations that
the long-range interaction between the grain boundary dislocations is so strong that an equilibrium state described by the Frank's formula \cite{Frank1950,Bilby1955,zhu2014continuum} is quickly reached during the evolution of the grain boundary. Here we follow the assumption made in two-dimensional case  that the Frank's formula always holds during the evolution of the grain boundary \cite{zhang2019new}. This leads to a new three dimensional formulation without long-range force:

\vspace{0.1in}
\noindent
\underline{\bf Continuum model without long-range force}
\vspace{0.05in}
\begin{flalign}
&v_n= M_\mathrm{d}\sum_{j=1}^J\frac{\|\nabla_S \eta_j\|}{\sum_{k=1}^J\|\nabla_S \eta_k\|}\mathbf{f}^{(j)}_{\mathrm{local}}\cdot\mathbf{n}, \label{eqn:mod2v}\\
&  \frac{\partial \eta_{ju}}{\partial t}=-M_\mathrm{r}\frac{\partial \gamma_{\rm gb}}{\partial \eta_{ju}}, \ \ \frac{\partial \eta_{jv}}{\partial t}=-M_\mathrm{r}\frac{\partial \gamma_{\rm gb}}{\partial \eta_{jv}},
\label{eqn:mod2e}\\
%&& \frac{\delta\eta_j}{\delta t}= -M_r\frac{\delta \gamma_{gb}}{\delta \eta_j},\label{eqn:mod2e}\\
&\text{subject to} \ \
\frac{\partial \eta_{ju}}{\partial v}-\frac{\partial \eta_{jv}}{\partial u}=0,\label{eqn:gcon1}\\
&\hspace{5em}\mathbf{h}=\theta(\mathbf{V}\times \mathbf{a}) - {\displaystyle \sum_{j=1}^J} \mathbf{b}^{(j)}(\nabla_S\eta_j\cdot\mathbf{V})=\mathbf 0.\label{eqn:mod2frank}
%&&\theta(\mathbf r_u ,\mathbf r_v,\eta_{ju},\eta_{jv})=\frac{1}{S}\int\int_S \sum_{j=1}^J \frac{ ( \eta_{ju}+  \eta_{jv})( \mathbf r_u + \mathbf r_v )\times\mathbf{a}{\cdot}\mathbf{b}^{(j)} }{\|(  \mathbf r_u+ \mathbf r_v) \times\mathbf{a}\|^2}  \mathrm{d}S.\hspace{0.1in}\label{eqn:mod2t}
\end{flalign}

Here, the constraint \eqref{eqn:mod2frank} is the Frank's formula that governs the equilibrium dislocation structure on a grain boundary~\cite{Frank1950,Bilby1955,zhu2014continuum}, in which $\theta$ is the misorientation angle of the grain boundary and is a constant over the grain boundary at any fixed time, $\mathbf{a}$ is the rotation axis, and $\mathbf{V}$ is any vector in the grain boundary's tangent plane. For a planar grain boundary,
the Frank's formula holds if and only if the long-range elastic fields generated by the grain boundary cancel out~\cite{Frank1950,Bilby1955}.  It has been shown in Ref.~\cite{zhu2014continuum} that this equivalence  also holds for a curved grain boundary.

%Since the equilibrium dislocation structure that satisfies the Frank's formula is stable \cite{xiang2018stability}. Thus, $M_d=0$ in the step of $\eta$ evolution \eqref{eqn:mod2e} is a good approximation.

Numerically, the constraint of Frank's formula in Eq.~\eqref{eqn:mod2frank} can also be implemented using a projection method, i.e., projecting in each time step the virtual evolution result without the constraint of the Frank's formula to a nearby configuration that satisfies the Frank's formula. This is a separate numerical treatment in addition to the project method discussed in the previous section for handling the constraint in \eqref{eqn:gcon1} for continuity of dislocation lines.

Specifically, in the evolution from $t_n$ to $t_{n+1}=t_n+\delta t$, in the virtual evolution of the grain boundary without the constraint of Frank's formulation in \eqref{eqn:mod2frank}, we have
\begin{flalign}
\mathbf r^*=&\mathbf r^n+\mathbf v^*\delta t,\\
\mathbf v^*=&v_n,\label{eqn:virtualv}
\end{flalign}
where
 $\mathbf v^*=(v_1^*,v_2^*,v_3^*)$ is the virtual velocity due to the local force without the constraint, i.e., $v_n$ in  Eq.~\eqref{eqn:mod2v}.   Evolution of dislocation structure represented by $\eta_j$'s remains the same as that given in the previous section.

%Eq.~\eqref{eqn:e2} is the evolution of dislocation structure due to dislocation reaction without the constraint.

In the projection step, the virtual profile of the grain boundary $\mathbf r^*$ is projected to a nearby configuration that satisfies the Frank's formula \eqref{eqn:mod2frank}. Note that misorientation angle $\theta$ is needed in \eqref{eqn:mod2frank} at time $t_{n+1}$. We calculate
the misorientation angle $\theta$ during the evolution by
\begin{equation}
\theta=\frac{1}{S_A}\int_S \sum_{j=1}^J \frac{ ( \eta_{ju}+  \eta_{jv})( \mathbf r_u + \mathbf r_v )\times\mathbf{a}{\cdot}\mathbf{b}^{(j)} }{\|(  \mathbf r_u+ \mathbf r_v) \times\mathbf{a}\|^2}  \mathrm{d}S.\label{eqn:mod1t}
\end{equation}
where $S_A$ is the area of the grain boundary $S$ that can be calculated by $S_A=\int_S\|\mathbf r_u\times\mathbf r_v\|\mathrm{d}u\mathrm{d}v$. This formulation of $\theta$ is obtained by taking average of the vector equation  \eqref{eqn:mod2frank} in the $\mathbf r_u$ and $\mathbf r_v$ directions; see  Appendix for details of the derivation. We calculate $\theta^{n+1}$ using this formula based on the virtual evolution result of $\mathbf r^*$, i.e., $\theta^{n+1}=\theta(\mathbf r^{*}_u ,\mathbf r^{*}_v,\eta^{n+1}_{ju},\eta^{n+1}_{jv})$. This means that we assume that the value of $\theta$ does not change in the projection step.
%The change of misorientation angle $\delta \theta$ during this time step is determined by this virtual evolution as given in Eq.~\eqref{eqn:dt}.
Based on this obtained $\theta^{n+1}$, the actual grain boundary  velocity $\mathbf v$ is obtained by projection of the virtual configuration of the grain boundary $\mathbf r^*$ to a state $\mathbf r^{n+1}$ that satisfies the constraint of Frank's formula \eqref{eqn:mod2frank}.

This projection procedure has been validated in the two dimensional case by comparisons with the full evolution with the long-range force and discrete dislocation dynamics simulation, and explicit formula of the velocity after projection has been obtained in the two dimensional case~\cite{zhang2019new}. Here we generalize the projection procedure to three dimensional case, based on the formulation of misorientation angle $\theta$ in the three dimensional case established in Eq.~\eqref{eqn:mod1t}.
The projection procedure in three dimensional case here can also be solved  similarly as in the two dimensional case, with extra treatments to handle the Frank's formula in three dimensions and  connectivity of dislocations.

Now we solve the projection procedure in three dimensional case. Without loss of generality, suppose that the rotation axis is in the $+z$ direction, i.e., $\mathbf a = (0,0,1)$.

Suppose that the grain boundary velocity is $\mathbf v$, and the Frank's formula \eqref{eqn:mod2frank} holds at the time $t_n$. After a small time step $\delta t$, if the Frank's formula still holds at $t_{n+1}=t_n+\delta t$, we have
\begin{flalign}
(\delta\theta \mathbf r_u+\theta \delta t \mathbf v_u)\times\mathbf{a}-\sum_{j=1}^{J}\mathbf{b}^{(j)}\delta\eta_{ju}=&0,\label{eqn:Fu}\\
(\delta\theta \mathbf r_v+\theta \delta t \mathbf v_v)\times\mathbf{a}-\sum_{j=1}^{J}\mathbf{b}^{(j)}\delta\eta_{jv}=&0.\label{eqn:Fv}
\end{flalign}
Here we have used  $\delta \mathbf r_u=\delta t \mathbf v_u$ and $\delta \mathbf r_v=\delta t \mathbf v_v$, where $\mathbf v_u=\frac{\partial \mathbf v}{\partial u}$ and $\mathbf v_v=\frac{\partial \mathbf v}{\partial v}$.

Integrating Eq.~\eqref{eqn:Fu} with respect to $u$, and Eq.~\eqref{eqn:Fv} with respect to $v$, we have
\begin{flalign}
&\left(\delta\theta \mathbf r(u,v)+\theta \delta t  \mathbf v(u,v)\right)\times\mathbf{a}-\sum_{j=1}^{J}\mathbf{b}^{(j)}\delta\eta_{j}(u,v)\nonumber\\
=&\left(\delta\theta \mathbf r(0,v)+\theta \delta t  \mathbf v(0,v))\right)\times\mathbf{a}-\sum_{j=1}^{J}\mathbf{b}^{(j)}\delta\eta_{j}(0,v), \label{Frank1}\\
&\left(\delta\theta \mathbf r(u,v)+\theta \delta t  \mathbf v(u,v)\right)\times\mathbf{a}-\sum_{j=1}^{J}\mathbf{b}^{(j)}\delta\eta_{j}(u,v)\nonumber \\
=&\left(\delta\theta \mathbf r(u,0)+\theta \delta t  \mathbf v(u,0))\right)\times\mathbf{a}-\sum_{j=1}^{J}\mathbf{b}^{(j)}\delta\eta_{j}(u,0). \label{Frank22}
\end{flalign}

Notice that the left-hand sides of Eqs.~\eqref{Frank1} and \eqref{Frank22} are equal, whereas the right-hand side of  Eq. \eqref{Frank1} depends only on $v$ and the right-hand side of  Eq. \eqref{Frank22} depends only on $u$. Thus the right-hand sides of Eqs.~\eqref{Frank1} and \eqref{Frank22} must equal to the same constant independent of $u$ and $v$, denoted by $\mathbf C=(c_1,c_2,c_3)$. That is,
\begin{equation}\label{eqn:C}
\left(\delta\theta \mathbf r(u,v)+\theta \delta t  \mathbf v(u,v)\right)\times\mathbf{a}-\sum_{j=1}^{J}\mathbf{b}^{(j)}\delta\eta_{j}(u,v)=\mathbf C.
\end{equation}

We want to solve for the actual velocity $\mathbf v=(v_1,v_2,v_3)$ such that the above vector equation holds. Since $\mathbf a=(0,0,1)$, the first two equations in \eqref{eqn:C} give
\begin{flalign}
v_1 =&-\frac{\delta \theta}{\theta\delta t}(x-c_1)-\frac{1}{\theta}\sum_{j=1}^J  b^{(j)}_2\frac{\delta\eta_{j}}{\delta t},
\label{velocity0}
\\
v_2 =&-\frac{\delta \theta}{\theta\delta t}(y-c_2)+\frac{1}{\theta}\sum_{j=1}^J  b^{(j)}_1\frac{\delta\eta_{j}}{\delta t}.
\label{velocity}
\end{flalign}
Note that in the projection procedure, we essentially adjust the local value of $\theta$ determined by the  Frank's formula in Eq.~\eqref{eqn:mod2frank} to achieve a uniform misorientation angle $\theta$ over the entire grain boundary. This procedure should not lead to additional rigid translation of the grain boundary. The two constants  $c_1$ and $c_2$ in the  projected velocity formula in Eqs.~\eqref{velocity0} and \eqref{velocity}  can be determined by this condition.
For some symmetric configuration of the grain boundary, e.g., when the top point of the grain boundary in the $+z$ direction always has a velocity in the $z$ direction due to some symmetry, we set the $z$ axis passing through that point, i.e., that point is $\mathbf r=(0,0,z)$ during the evolution. In this case, at that point, we have $(\delta\theta \mathbf r+\theta \delta t  \mathbf v)\times\mathbf{a}=\mathbf 0$, and we set  $\eta_j=0$, $j=1,2,\cdots,J$, at that point.
Thus, we have $c_1=c_2=0$.  Eqs.~\eqref{velocity0} and \eqref{velocity} actually hold in the continuum model, i.e.,
$v_1 =-\frac{1}{\theta} \frac{d \theta}{d t}(x-c_1)-\frac{1}{\theta}\sum_{j=1}^J  b^{(j)}_2\frac{d\eta_{j}}{d t}$ and
$v_2 =-\frac{1}{\theta} \frac{d \theta}{d t}(y-c_2)+\frac{1}{\theta}\sum_{j=1}^J  b^{(j)}_1\frac{d\eta_{j}}{d t}$,
by letting $\delta t\rightarrow 0$.

The condition in Eq.~\eqref{eqn:C} does not impose any restriction on the velocity in the direction of the rotation axis, i.e., the $z$ direction. Thus we simply keep the $z$-component $v_3=v^*_3$, where $\mathbf v^*=(v^*_1,v^*_2,v^*_3)$ is the virtual
velocity  in Eq.~\eqref{eqn:virtualv} without the constraint of the Frank's formula.

In summary, combining  with the algorithm to maintain the dislocation continuity presented in previous section,
we have the following efficient numerical algorithm  without calculation of the long-range force:

\vspace{0.1in}
\noindent
\underline{\bf Numerical Algorithm}
\vspace{0.05in}

From $t_n$ to $t_{n+1}=t_n+\delta t$,
\begin{flalign}
\mathbf v^*=&\left(M_{\rm d}\sum_{j=1}^J\frac{\|\nabla_S\eta_j\|}{\sum_{k=1}^J \|\nabla_S\eta_k\| }\mathbf f_{\rm local}^{(j)}\cdot \mathbf n \right) \mathbf n,\label{eqn:nav1}\\
\mathbf r^*=&\mathbf r^n +\mathbf v^* \delta t,\\
\eta^*_{ju}=&\eta^{n}_{ju} -\left.M_{\rm r}\frac{\partial \gamma_{\rm gb}}{\partial \eta_{ju}} \right|_{t_n} \delta t, \ \ \eta^*_{jv}=\eta^{n}_{jv} -\left.M_{\rm r}\frac{\partial\gamma_{\rm gb}}{\partial\eta_{jv}}\right|_{t_n} \delta t, \label{eqn:mod3en}\\
\bigtriangleup \lambda_{j}^{n+1}=&\frac{1}{\delta t}\left(\frac{\partial \eta_{jv}^{*}}{\partial u}- \frac{\partial \eta_{ju}^{*}}{\partial v}\right),\label{eqn:Poisson}\\
  \eta_{ju}^{n+1}=&\eta_{ju}^*+\frac{\partial\lambda_{j}^{n+1}}{\partial v}\delta t, \ \ \eta_{jv}^{n+1}=\eta_{jv}^*-\frac{\partial\lambda_{j}^{n+1}}{\partial u}\delta t,\\
\delta \theta=&\theta(\mathbf r^{*}_u ,\mathbf r^{*}_v,\eta^{n+1}_{ju},\eta^{n+1}_{jv}  )-\theta(\mathbf r^{n}_u ,\mathbf r^{n}_v,\eta^{n}_{ju},\eta^{n}_{jv}),\label{eqn:deltatheta3}\\
\mathbf v=&\left(-\frac{\delta \theta}{\theta \delta t}(x-c_1),-\frac{\delta \theta}{\theta \delta t}(y-c_2),  v^*_3\right)\nonumber\\
&+\left(-\frac{1}{\theta}\sum_{j=1}^J  b^{(j)}_2\frac{\delta \eta_{j}}{\delta t},\frac{1}{\theta}\sum_{j=1}^J  b^{(j)}_1\frac{\delta \eta_{j}}{\delta t},0\right), \label{eqn:fv2}\\
\mathbf r^{n+1}=&\mathbf r^n +\mathbf v\rm \delta t.\label{eqn:nav2}
\end{flalign}
Here  constanta $c_1$ and $c_2$ can be determined by the condition that the projection procedure alone does not lead to extra rigid translation of the grain boundary as discussed above.

Note that in the projected velocity formula in Eq.~\eqref{eqn:fv2},  the first term describes the pure coupling motion of the grain boundary, the second term describes the additional effect of the sliding motion  of the grain boundary due to dislocation reaction.

\vspace{0.1in}
\noindent
\underline{\bf Initial dislocation structure}
\vspace{0.05in}

We assume that the initial grain boundary has an
equilibrium dislocation structure that satisfies the Frank's formula and has the lowest energy. See Ref.~\cite{Qin2020} for  the method based on constrained energy minimization to find the equilibrium dislocation structure on a curved low angle grain boundary, which is a generalization of the model for planar low angle grain boundaries \cite{zhang2017energy} examined extensively by comparisons with atomistic simulation results.

\section{Analysis of the continuum simulation method}\label{sec:analysis}

In this section, we summarize some analysis results on the derivation and properties of the continuum model and numerical method.

A simplified form of the local grain boundary energy density in Eq.~\eqref{eqn:gamma_gb}, neglecting the orientation-dependent factor, is $\tilde\gamma_{\rm gb}(\|\nabla_S \eta_j\|)=-\|\nabla_S \eta_j\|\log\|\nabla_S \eta_j\| $, which is a concave function
 of $\|\nabla_S \eta_j\|$. In fact,  $\tilde\gamma'_{\rm gb}(\|\nabla_S \eta_j\|)=-\log\|\nabla_S \eta_j\| -1$ and $\tilde\gamma''_{\rm gb}(\|\nabla_S \eta_j\|)= -\frac{1}{\|\nabla_S \eta_j\|}<0$. We have pointed out in Sec.~\ref{sec:illposed} that such an energy functional will lead to illposed gradient flow. We prove this  rigorously in the following theorem.

\begin{theorem}
Consider the energy
\begin{equation}
E=\int_S f(\|\nabla_S \eta\|) \mathrm{d}S,
\end{equation}
where $\eta$ is a smooth function defined on the surface $S$ and $f$ is a smooth concave function, i.e., $f''<0$. The gradient flow due to this energy is
\begin{flalign}
\frac{\partial \eta}{\partial t}=M_{\eta}&\left[f''(\|\nabla_S \eta\|)  \left(\frac{\nabla_S \eta}{\|\nabla_S \eta\|}\right)^T(\nabla_S\nabla_S \eta)\frac{\nabla_S \eta}{\|\nabla_S \eta\|}\right.\nonumber\\
& \left.+ f^{'}(\|\nabla_S \eta\|) \nabla_S\cdot\left(\frac{\nabla_S \eta}{\|\nabla_S \eta\|}\right)\right],\label{eqn:gra_flow_eta}
\end{flalign}
where $\nabla_S\nabla_S \eta$ is the Hessian of $\eta$ and mobility $M_\eta>0$. This gradient flow equation is illposed.
\end{theorem}

\begin{proof}
Consider the energy $E$ due to $\eta$  with a small perturbation $\delta \eta$. The energy change is
\begin{flalign}
\delta E=&E[\eta+\delta \eta]-E[\eta]\nonumber\\
=&\int_S f' (\|\nabla_S \eta\|)\frac{\nabla_S \eta\cdot \nabla_S \delta \eta}{\|\nabla_S \eta\|}dS\nonumber\\
=&-\int_S \nabla_S \cdot \left(f' (\|\nabla_S \eta\|)\frac{\nabla_S \eta}{\|\nabla_S \eta\|}\right)\delta \eta dS.
\end{flalign}
Thus
\begin{flalign}
\frac{\delta E}{\delta \eta}=&-\nabla_S \cdot \left(f' (\|\nabla_S \eta\|)\frac{\nabla_S \eta}{\|\nabla_S \eta\|}\right)\nonumber\\
=&-\nabla_S  f' (\|\nabla_S \eta\|)\cdot \frac{\nabla_S \eta}{\|\nabla_S \eta\|}-f' (\|\nabla_S \eta\|)\nabla_S \cdot \left(\frac{\nabla_S \eta}{\|\nabla_S \eta\|}\right)\nonumber\\
=&-f''(\|\nabla_S \eta\|)  \left(\frac{\nabla_S \eta}{\|\nabla_S \eta\|}\right)^T(\nabla_S\nabla_S \eta)\frac{\nabla_S \eta}{\|\nabla_S \eta\|} \nonumber\\
&-f' (\|\nabla_S \eta\|)\nabla_S \cdot \left(\frac{\nabla_S \eta}{\|\nabla_S \eta\|}\right).
\end{flalign}

The gradient flow $\frac{\partial \eta}{\partial t}=-M_{\eta}\frac{\delta E}{\delta \eta}$, where $M_\eta>0$, gives Eq.~\eqref{eqn:gra_flow_eta}. Recall that $f''(\|\nabla_S \eta\|)<0$ in this equation.

We show that the evolution equation \eqref{eqn:gra_flow_eta} is illposed by
proof by contradiction.

Assume that the grain boundary $S$ is expressed by $\mathbf r(u,v)$, where $(u,v)$ is an orthogonal parametrization. When $\eta$ depends only on the parameter $u$ and $\eta_u>0$, Eq.~\eqref{eqn:gra_flow_eta} is reduced to the one-dimensional equation
\begin{equation}\label{eqn:ill-1D}
\eta_t=f''(|\eta_u|)\eta_{uu}.
\end{equation}
Here without loss of generality, we let $M_{\eta}=1$. Since $f''<0$, this equation is a backward diffusion equation with variable coefficient.

 We consider $C^2$ solution of the initial value problem with periodic boundary condition in $u$, and without loss of generality, let the period be $2\pi$. Suppose that Eq.~\eqref{eqn:ill-1D} is wellposed for time $t\in[0,T]$.
    There exists a constant $M>0$, such that for any two solutions $\eta^I$ and $\eta^{II}$  of  Eq.~\eqref{eqn:ill-1D} with different initial conditions, we have
\begin{equation}\label{eqn:wellposed}
\|\eta^I(\cdot,t)-\eta^{II}(\cdot,t)\|_{C^2}\leq M\|\eta^I(\cdot,0)-\eta^{II}(\cdot,0)\|_{C^2}.
\end{equation}

Consider two solutions with initial conditions $\eta^I(u,0)=pu$ and
$\eta^{II}(u,0)=pu+\frac{\varepsilon}{k^2} \exp^{iku}$, where $p>0$ is a constant, $k\geq1$, and $\varepsilon$ is small. We have %$\|\eta^I(\cdot,0)-\eta^{II}(\cdot,0)\|_{C^2}=\varepsilon$.
\begin{equation}\label{eqn:contradic2}
\|\eta^I(\cdot,0)-\eta^{II}(\cdot,0)\|_{C^2}=\varepsilon.
\end{equation}
%Using the wellposed condition in Eq.~\eqref{eqn:wellposed}, the equation for $\eta^{II}$ can be simplified as

We write Eq.~\eqref{eqn:ill-1D} as
\begin{equation}\label{eqn:ill-linear}
\eta_t=f''(p)\eta_{uu}+g(u,t),
\end{equation}
where $g(u,t)=(f''(|\eta_u|)-f''(p))\eta_{uu}$. %=O(\varepsilon^2)$ in terms of $C^2$ norm.
Note that $p+\varepsilon\geq \eta^{II}_u(u,0)\geq p-\varepsilon$. We choose $\varepsilon$ to be small such that $p-M\varepsilon>0$. Using the wellposedness condition in Eq.~\eqref{eqn:wellposed}, we have, for $t\in [0,T]$,
\begin{equation}\label{eqn:eta-bounds}
p+M\varepsilon\geq \eta^{II}_u(u,t)\geq p-M\varepsilon>0.
\end{equation}

Consider Fourier transform of these functions, i.e., $\eta^{I,II}(u,t)=pu+\sum_k A_k^{I,II}(t)\exp^{iku}$ and $g(u,t)=\sum_k g_k(t)\exp^{iku}$. Here $A_k^{I}(t)=0$ for all $k$. Using definition of Fourier transform and the wellposedness condition in Eq.~\eqref{eqn:wellposed}, we have
\begin{equation}\label{eqn:well-k}
|A^I_k(t)-A^{II}_k(t)|\leq \|\eta^I(\cdot,t)-\eta^{II}(\cdot,t)\|_{C^0} \leq M \|\eta^I(\cdot,0)-\eta^{II}(\cdot,0)\|_{C^2}.
\end{equation}

 Since $\eta^{II}$ is a solution of Eq.~\eqref{eqn:ill-linear}, the Fourier coefficient  $A^{II}_k(t)$ of $\eta^{II}$ satisfies
\begin{equation}\label{eqn:ode-eta2}
{A^{II}_k}'(t)=-k^2f''(p)A^{II}_k(t)+g_k(t).
\end{equation}
where $g(u,t)=(f''(|\eta^{II}_u|)-f''(p))\eta^{II}_{uu}$.
Using the initial condition ${A^{II}_k}(0)=\frac{\varepsilon}{k^2}$, the solution of Eq.~\eqref{eqn:ode-eta2} is
\begin{equation}\label{eqn:AII}
A^{II}_k(t)=\frac{\varepsilon}{k^2} e^{-k^2f''(p)t}+e^{-k^2f''(p)t}\int_0^tg_k(\tau)e^{k^2f''(p)\tau} d\tau.
\end{equation}

Now consider $g(u,t)=(f''(|\eta^{II}_u|)-f''(p))\eta^{II}_{uu}$. First,
$f''(|\eta^{II}_u|)-f''(p))=f'''(\xi)(\eta^{II}_u-p)$, where $\xi$ is between $\eta^{II}_u$ and $p$. Using the bounds in Eq.~\eqref{eqn:eta-bounds}, we have $|\eta^{II}_u-p|\leq M\varepsilon $,
and $|f'''(\xi)|\leq C_p$, where $C_p$ is a constant depending on $p$.
Moreover, since $\eta^{II}_{uu}=\eta^{II}_{uu}-\eta^{I}_{uu}$,
 we have $|\eta^{II}_{uu}|=|\eta^{II}_{uu}-\eta^{I}_{uu}|\leq M \|\eta^I(\cdot,0)-\eta^{II}(\cdot,0)\|_{C^2}=M\varepsilon $. Using these results, we have
\begin{flalign}
|g_k(t)|\leq &\max_u|g(u,t)| \leq C_pM^2\varepsilon^2.
\end{flalign}
Thus, the integral in second term in $A^{II}_k(t)$ in Eq.~\eqref{eqn:AII} can be bounded as $|\int_0^tg_k(\tau)e^{k^2f''(p)\tau} d\tau|
%\leq e^{-k^2f''(p)t}\int_0^t|g_k(\tau)|e^{k^2f''(p)\tau} d\tau
\leq \int_0^t\varepsilon^2 C_pM^2e^{k^2f''(p)\tau} d\tau=\frac{\varepsilon^2 C_pM^2}{k^2f''(p)}\left(e^{k^2f''(p)t}-1\right)$. Therefore, we have
\begin{equation}\label{eqn:contradict1}
|A^I_k(t)-A^{II}_k(t)|=|A^{II}_k(t)|\geq \frac{\varepsilon}{k^2} e^{-k^2f''(p)t}\left[1-\frac{C_pM^2}{-f''(p)}\varepsilon\left(1-e^{k^2f''(p)t}\right)\right].
\end{equation}
We choose $\varepsilon$ to be small enough such that $1-\frac{C_pM^2}{-f''(p)}\varepsilon>0$.

Since $f''(p)<0$, Eqs.~\eqref{eqn:contradict1} and \eqref{eqn:contradic2} contradict with Eq.~\eqref{eqn:well-k}
as $k\rightarrow \infty$.

\end{proof}

\begin{remark}
 With the actual local energy density in Eq.~\eqref{eqn:gamma_gb}, we have
$\frac{\delta E_{\rm local}}{\delta \eta_j}= \mathbf{f}^{(j)}_{\mathrm{local}}\cdot\frac{\nabla_S \eta_j}{\|\nabla_S \eta_j\|}$, where $\mathbf{f}^{(j)}_{\mathrm{local}}$ is given by Eq.~\eqref{eqn:flt}. Illposedness of the gradient flow $\frac{\partial \eta_j}{\partial t}=-M_{\eta}\frac{\delta E}{\delta \eta_j}$ associated with this energy formula can be proved similarly.
\end{remark}

\begin{remark}
Illposedness of the gradient flow comes from the Hessian term in Eq.~\eqref{eqn:gra_flow_eta},  or in Eq.~\eqref{eqn:flt} for the actual evolution equation. In the previous models \cite{Zhu-Xiang2012,zhu2014continuum}, this term was removed when the driving force is dominated by the long-range force (Eqs.~\eqref{eqn:c3flong} and \eqref{eqn:lsig}). However, in the grain boundary dynamics problem, the long-range force is essentially canceled during the evolution, and this Hessian term due the local energy plays critical roles and cannot be removed from the equation.
\end{remark}

Next, in the following theorem, we show existence and uniqueness of the projection operation used in Sec.~\ref{sec:cm}. This theorem plays the same role as the Helmholtz–Hodge decomposition theorem in  the projection method for solving fluid dynamics problems \cite{chorin1990mathematical}.

\begin{theorem}
Given a smooth vector function $\bm{\zeta}=(\zeta_1,\zeta_2)$ in a periodic $(u,v)$ domain, there exist a unique periodic vector function $\bm{\zeta}^{\tilde{ D}}=(\zeta^{\tilde{ D}}_1,\zeta^{\tilde{ D}}_2)$ and a periodic function $\lambda$ such that
	\begin{equation}
	\bm{\zeta}=\bm{\zeta}^{\tilde { D}}+\tilde{\mathbf{G}}\lambda,
	\end{equation}
where
\begin{equation}\label{eqn:projectionproof1}
	\tilde{{D}}\bm{\zeta}^{\tilde { D}}=\frac{\partial \zeta^{\tilde { D}}_1}{\partial v}-\frac{\partial \zeta^{\tilde { D}}_2}{\partial u}=0,
	\end{equation}
and
	\begin{equation}
	\tilde{\mathbf{G}}\lambda=\left(\frac{\partial \lambda}{\partial v},-\frac{\partial \lambda}{\partial u} \right).
	\end{equation}	
\end{theorem}

\begin{proof}	
	We first prove existence of $ \bm{\zeta}^{\tilde { D}} $. If $ \bm{\zeta}=\bm{\zeta}^{\tilde { D}}+\tilde{\mathbf{G}}\lambda$ holds, we have $\tilde { D} \bm{\zeta}= \bigtriangleup \lambda$, where $\bigtriangleup$ is the Laplacian operator. Under periodic boundary condition, the solution $\lambda$ is unique up to addition of a constant. With the solved $\lambda$, we can define $\bm{\zeta}^{\tilde { D}}=\bm{\zeta}-\tilde{\mathbf{G}}\lambda $.	

	Now we proof uniqueness of $ \bm{\zeta}^{\tilde { D}} $.  If $ \bm{\zeta}^{\tilde { D}} $ exists,
  we have
  \begin{equation*}
  <\bm{\zeta}^{\tilde { D}}, \tilde{\mathbf{G}}\lambda >=-<\tilde { D}\bm{\zeta}^{\tilde { D}}, \lambda > =0,
  \end{equation*}
  where the inner product $<f,g>=\int_D fg \, \mathrm{d}u \mathrm{d}v$ with $D$ being the periodic domain.
  This gives $\|\bm{\zeta}\|^2= \|\bm{\zeta}^{\tilde { D}}\|^2+\|\tilde{\mathbf{G}}\lambda\|^2$, where $\|\cdot\|$ is the $L_2$-norm over $D$. Thus, we have $\bm{\zeta}^{\tilde { D}}=\mathbf 0 $ when $\bm{\zeta}=\mathbf 0 $, from which  uniqueness of $ \bm{\zeta}^{\tilde { D}} $ follows.
\end{proof}

\begin{remark}
From the proof of Theorem 2, we have $\|\bm{\zeta}\|^2= \|\bm{\zeta}^{\tilde { D}}\|^2+\|\tilde{\mathbf{G}}\lambda\|^2$ and $<\bm{\zeta}^{\tilde { D}}, \tilde{\mathbf{G}}\lambda >=0$.
These mean that $\bm{\zeta}^{\tilde{ D}}$ is the projection of $\bm{\zeta}$ that satisfies Eq.~\eqref{eqn:projectionproof1}.
\end{remark}

\begin{remark}
In the second step in the projection method used in Sec.~\ref{sec:cm}, i.e., Eq.~\eqref{ddd}, we project the result $(\eta^*_{ju},\eta^*_{jv})$ in Eqs.~\eqref{eqn:dp01} and \eqref{eqn:dp1} obtained in  the first step without constraint, to the result $(\eta^{n+1}_{ju},\eta^{n+1}_{jv})$ that satisfies the constraint $\tilde{{D}}(\eta^{n+1}_{ju},\eta^{n+1}_{jv})=\frac{\eta^{n+1}_{ju}}{\partial v}-\frac{\eta^{n+1}_{jv}}{\partial u} =0$, and $-\lambda^{n+1}_j\delta t$ in Eq.~\eqref{ddd} is the function $\lambda$ in Theorem 2.
\end{remark}

\section{Numerical simulations}\label{sec:nr}
In this section, we perform numerical simulations of grain boundary dynamics using our numerical algorithm in Eqs.~\eqref{eqn:nav1}-\eqref{eqn:nav2}, which is a numerical implementation of the continuum model of constrained evolution in Eqs.~\eqref{eqn:mod2v}--\eqref{eqn:mod2frank}. The numerical simulation results are compared extensively with those obtained by atomistic-level simulations using phase field crystal model~\cite{yamanaka2017phase} and amplitude expansion phase field crystal model~\cite{salvalaglio2018defects} for various properties of coupling and sliding motions of the grain boundary to validate our continuum model. Convergence tests show that the proposed continuum simulation algorithm indeed fixes the problem of illposedness and that the projection algorithms converge.

%\subsection{fcc spherical grain boundaries with rotation axis [111]}\label{sec:fcc}
We consider grain boundaries in fcc Al. We choose the directions $[\bar{1}10]$, $[\bar{1}\bar{1}2]$, $[111]$ to be the $x$, $y$ and $z$ directions, respectively. In this coordinate system, the six Burgers vectors are $\mathbf{b}^{(1)}=(1,0,0)b$, $\mathbf{b}^{(2)}=\left(\frac{1}{2},\frac{\sqrt{3}}{2},0\right)b$, $\mathbf{b}^{(3)}=\left(\frac{1}{2},-\frac{\sqrt{3}}{2},0\right)b$,
$\mathbf{b}^{(4)}=\left(0,\frac{\sqrt{3}}{3},-\frac{\sqrt{6}}{3}\right)b$,
$\mathbf{b}^{(5)}=\left(\frac{1}{2},\frac{\sqrt{3}}{6},\frac{\sqrt{6}}{3}\right)b$, and
$\mathbf{b}^{(6)}=\left(-\frac{1}{2},\frac{\sqrt{3}}{6},\frac{\sqrt{6}}{3}\right)b$, where $b$ is the magnitude of the Burgers vectors. In Al, $b=0.286\rm nm$ and the Poisson ratio is $\nu=0.347$. The rotation axis $\mathbf a$ is in the $[111]$ direction, i.e., $+z$ direction.

We study the evolution of an initially spherical grain boundary, whose radius is $R=20b$ and misorientation angle is $\theta = 5^\circ $.
There are three sets of dislocations with Burgers vectors $\mathbf b^{(1)}$, $\mathbf{b}^{(2)}$, and $\mathbf{b}^{(3)}$, respectively, in the equilibrium dislocation structure on this initial, spherical grain boundary; see the top image in Fig.~\ref{fig:fccfigure0}(a).

In the dynamics simulation, the grain boundary is parameterized using spherical coordinates $R=R(\alpha,\beta)$, for $0\leq \alpha < 2\pi$ and $0\leq \beta\leq \pi$. Here $\alpha$ is the angle between the position vector of a point on the grain boundary and the $x$ axis, and $\beta$ is the angle between the position vector of the point  and the $z$ axis.
Initially, $R(\alpha,\beta)=20b$. The $(\alpha,\beta)$ domain is discretized into $40\times20$ uniform grids during the evolution. The center of the spherical grain boundary is the origin $(0,0,0)$ in the coordinate system. Due to symmetry, the two constants $c_1=c_2=0$ in the projected velocity formula in Eq.~\eqref{eqn:fv2}.  Simulation of one example  took less than two minutes on a laptop with a single i7-6500u processor.

\subsection{Convergence of the numerical algorithm}

We perform convergence tests in time for our numerical algorithm to show that the problem of illposedness has been fixed and the projection algorithms converge.

\begin{figure}[htbp]
	\centering
	\subfigure[]{\includegraphics[width=0.45\linewidth]{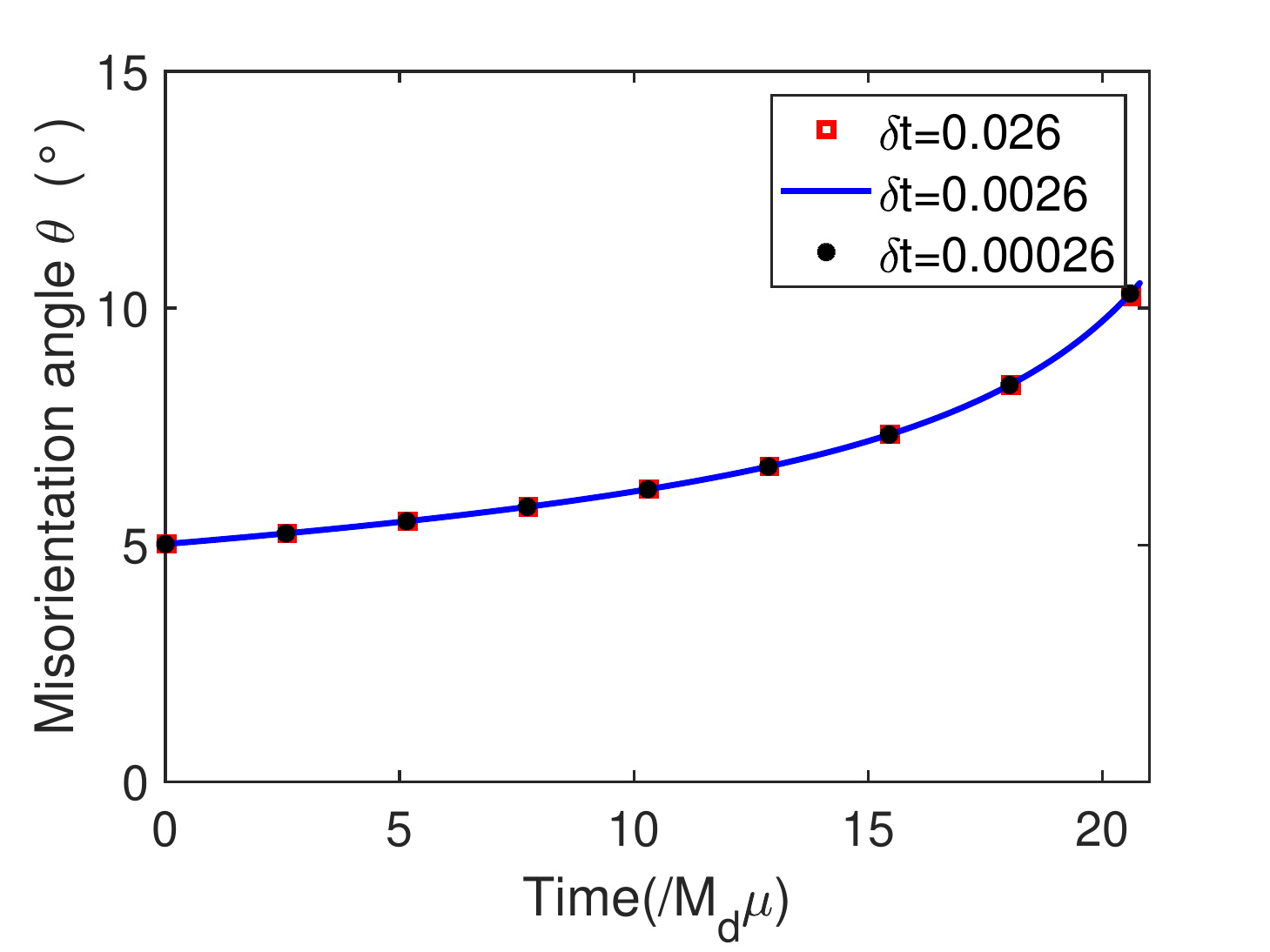}}
	\subfigure[]{\includegraphics[width=0.45\linewidth]{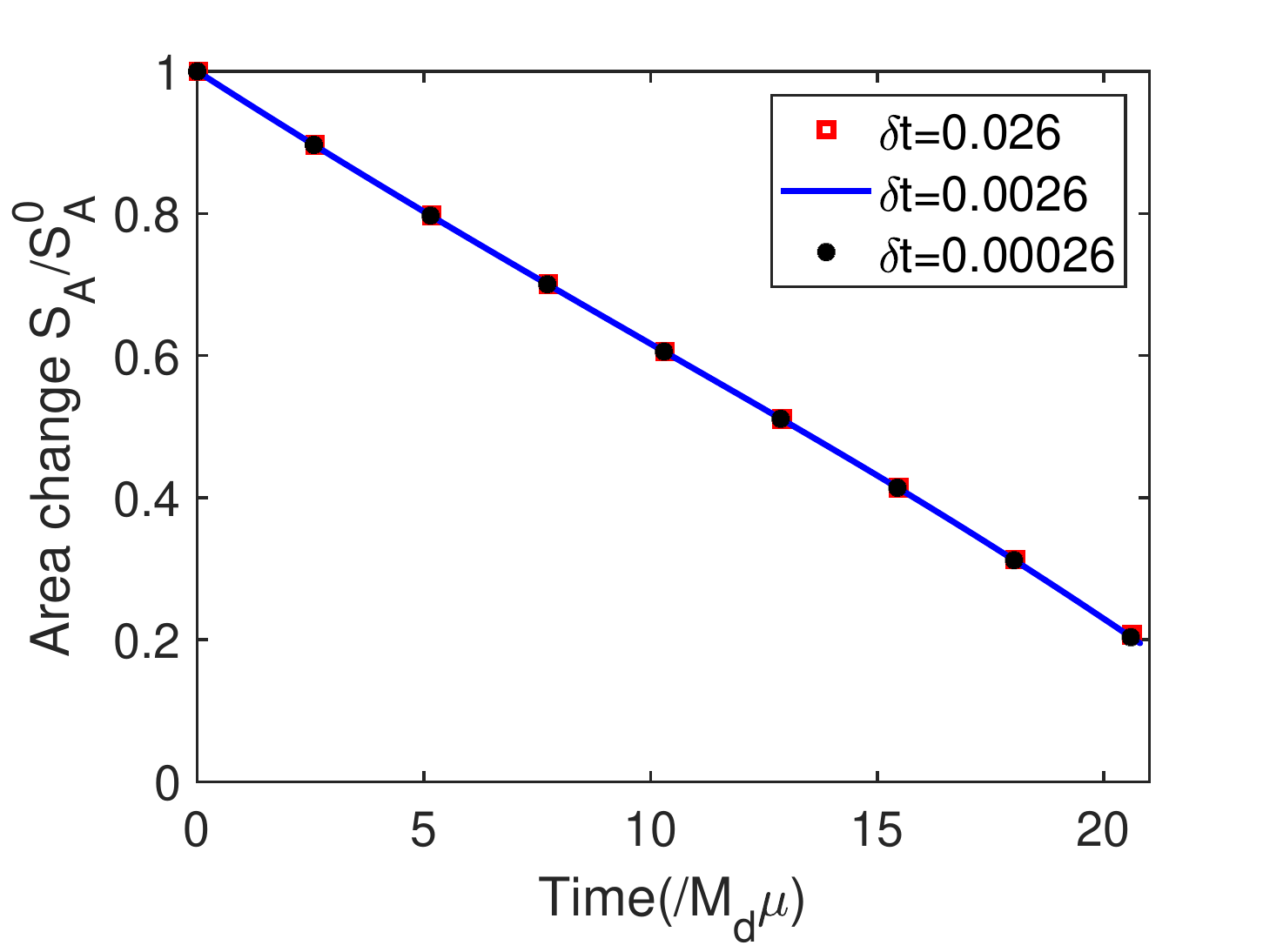}}
	\caption{ Evolutions of (a) misorientation angle $\theta$  and (b) surface area  $S_A$ of the grain boundary during the evolution with different values of time step $\delta t$.
 }\label{fig:convergencedt}
\end{figure}

We examine the misorientation angle $\theta$  and the surface area  $S_A$ of the grain boundary during the evolution up to time $t=20.8 /M_d\mu$ with different values of time step $\delta t$.
Recall that the misorientation angle $\theta$ is calculated using Eq.~\eqref{eqn:mod1t} and surface area $S_A=\int_S\|\mathbf r_u\times\mathbf r_v\|\mathrm{d}u\mathrm{d}v$. The surface area of the initial grain boundary is denoted as $S_A^0$.  Evolutions of these two quantities are shown in Fig.~\ref{fig:convergencedt}, from which convergence can be seen with  different values of time step $\delta t$.

\begin{table}[htbp]
    \caption{\label{tab:table-1}         \protect\\
Misorientation angle $\theta$ and surface area $S_A$ at time $t=20.8 /M_d\mu$.}
\centering
\begin{tabular}{|c|c|c|c|c|c}
\hline
%\diagbox{$dt(/M_d\mu)$}{b}& Misorientation angle $ \theta (^\circ)$&Area change $(S/S_0)$\\
$\delta t (/M_d\mu)$& Misorientation Angle $\theta (^\circ)$  & $Q_\theta$ & Surface Area $(S_A/S_A^0)$ & $Q_{S_A}$ \\
\hline
$0.416000$&9.18428427 & 1.6214   &0.25978820 & 1.7900\\
\hline
$0.208000$&9.78932770 & 2.0195     &0.22902276 & 2.0263\\
\hline
$0.104000$&10.16249512 &1.9196 &0.21183550 & 1.9543\\
\hline
   $0.052000$&10.34727401 &1.9422 &0.20335333 &1.9691 \\
\hline
$0.026000$&10.44353091 &  1.9794 &0.19901303 &  1.9929 \\
\hline
$0.013000$&10.49309176 & &0.19680884 & \\
\hline
$0.006500$&10.51813002 &  &0.19570282 & \\
\hline
%$0.003250$&10.53079239 & 1.9774 &0.19514699 & 1.9899\\
%   \hline
%$0.001625$& 10.53720951 & 1.9732 &0.19486724 & 1.9869\\
%\hline
\end{tabular}
\end{table}

We further examine the orders of convergence of $\theta$ and $S_A$, and the results are shown in Table~\ref{tab:table-1}.  The ratio $Q_g=\frac{g_{\delta t}-g_{\frac{\delta t}{2}}}{g_{\frac{\delta t}{2}}-g_{\frac{\delta t}{4}}}$, where $g_{\delta t}$ is the numerical value of $g$ at time $\delta t$.
These results show a first order convergence of the numerical algorithm.
 These validate our numerical algorithm, and especially,  there is no numerical instability and the  projection algorithms that we employ converge.

%\begin{table}[htbp]
%    \caption{\label{tab:table-1}         \protect\\
%	Misorientation angle $\theta$ and the area ratio $S/S_0$ at  $t=20.8 /M_d\mu$.}
%	\centering
%	\begin{tabular}{|c|c|c|}
%		\hline
%		%\diagbox{$dt(/M_d\mu)$}{b}& Misorientation angle $ \theta (^\circ)$&Area change $(S/S_0)$\\
%		$dt(/M_d\mu)$& Misorientation angle $ \theta (^\circ)$&Area ratio $(S/S_0)$\\
%		\hline
%		$0.02600$&10.4435&0.1990\\
%		\hline
%		$0.01300$&10.4931&0.1968\\
%		\hline
%		$0.00260$&10.5334&0.1950\\
%	    \hline
%		$0.00130$&10.5385 &0.1948\\
%		\hline
%		$0.00026$&10.5427&0.1946\\
%		\hline
%	\end{tabular}
%\end{table}

\subsection{Pure coupling motion} \label{subsec:coup}
We first consider the grain boundary motion without dislocation reaction, i.e. the reaction mobility $M_{\rm r}=0$ in Eq.~\eqref{eqn:mod3en}, and accordingly $\delta \eta_j=0$ in Eq.~\eqref{eqn:fv2}. This is the pure coupling motion.

\begin{figure}[htbp]
	\centering
	\centering
	\subfigure[]{
		\begin{minipage}{0.21\linewidth}
			\includegraphics[width=\linewidth]{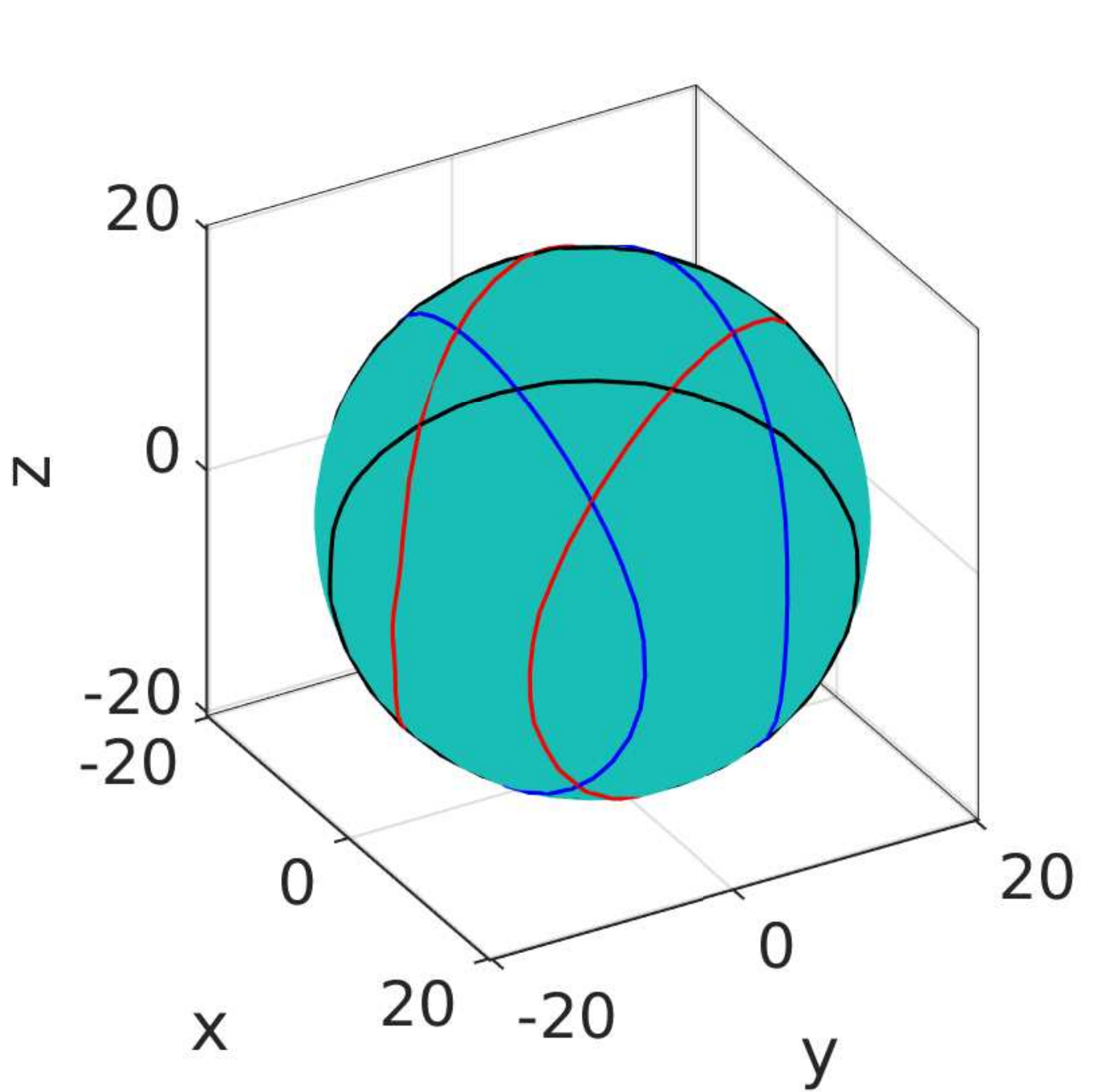}
			\includegraphics[width=\linewidth]{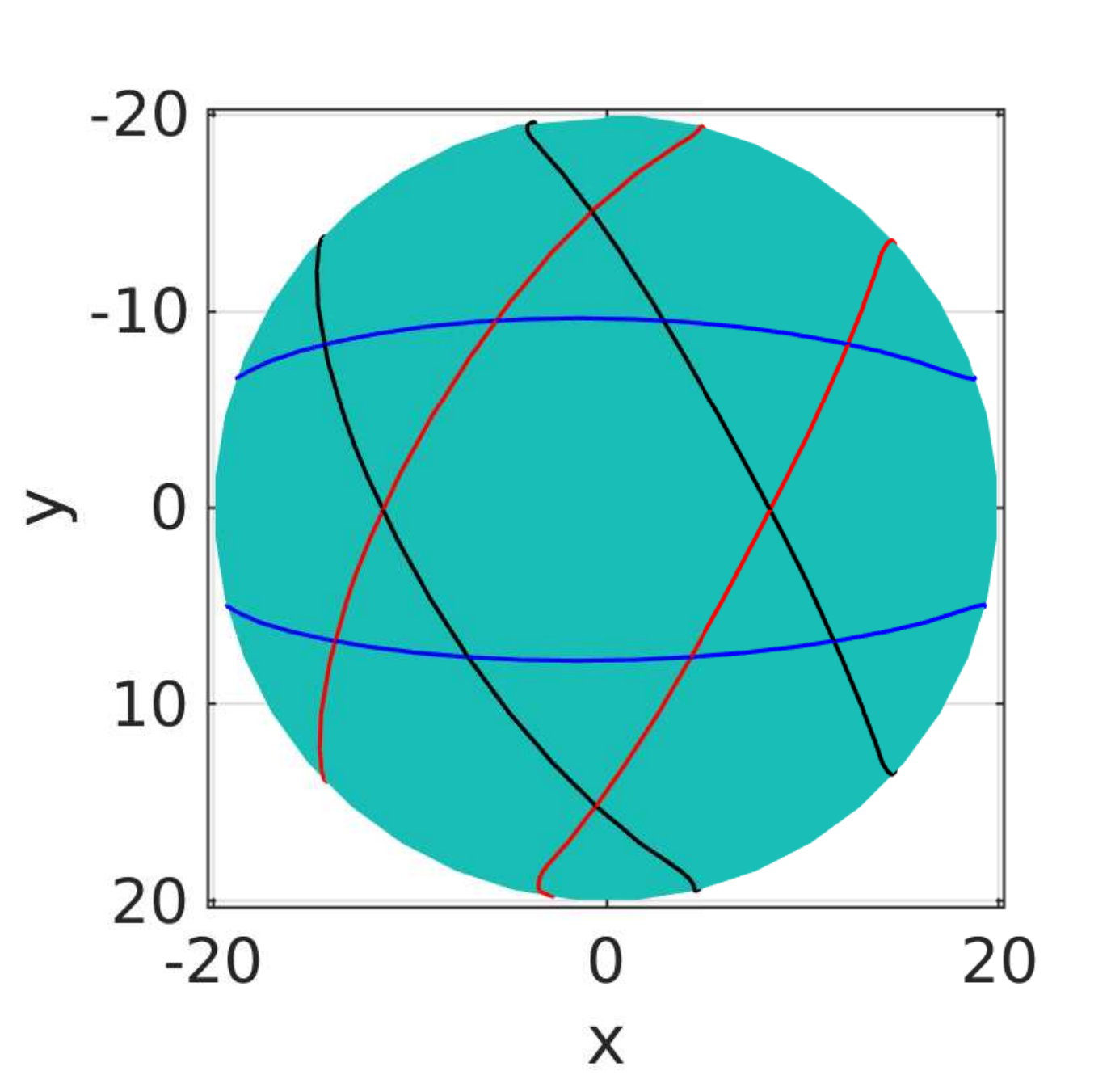}
			\includegraphics[width=\linewidth]{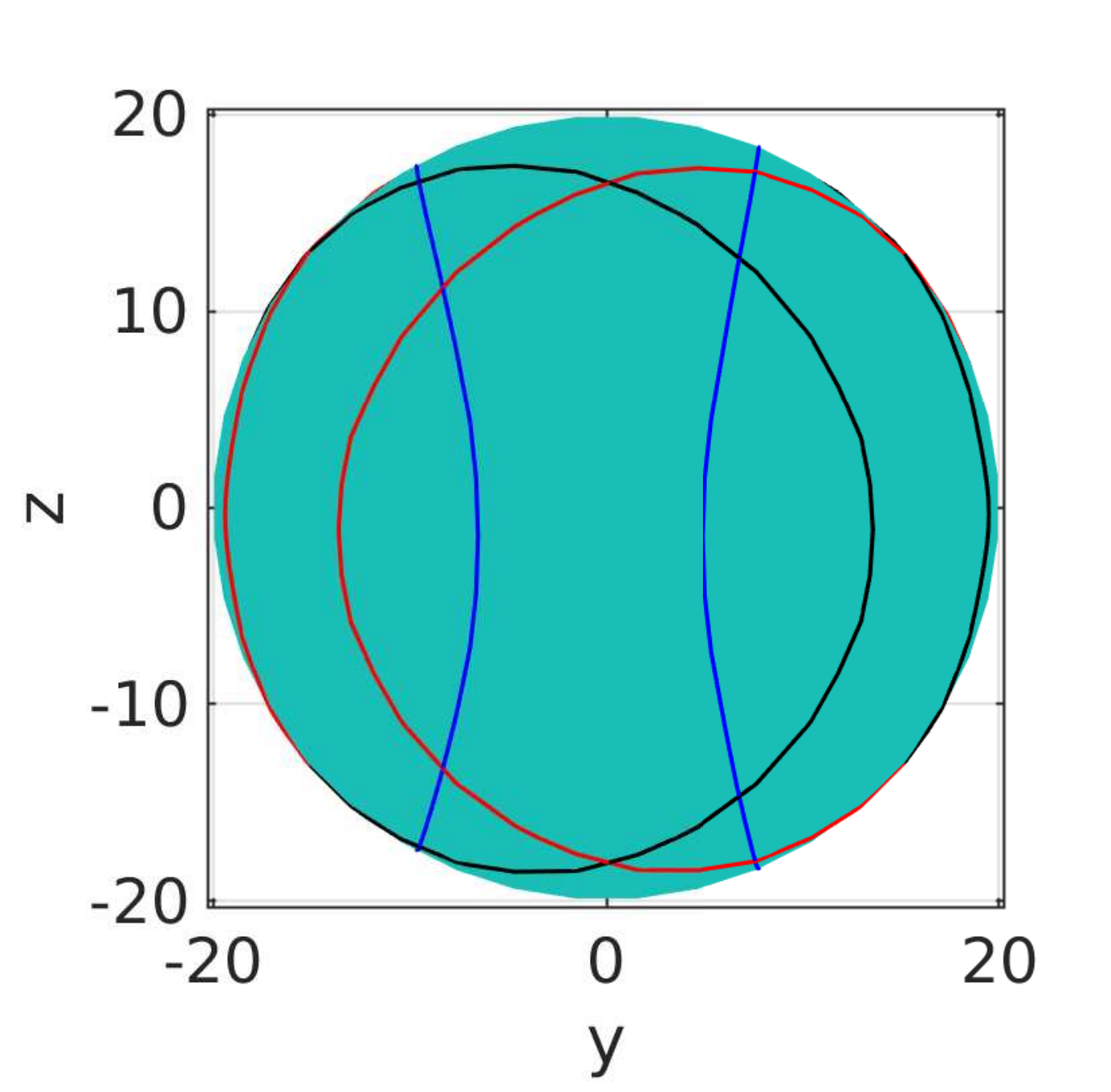}
		\end{minipage}
	}
	\subfigure[]{
		\begin{minipage}{0.21\linewidth}			
			\includegraphics[width=\linewidth]{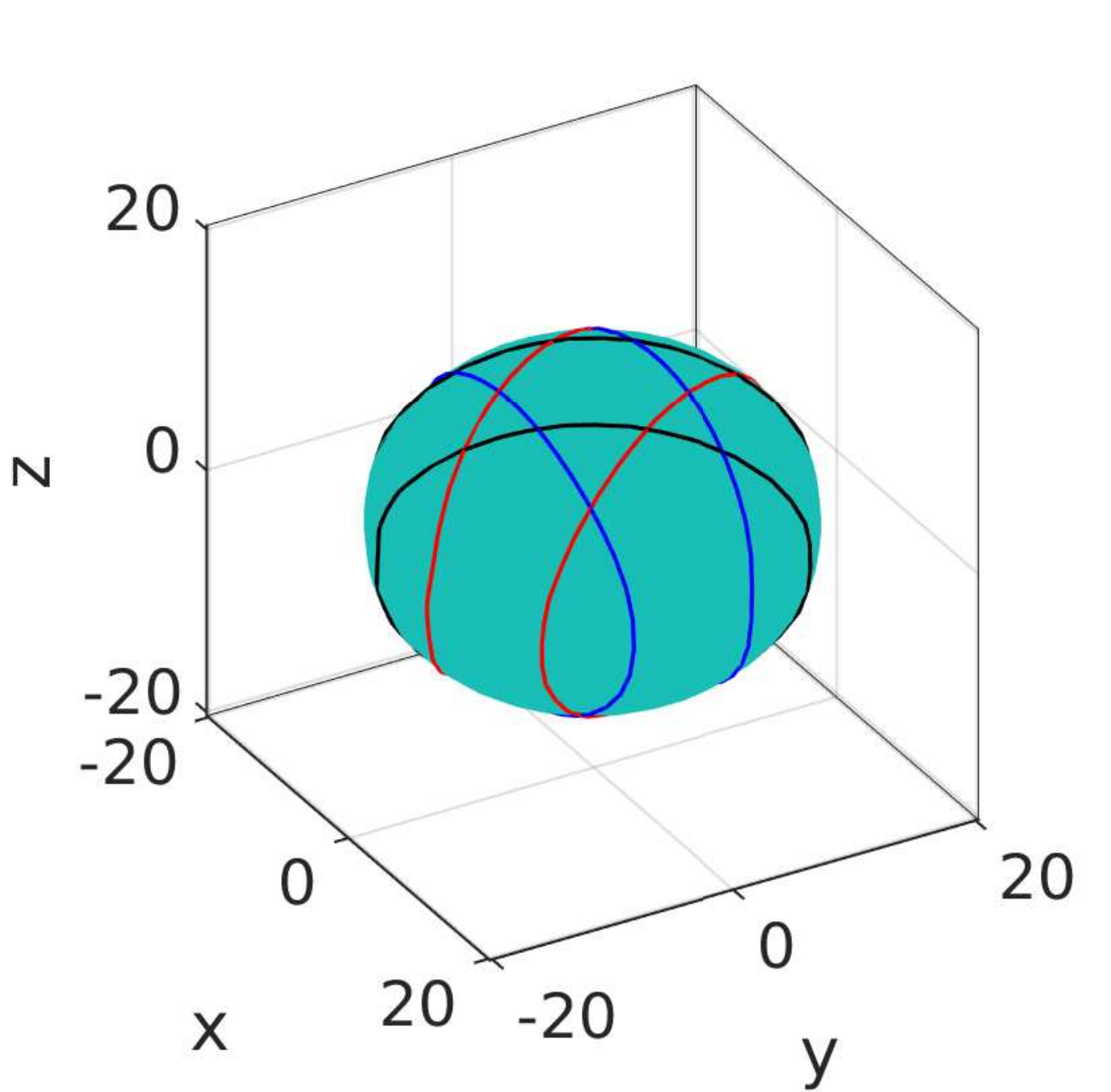}
			\includegraphics[width=\linewidth]{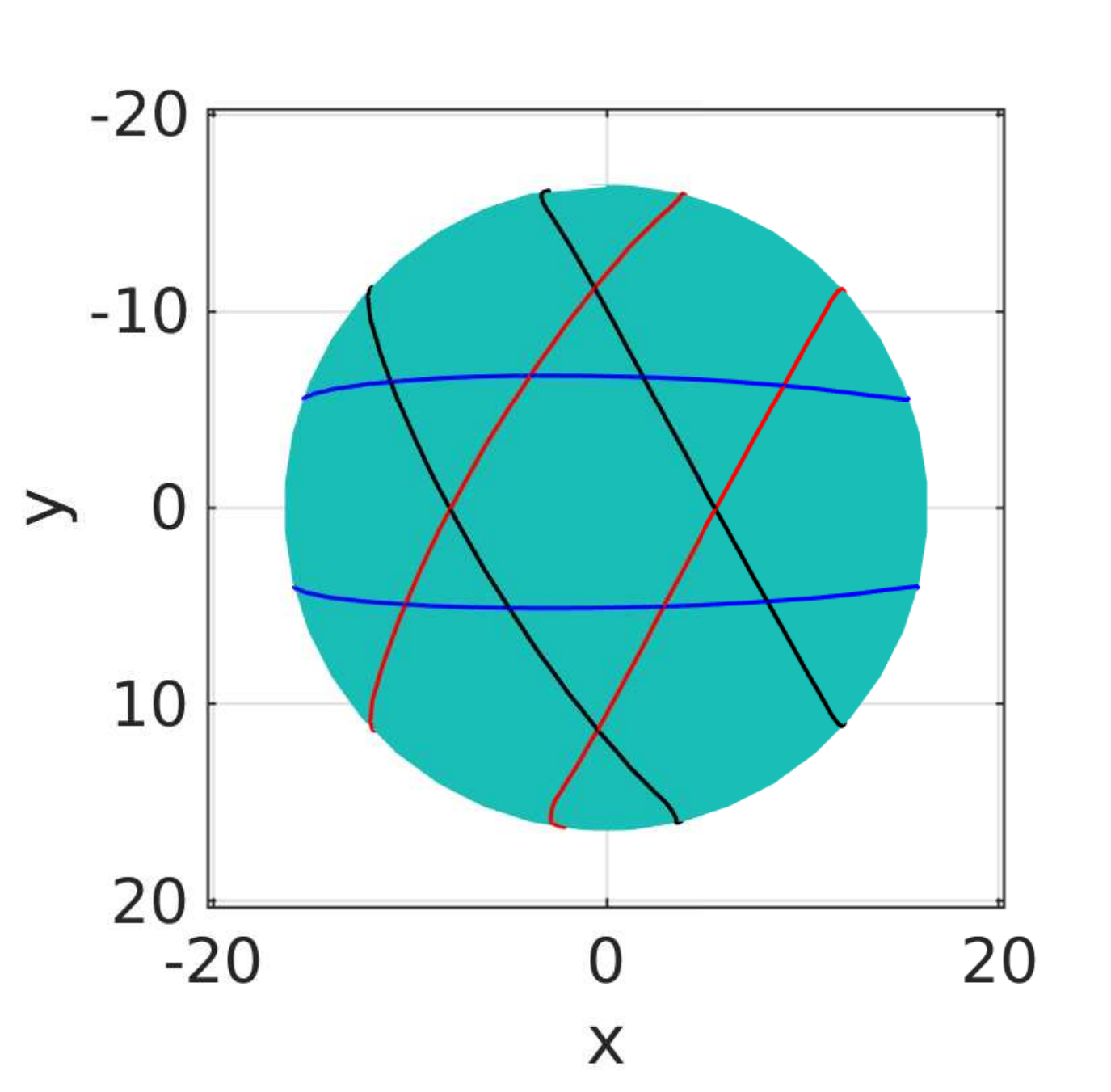}
			\includegraphics[width=\linewidth]{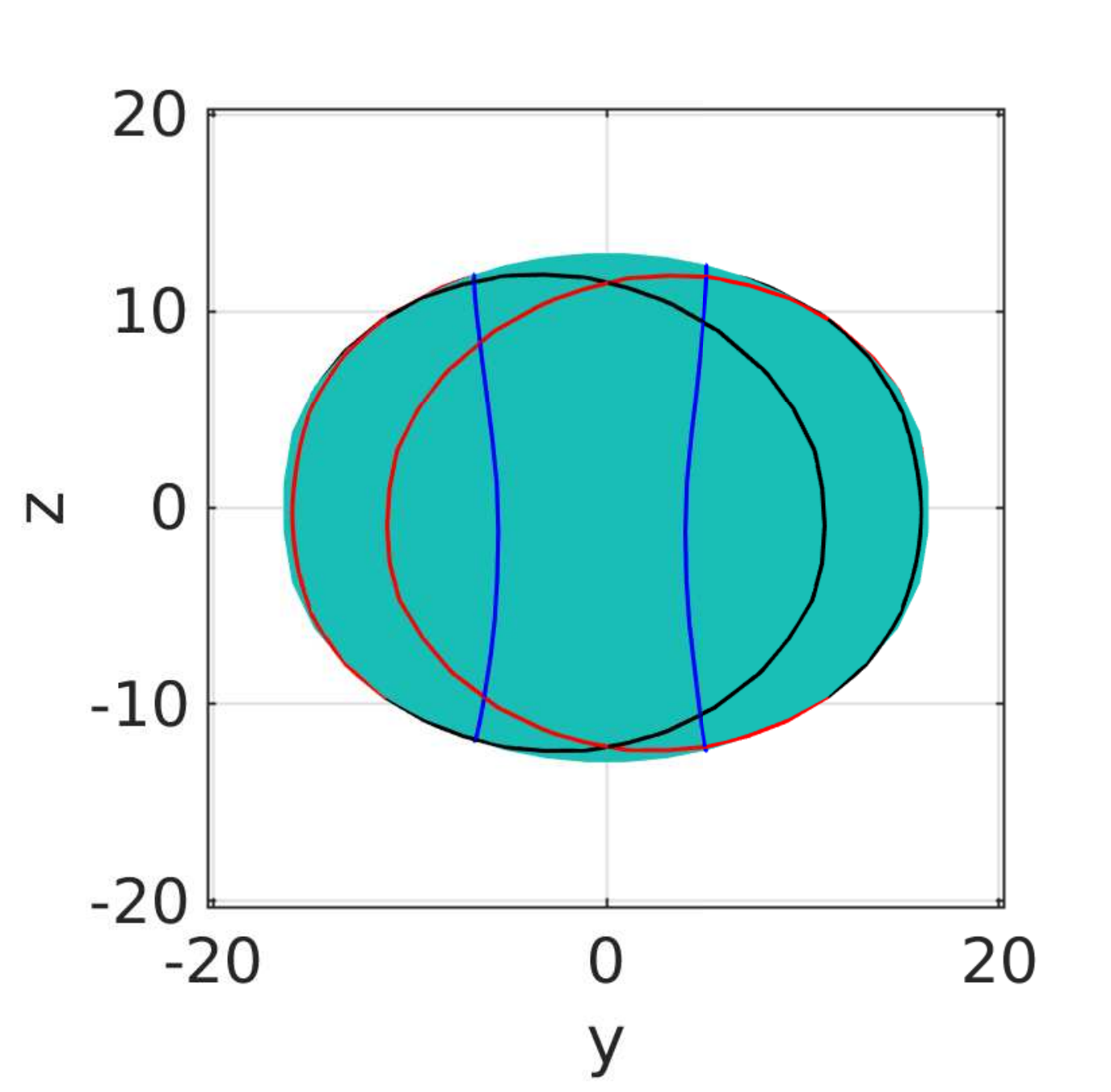}
		\end{minipage}
	}	
	\subfigure[]{
		\begin{minipage}{0.21\linewidth}			
			\includegraphics[width=\linewidth]{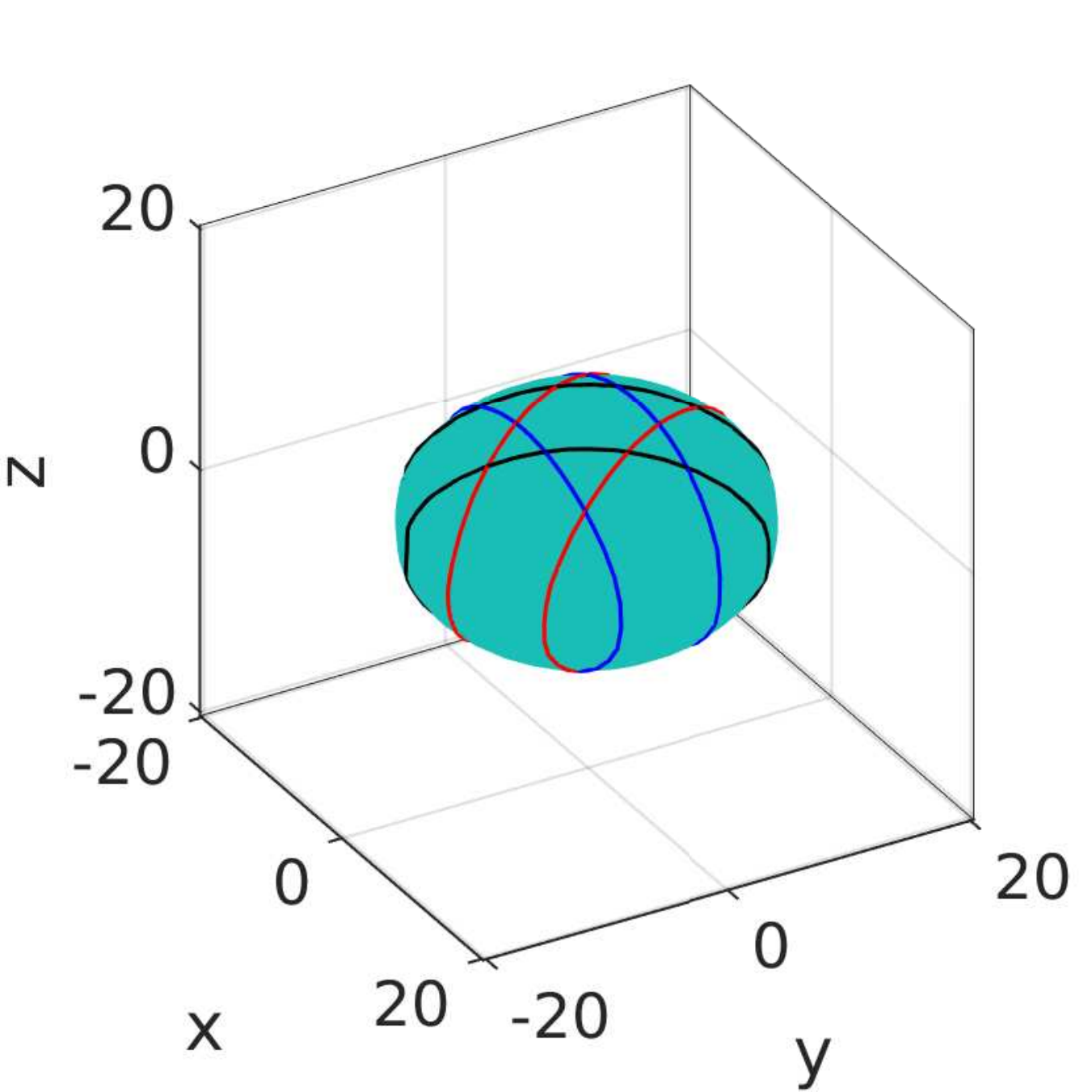}
			\includegraphics[width=\linewidth]{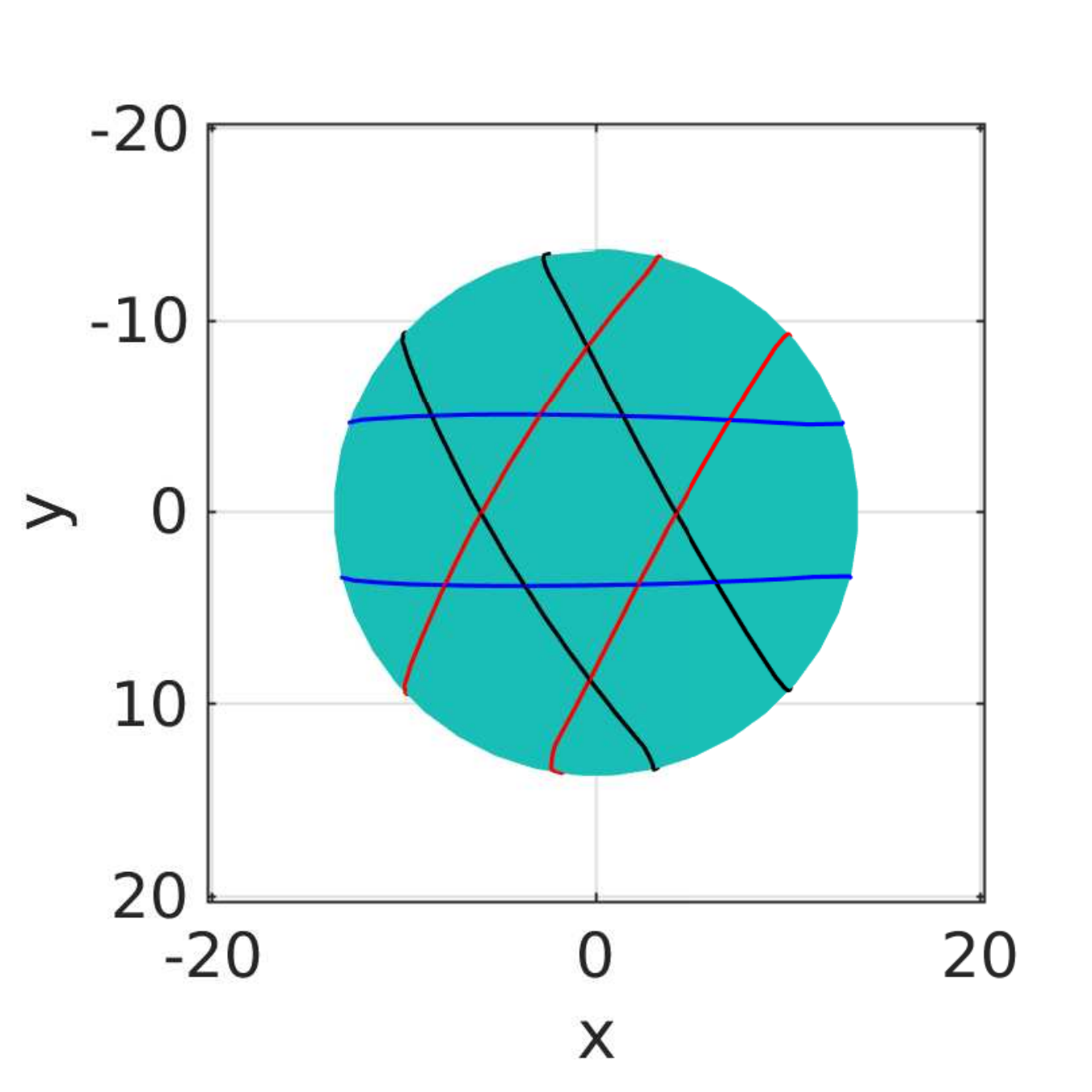}
			\includegraphics[width=\linewidth]{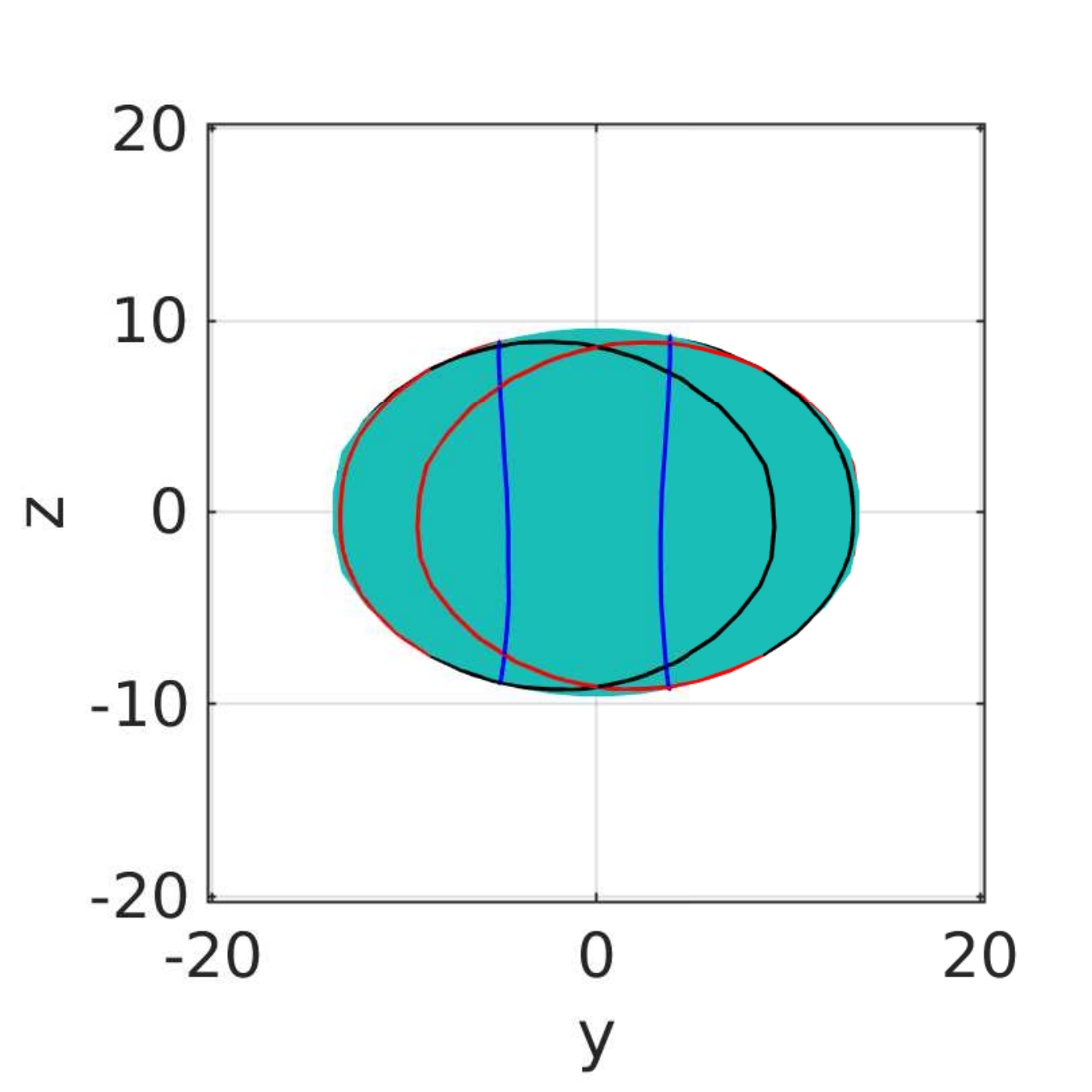}
		\end{minipage}
	}	
	\subfigure[]{
		\begin{minipage}{0.21\linewidth}			
			\includegraphics[width=\linewidth]{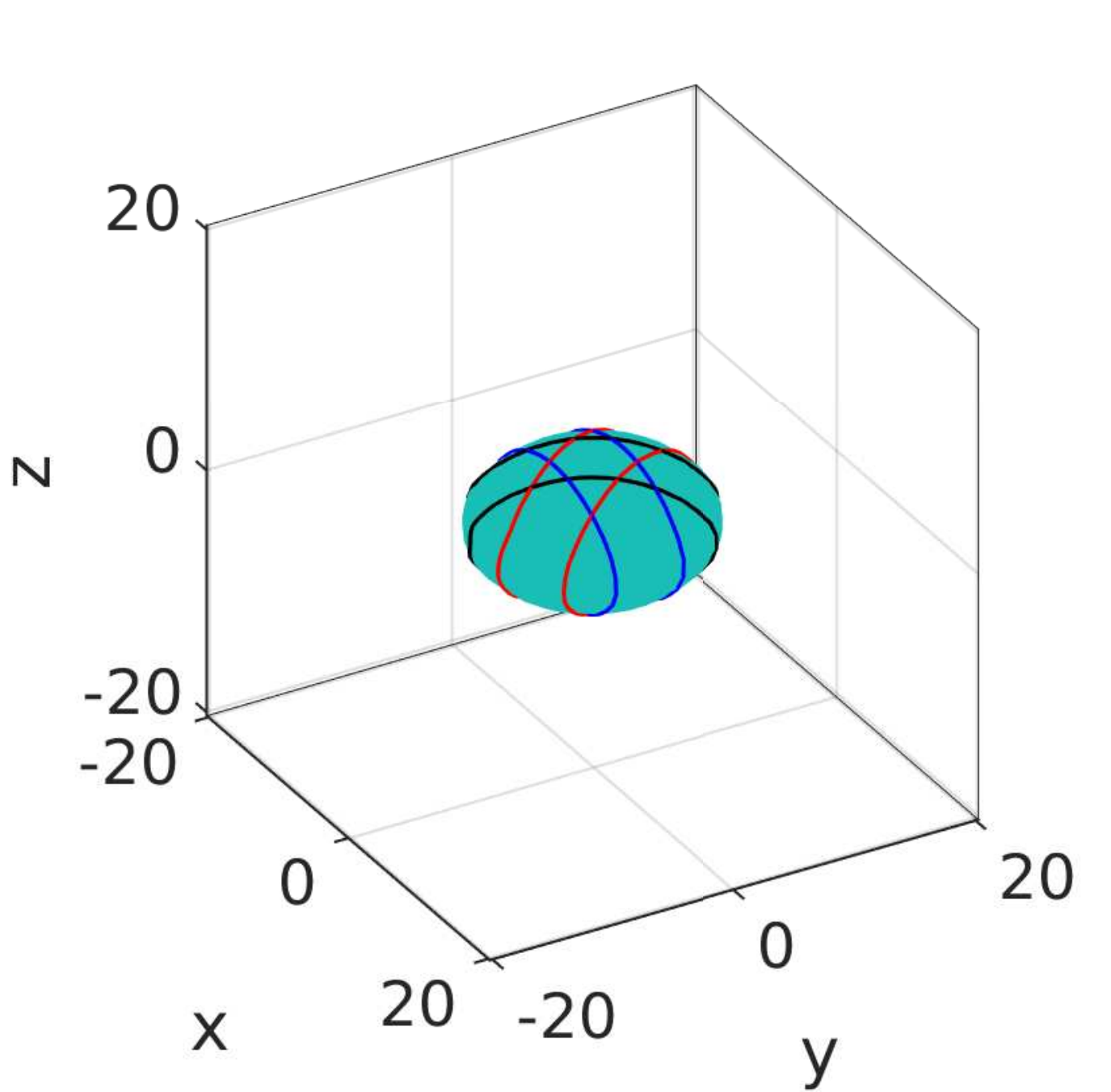}
			\includegraphics[width=\linewidth]{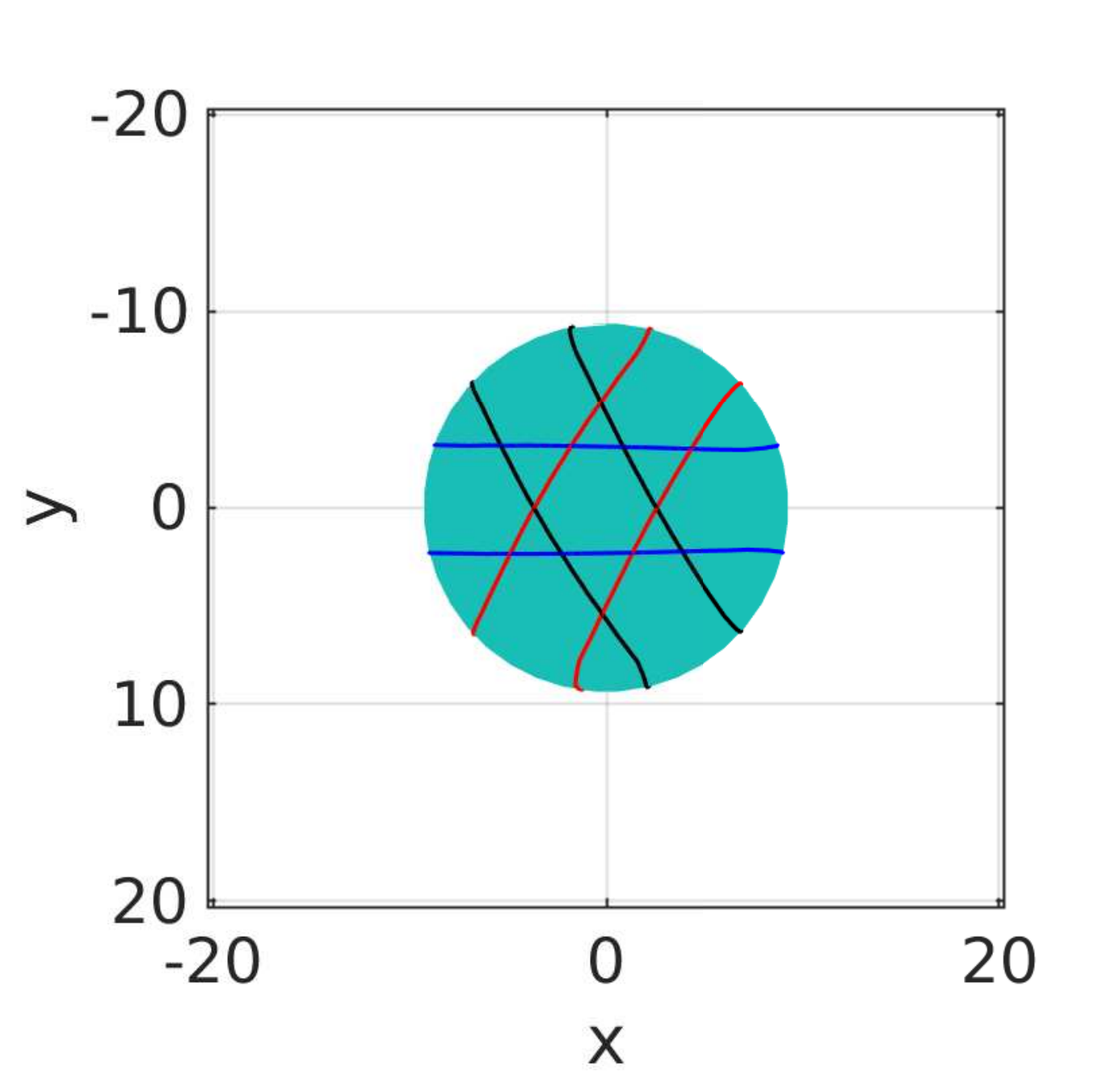}
			\includegraphics[width=\linewidth]{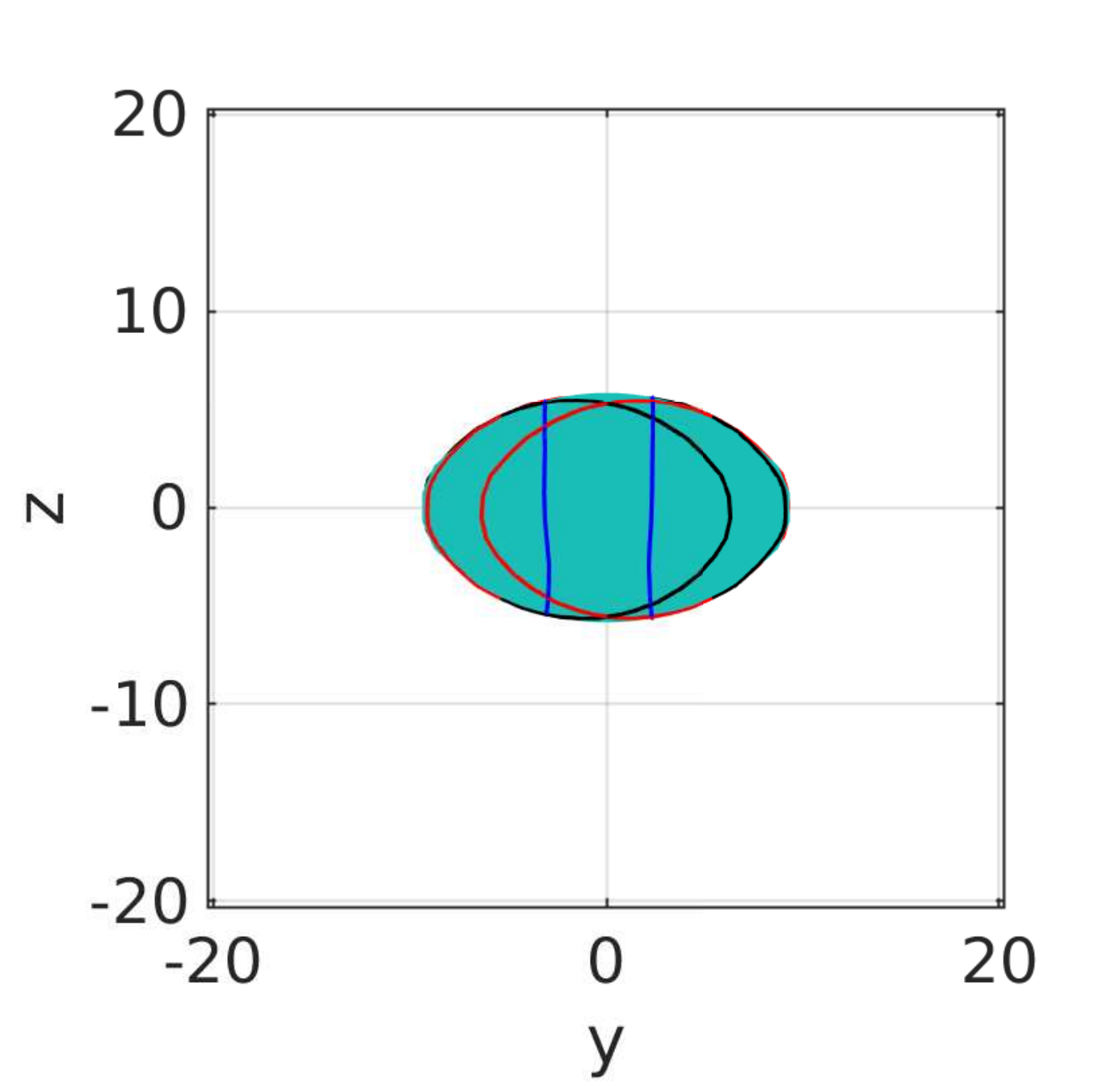}
		\end{minipage}
	}
	\caption{Shrinkage of an initially spherical grain boundary in fcc under pure coupling motion, i.e., without dislocation reaction. The rotation axis is the $z$ direction ($[111]$), and the initial misorientation angle $\theta=5^\circ$. The upper panel of images show the three-dimensional view of the grain boundary during evolution. The middle panel of images show the grain boundary during evolution viewed from the $+z$ direction ($[111]$), and the lower panel of images show the grain boundary during evolution viewed from the $+x$ direction ($[\bar{1}10]$). Dislocations with Burgers vectors $\mathbf b^{(1)}$, $\mathbf b^{(2)}$ and $\mathbf b^{(3)}$ are shown by blue, black and red lines, respectively. Length unit: $b$. (a) The initial spherical grain boundary. (b), (c), and (d) Configurations at time $t=10/M_{\rm d}\mu, 15/M_{\rm d}\mu, 20/M_{\rm d}\mu$, respectively.}\label{fig:fccfigure0}
\end{figure}

Fig.~\ref{fig:fccfigure0} shows the shrinkage of the spherical grain boundary under this pure coupling motion. The grain boundary eventually disappears. In this case, since $M_{\rm r}=0$ and $\delta \eta_j=0$, the grain boundary velocity in Eq.~\eqref{eqn:fv2} becomes $\mathbf v =-\frac{\delta \theta}{\theta\delta t}(x,y,0)+(0,0, v^*_3)$. In the direction normal the rotation axis, i.e., in the $xy$ plane, the velocity component is in the inward radial direction, as in the two-dimensional model \cite{zhang2018motion,zhang2019new}; this is adjusted from the velocity component due to curvature flow in order to satisfy the Frank's formula. Whereas in the direction of the rotation axis, i.e., the $z$ direction, there is no constraint imposed by the Frank's formula, and the velocity component is
the same as that in the curvature flow.

As an example, we consider the cross-section of the grain boundary with the $z=0$ plane (i.e., cross-section normal to the $[111]$ rotation axis), which is the equator of the grain boundary in the three dimensional view in the upper panel in Fig.~\ref{fig:fccfigure0} and is a circle (the outer circle) as shown  in the second panel in Fig.~\ref{fig:fccfigure0} for the view from $+z$ direction. The grain boundary along this circular cross-section is pure tilt, which is similar to the two-dimensional grain boundary discussed in Ref.~\cite{zhang2018motion,zhang2019new}. Along this circle, during the evolution, we have $\mathbf v^*_3=0$, and the grain boundary velocity is $\mathbf v =-\frac{\delta \theta}{\theta\delta t}(x,y,0)$, which is completely in the inward radial direction in the $z=0$ plane. Thus the cross-section keeps the circular shape as it shrinks during the evolution, as shown in the  second panel in Fig.~\ref{fig:fccfigure0}.
This shape-preserving evolution agrees with the results of the two-dimensional grain boundary dynamics models \cite{Taylor2007,zhang2018motion,zhang2019new} and shrinkage of circular grain boundaries in two dimensions by molecular dynamics \cite{srinivasan2002challenging} and phase field crystal \cite{wu2012phase} simulations. However, here the changing rate of misorientation angle $\frac{\delta \theta}{\delta t}$ in the velocity formula is depending on the entire grain boundary in three dimensions by Eq.~\eqref{eqn:deltatheta3}, and is not just depending on the circular cross-section itself as in the two dimensional continuum model \cite{zhang2019new}.

Next, we consider the cross-section of the grain boundary with the $x=0$ plane (i.e., cross-section normal to the $[\bar{1}10]$ direction); see the lower panel in Fig.~\ref{fig:fccfigure0} (the outer boundary of the projected grain boundary surface).
Initially, the cross-section is a circle, and  it gradually changes to an ellipse as it shrinks during the evolution. This shows that the velocity in the rotation axis direction,  i.e. $z$ direction is larger than that in the $x$ and $y$ directions.
 The reason for this anisotropic motion is that there is no constraint of Frank's formula in the $z$ direction which is the direction of the rotation axis, and the velocity at the  two poles on the grain boundary with respect to the $z$ direction (where the grain boundary is pure twist) is the same as that in the curvature flow; whereas the velocity components in the $x$ and $y$ directions are adjusted from those in the curvature flow by the constraint of the Frank's formula, and the resulting velocity in the $xy$ plane are depending on the entire grain boundary through the coefficient $\delta \theta$, as discussed above.
 Evolution of this initially spherical grain boundary and its dislocation structure, especially the property that the shrinkage of the grain boundary is faster in the direction of the rotation axis than in other directions, agree with the results of atomistic-level simulations using phase field crystal model~\cite{yamanaka2017phase} and amplitude expansion phase field crystal model~\cite{salvalaglio2018defects}.

%Notice that
% with respect to the rotation axis of the $z$ direction,  at the two poles on the grain boundary,  the grain boundary is pure twist, and the constituent dislocations are screw dislocations;
%along the equator of the grain boundary, the grain boundary is pure tilt, and the dislocations are edge dislocations. It was suggested in Ref.~\cite{yamanaka2017phase} that such an anisotropic motion may be due to the difference in dislocation densities or that in the mobilities of screw and edge dislocations.  Here, our continuum model provides a further explanation that this anisotropic motion of the grain boundary (and accordingly, the anisotropic motion of the screw and edge portions of the constituent dislocations) is due to the constraint of Frank's formula in order to maintain an equilibrium dislocation structure.

%  This results agrees with the high temperature case of bcc iron spherical grain boundary in \cite{yamanaka2017phase} and the fcc, bcc spherical grain boundaries in \cite{salvalaglio2018defects} obtained using phase-field crystal model.

\begin{figure}[htbp]
	\centering
	\subfigure[]{\includegraphics[width=0.45\textwidth]{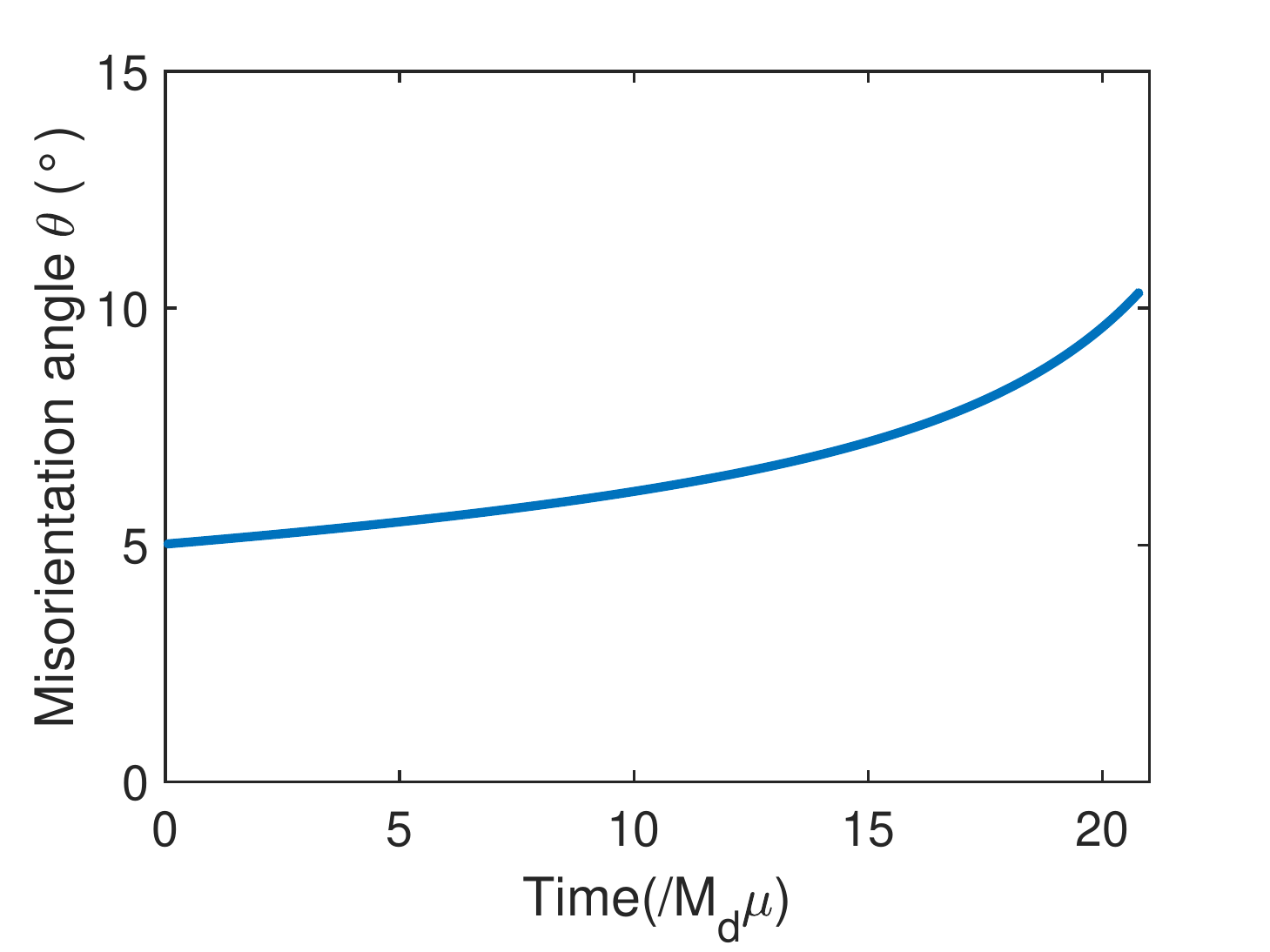}}
	\subfigure[]{\includegraphics[width=0.45\textwidth]{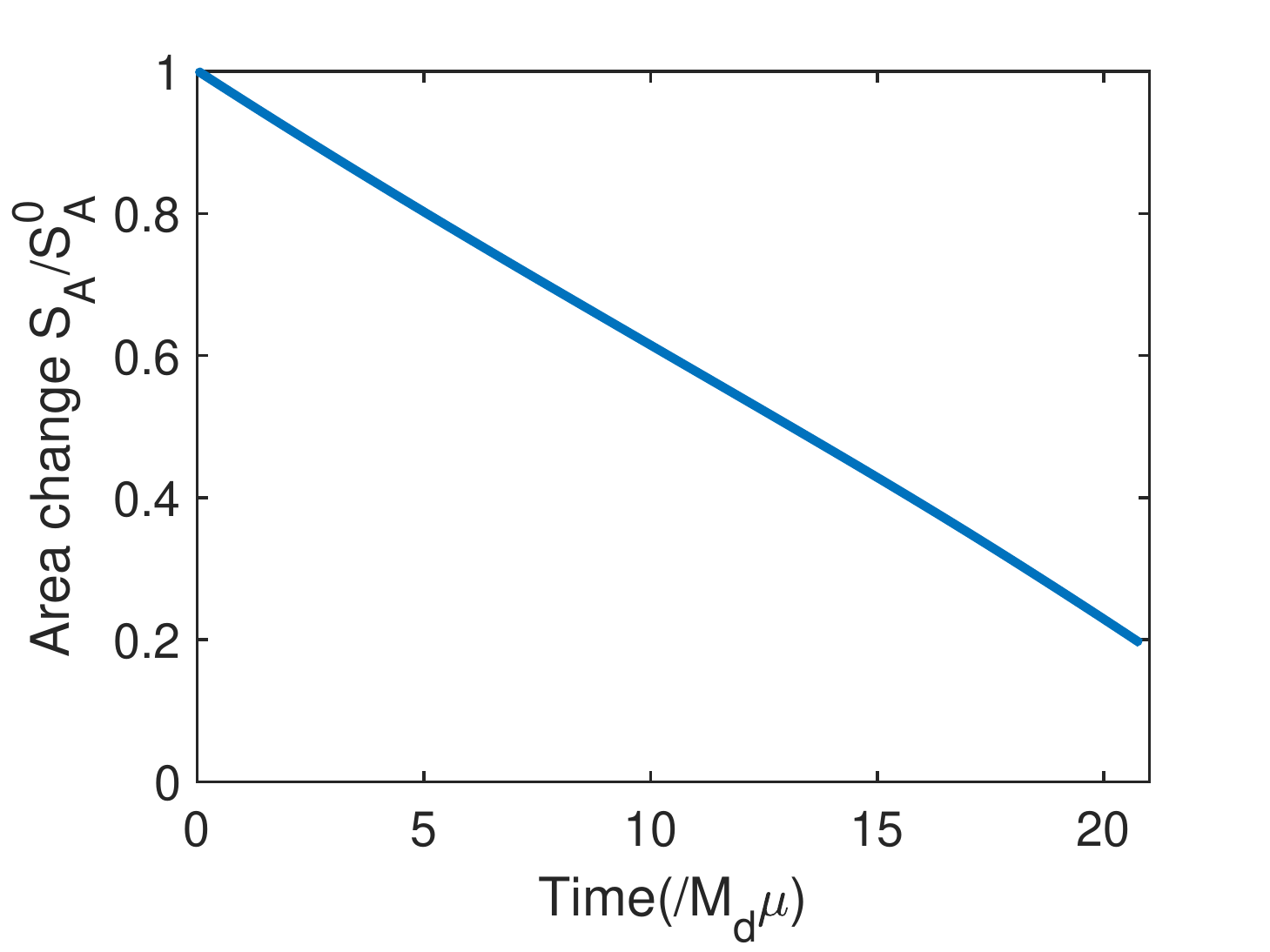}}
	\subfigure[]{\includegraphics[width=0.45\textwidth]{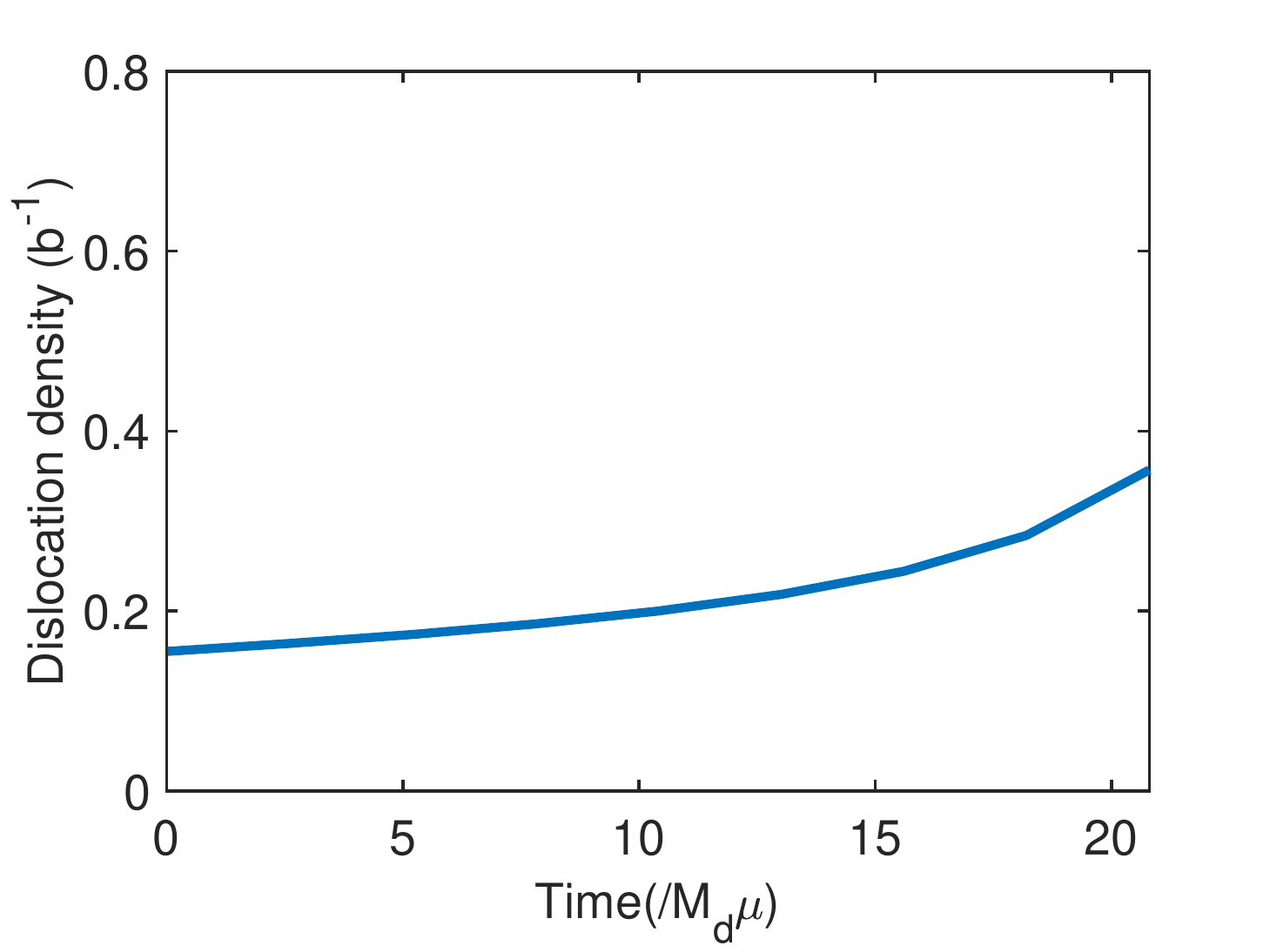}}
	\subfigure[]{\includegraphics[width=0.45\textwidth]{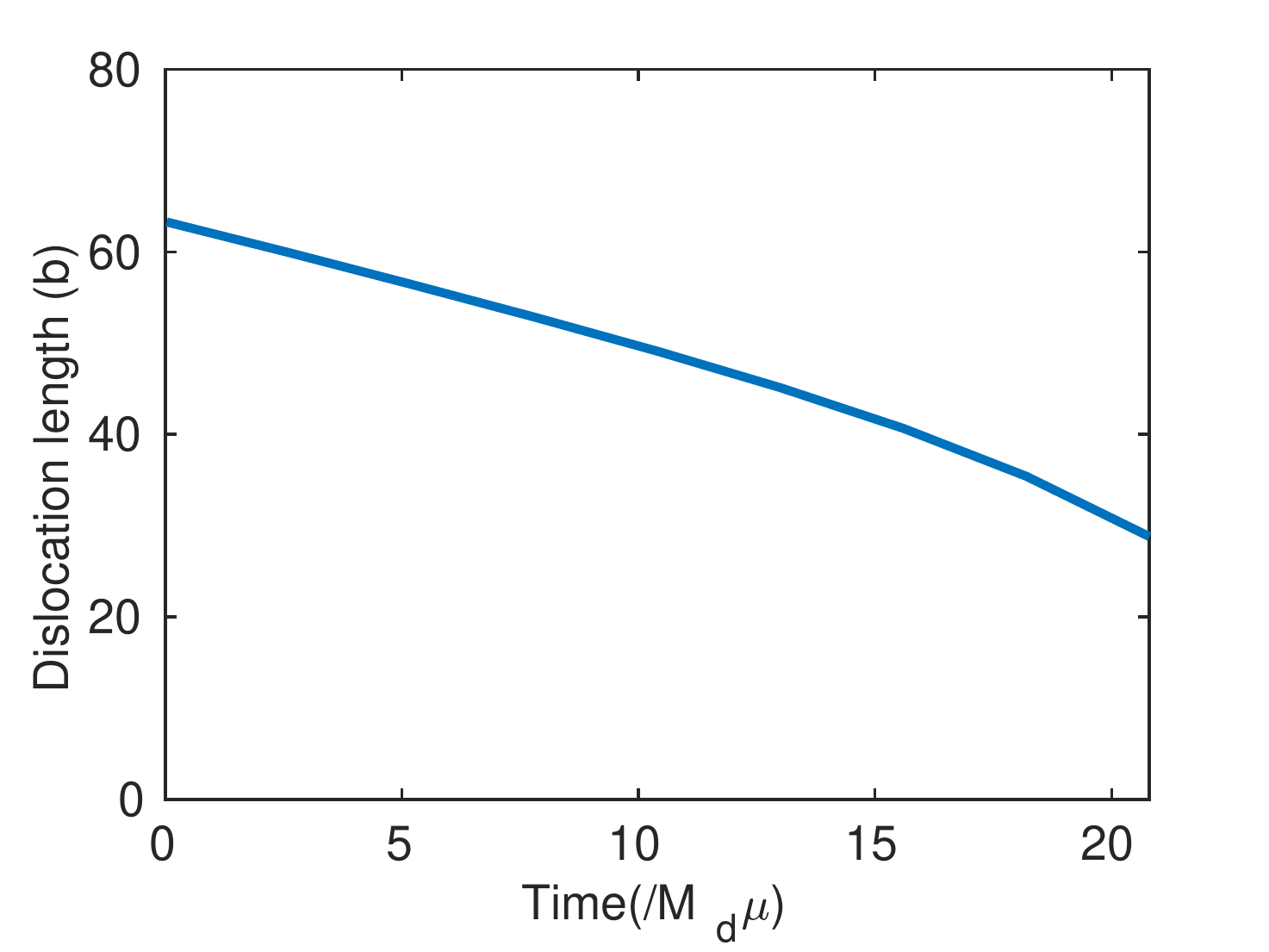}}
	\caption{Shrinkage of an initially spherical grain boundary in fcc under pure coupling motion. The rotation axis is the $z$ direction ($[111]$), and the initial misorientation angle $\theta=5^\circ$.  (a) Evolution of  misorientation angle $\theta$. (b) Evolution of grain boundary area $S_A$, where $S_A^0$ is the area of the initial grain boundary. (c) Evolution of density of dislocations with Burgers vector $\mathbf b^{(1)}$/ $\mathbf b^{(2)}$/$\mathbf b^{(3)}$ on the grain boundary. (d) Evolution of the total length of dislocations with Burgers vector $\mathbf b^{(1)}$/ $\mathbf b^{(2)}$/$\mathbf b^{(3)}$ on the grain boundary. In (c) and (d), the densities and total lengths of dislocations with these three Burgers vectors are almost identical. }\label{fig:fcctheta0}
\end{figure}

Fig. \ref{fig:fcctheta0}(a) shows the change of misorientation angle $\theta$ during the evolution, which is continuously  increasing. This behavior agrees with Cahn-Taylor theory \cite{cahn2004unified}, three dimensional phase-field crystal simulations \cite{yamanaka2017phase}, and  two-dimensional atomistic \cite{srinivasan2002challenging,trautt2012grain}, phase field crystal \cite{wu2012phase}, and continuum \cite{zhang2018motion,zhang2019new} simulations. Such increasing  of misorientation angle cannot be obtained by the classical motion by mean curvature models or pure sliding models, in which the misorientation angle is constant or is decreasing during the evolution.

Fig.~\ref{fig:fcctheta0}(b) shows evolution of the area of the grain boundary, which  reveals the relation:
\begin{equation}
\frac{S_A(t)}{S_A^0}=1-At,
\label{eqn:area}
\end{equation}
where $A$ is some constant, and $S_A(t)$ and $S_A^0$ are the grain boundary area at time $t$ and that of the initial configuration, respectively. This agrees with the results of nearly linear decrease of the grain boundary area using amplitude expansion phase field crystal model for grain boundaries in both fcc and bcc crystals \cite{salvalaglio2018defects}.
The phase field crystal simulations in Ref.~\cite{yamanaka2017phase} showed that the decrease of the volume of the grain enclosed by an initially spherical low angle grain boundary in a bcc crystal approximately follows the relation $\frac{V^{2/3}(t)}{V_0^{2/3}}=1- A_1t$, where $A_1$ is some constant,   and $V(t)$ and $V_0$ are the volume of the grain enclosed by the grain boundary at time $t$ and that of the initial configuration, respectively. It was argued in  Ref.~\cite{yamanaka2017phase} that their results are consistent with the result of classical Von Neumann-Mullins relation \cite{mullins1956two} for a two dimensional grain boundary driven by curvature with constant energy density, i.e., Eq.~\eqref{eqn:area} if $S_A$ denotes the area enclosed by the grain boundary in two dimensions, considering the approximate relation $V^{2/3}\sim S_A$.
In this sense, simulation results using our continuum model and the amplitude expansion phase field crystal simulations in \cite{salvalaglio2018defects} are consistent with the results in  Ref.~\cite{yamanaka2017phase} as well as the result of the classical Von Neumann-Mullins relation. The nearly linear decrease of the grain boundary area in Eq.~\eqref{eqn:area} obtained by our continuum model and the amplitude expansion phase field crystal model in Ref.~\cite{salvalaglio2018defects} is also in consistent with the result that the area enclosed by a two dimensional grain boundary is linearly decreasing in the two dimensional phase field crystal simulations for circular grain boundaries \cite{wu2012phase} and continuum model simulations for circular \cite{zhang2018motion} and general shape \cite{zhang2019new} grain boundaries in two dimensions.

%%\begin{comment}
%We approximately rewrite the area formula in Eq. \eqref{eqn:area} as $\frac{dR}{dt} \simeq -\frac{A}{2R}\simeq v_{\perp} $. A the grain boundary keeps shrinkage, the radius $R$ is decreasing which implies the normal velocity is increasing. This relation explains the increasing tangential velocity in Fig. \ref{fig:fcctheta0} (a).
%%\end{comment}

Evolutions of dislocation densities on the grain boundary and total length of dislocations are shown in Figs.~\ref{fig:fcctheta0}(c) and (d). It can be seen from Fig.~\ref{fig:fcctheta0}(c) that the densities of the dislocations with all the three Burgers vectors are increasing during the evolution. This is consistent with the increase of misorientation angle $\theta$ during the evolution.
%Recall that in the two dimensional case with dislocation conservation, the inter-dislocation distance is decreasing as the grain boundary shrinks,  which leads to the increases of both the dislocation densities and the misorientation angle during the evolution \cite{cahn2004unified}.
The total length of dislocations is decreasing during the evolution as shown in Fig.~\ref{fig:fcctheta0}(d). This is in agreement with the phase field crystal simulation results in Ref.~\cite{yamanaka2017phase}.
Unlike in the two dimensional case with dislocation conservation \cite{srinivasan2002challenging,cahn2004unified,trautt2012grain,wu2012phase,zhang2018motion,zhang2019new} where dislocations are infinite straight lines, in three dimensions without dislocation reaction, the constituent dislocations are closed loops,
 and all the dislocation loops are shrinking and the total length of dislocations is decreasing as the grain boundary shrinks.

\subsection{Motion with dislocation reaction}\label{subsec:fccreact}

Now we perform simulations using our continuum model considering dislocation reaction, i.e. $M_{\rm r}\neq 0$. Dislocation reaction leads to removal of dislocations,  resulting in the coupling motion of the grain boundary \cite{srinivasan2002challenging,cahn2004unified,trautt2012grain,yamanaka2017phase,zhang2018motion,zhang2019new}. The mobility $M_{\rm r}$ is a temperature-dependent material parameter, and it may also depend on the local dislocation reaction mechanism \cite{trautt2012grain,yamanaka2017phase}. We set $M_{\rm r}$ to be constant in our simulations to examine the effect of dislocation reaction. We use the same initial spherical grain boundary as in Sec.~\ref{subsec:coup} without dislocation reaction.

\begin{figure}[htbp]
	\centering
	\centering
	\subfigure[]{
		\begin{minipage}{0.21\linewidth}
			\includegraphics[width=\linewidth]{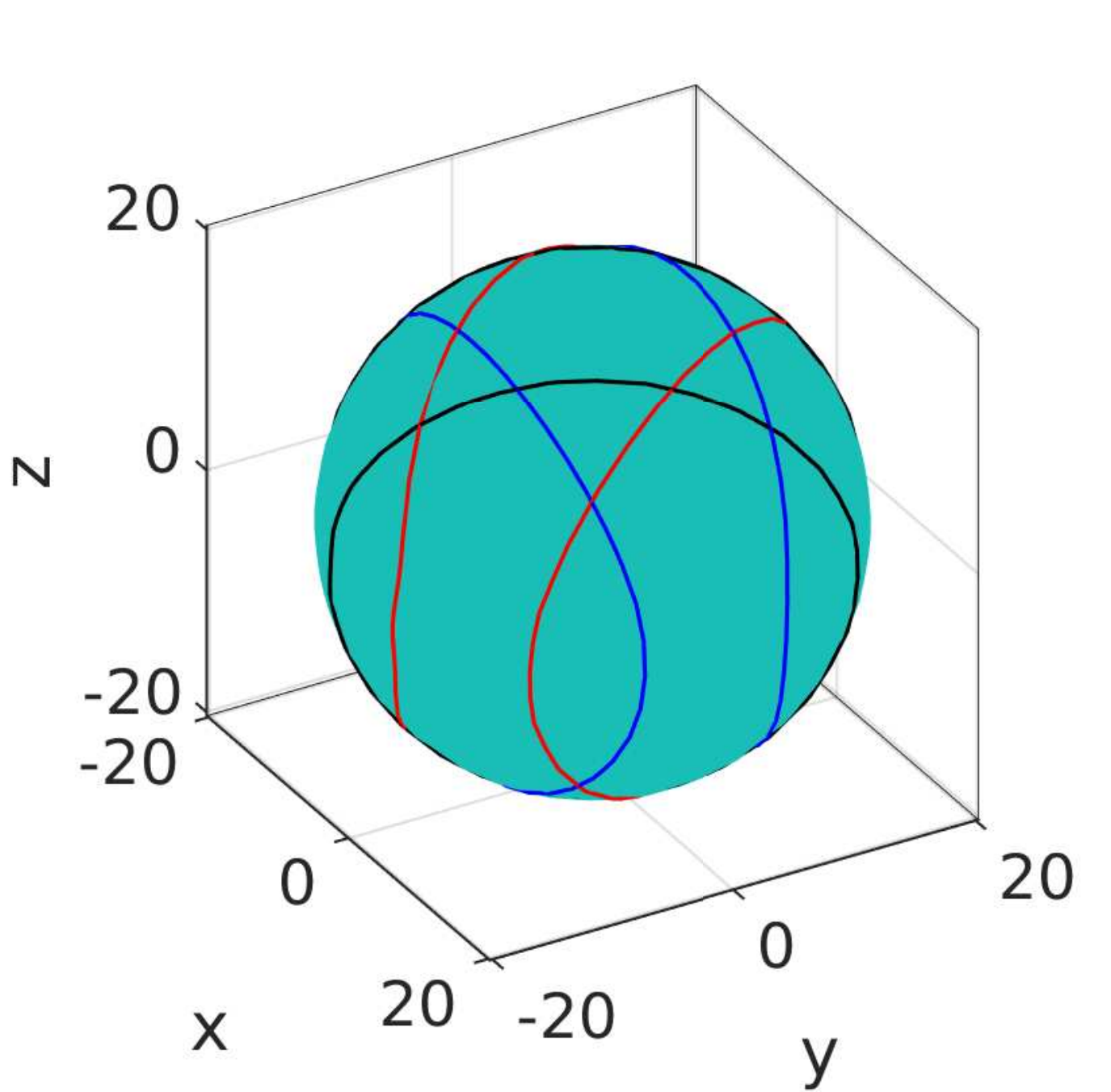}
			\includegraphics[width=\linewidth]{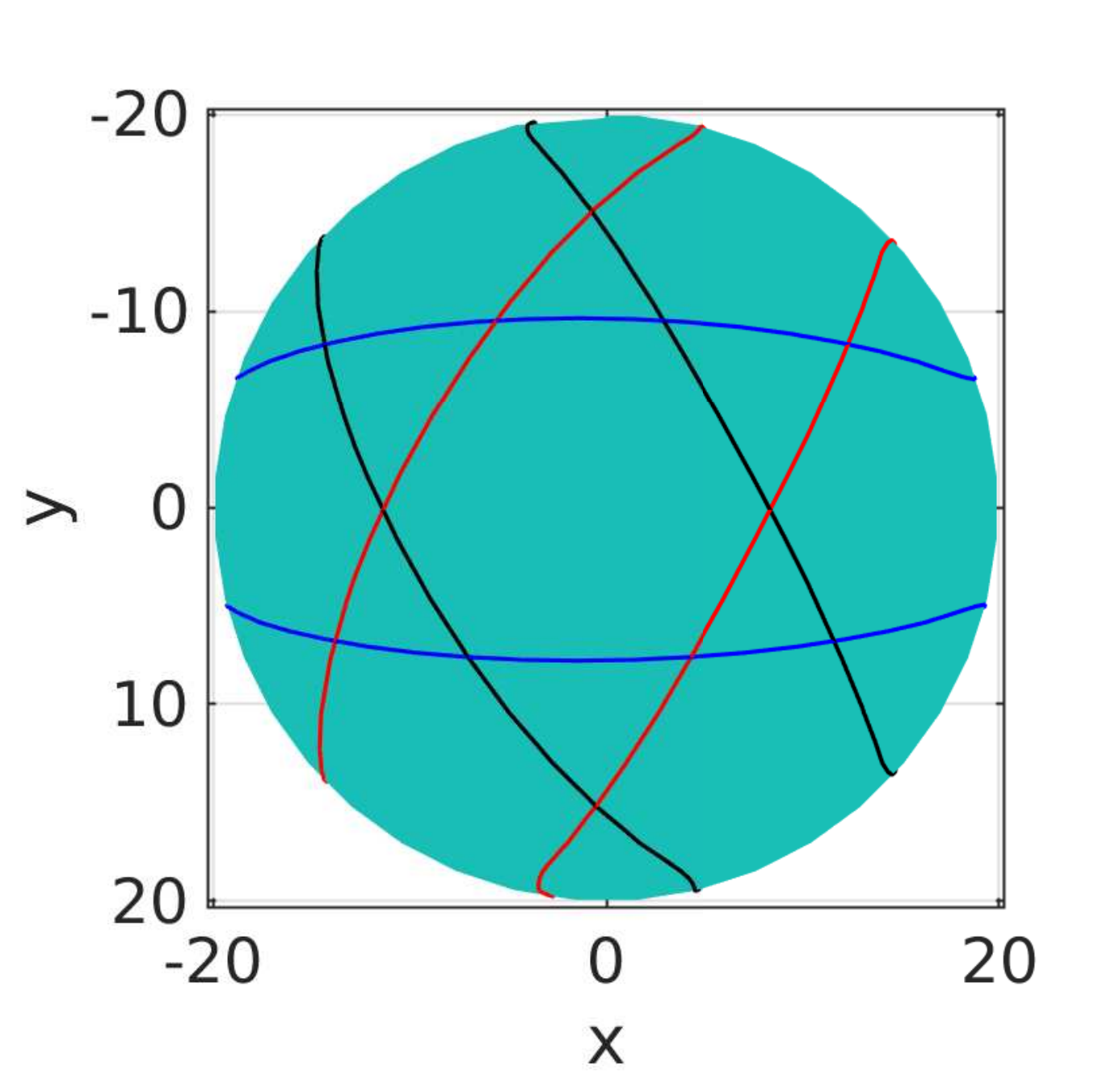}
			\includegraphics[width=\linewidth]{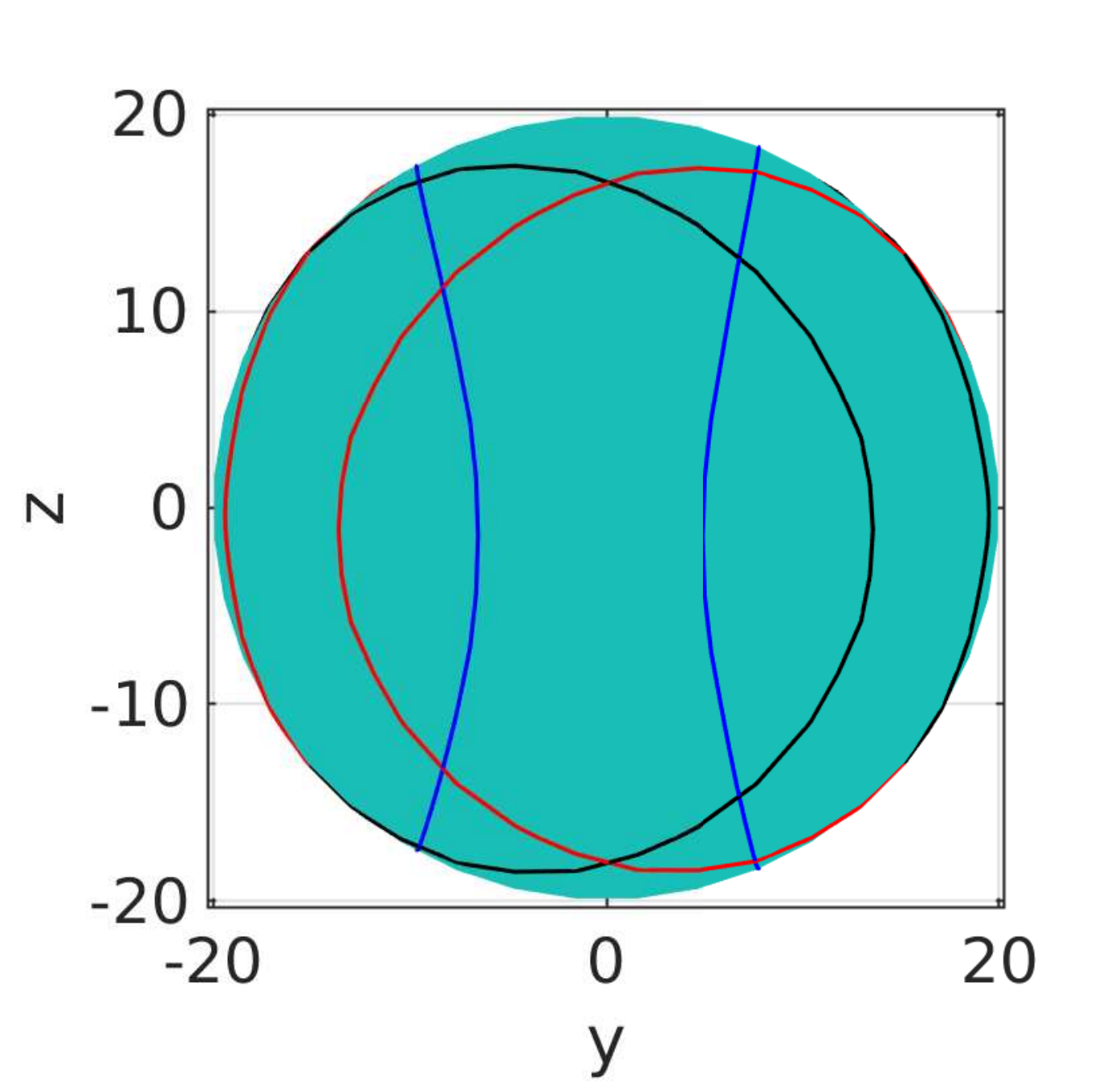}
		\end{minipage}
	}
	\subfigure[]{
		\begin{minipage}{0.21\linewidth}			
			\includegraphics[width=\linewidth]{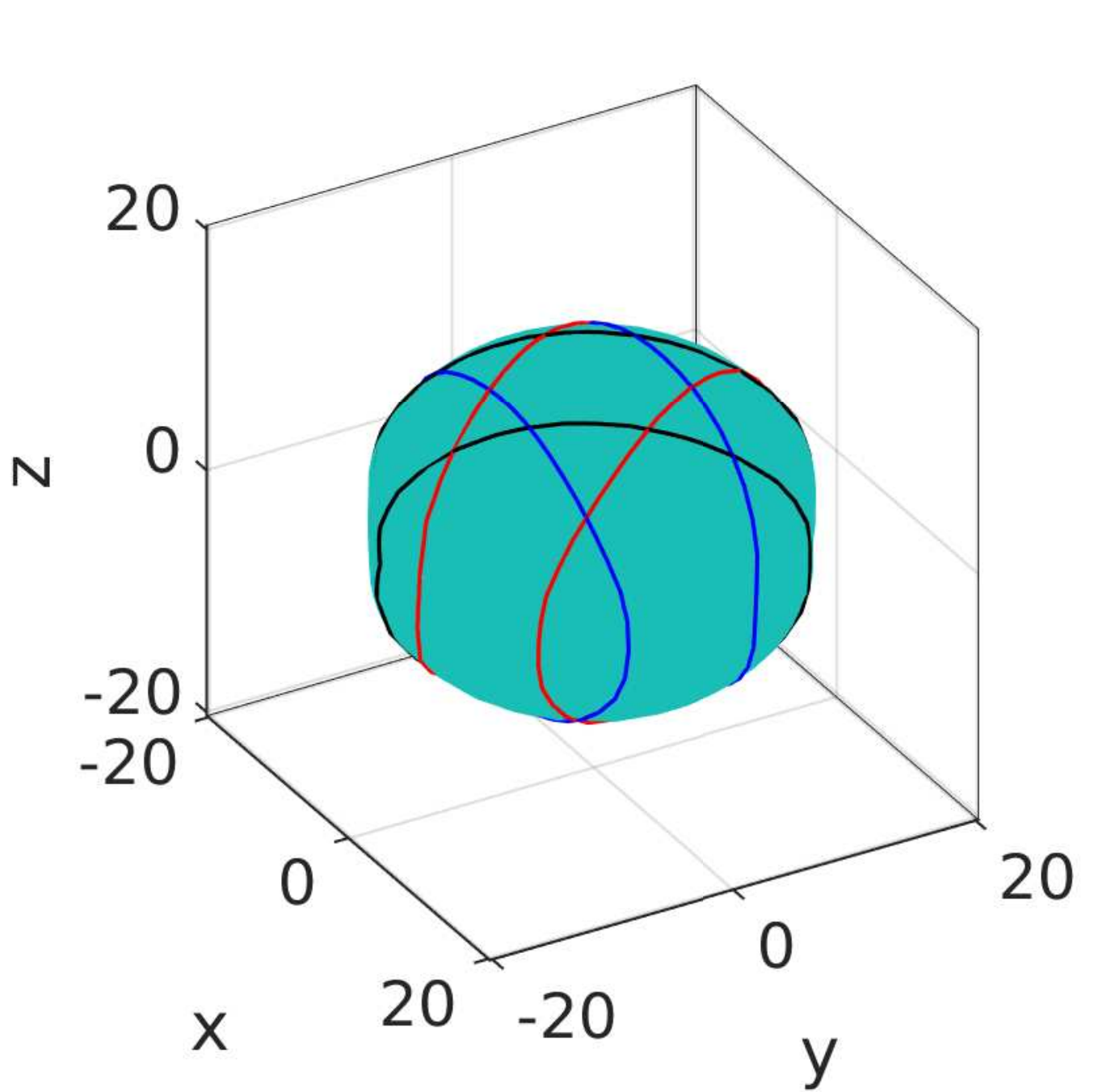}
			\includegraphics[width=\linewidth]{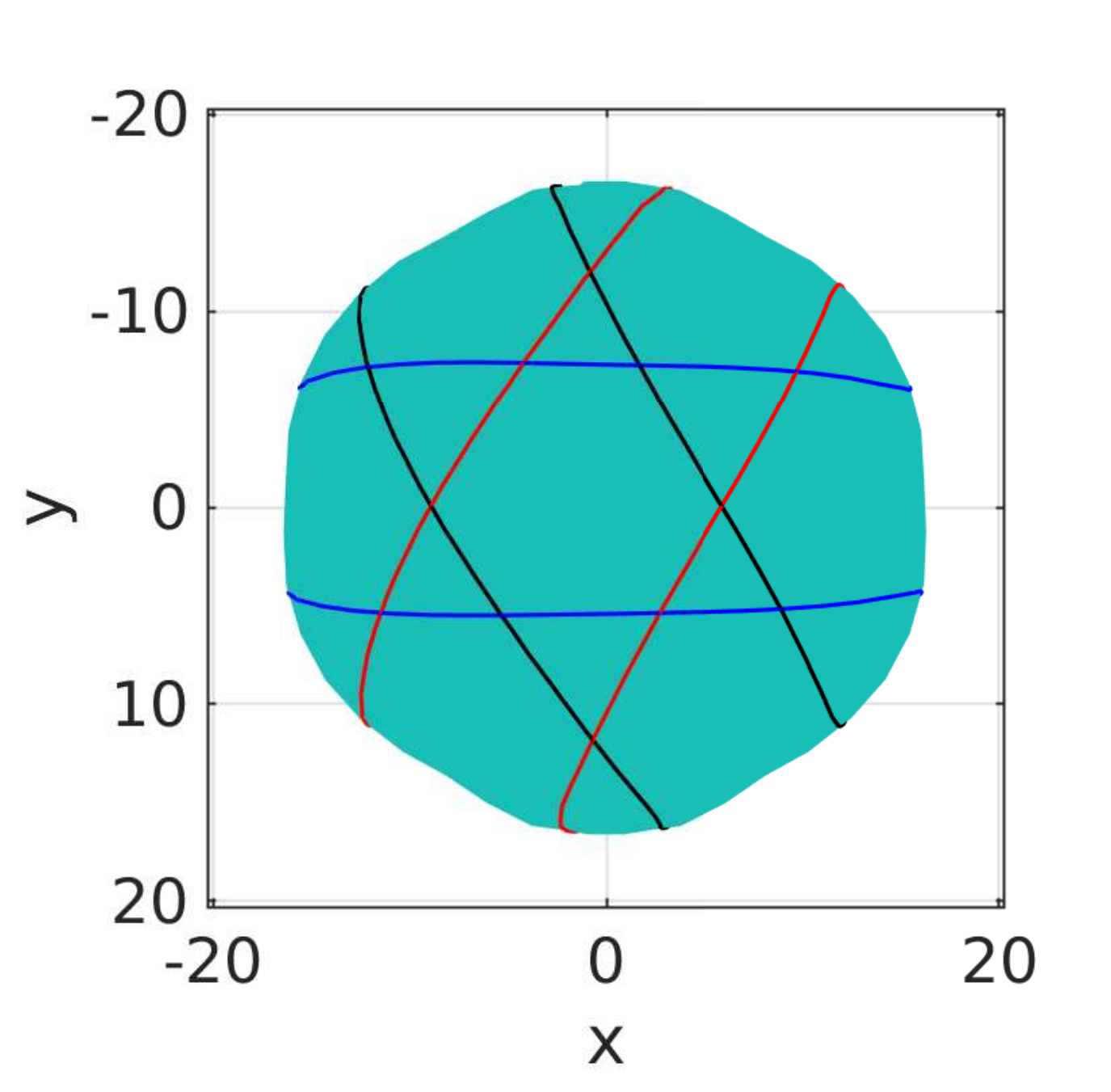}
			\includegraphics[width=\linewidth]{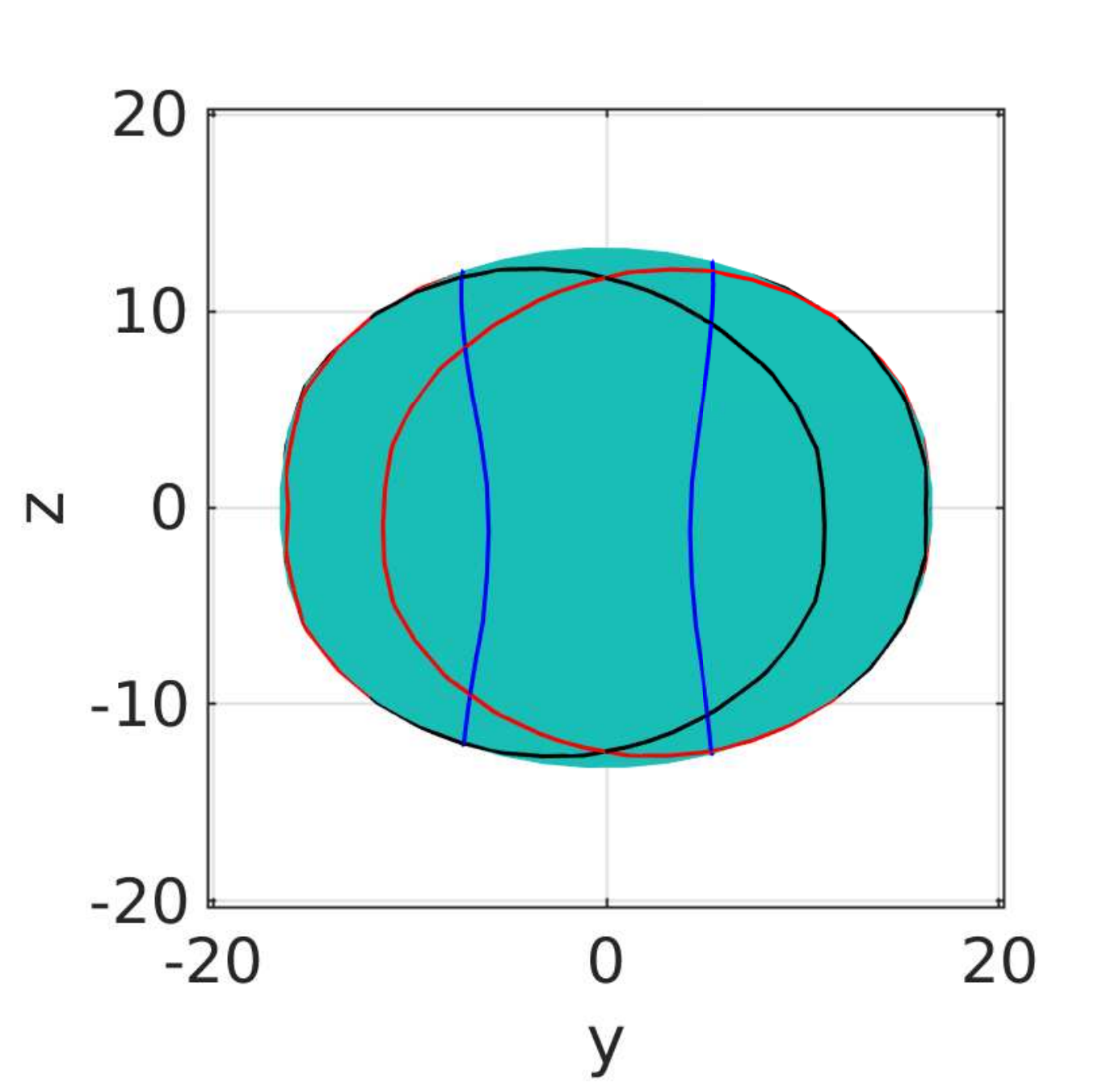}
		\end{minipage}
	}	
	\subfigure[]{
		\begin{minipage}{0.21\linewidth}			
			\includegraphics[width=\linewidth]{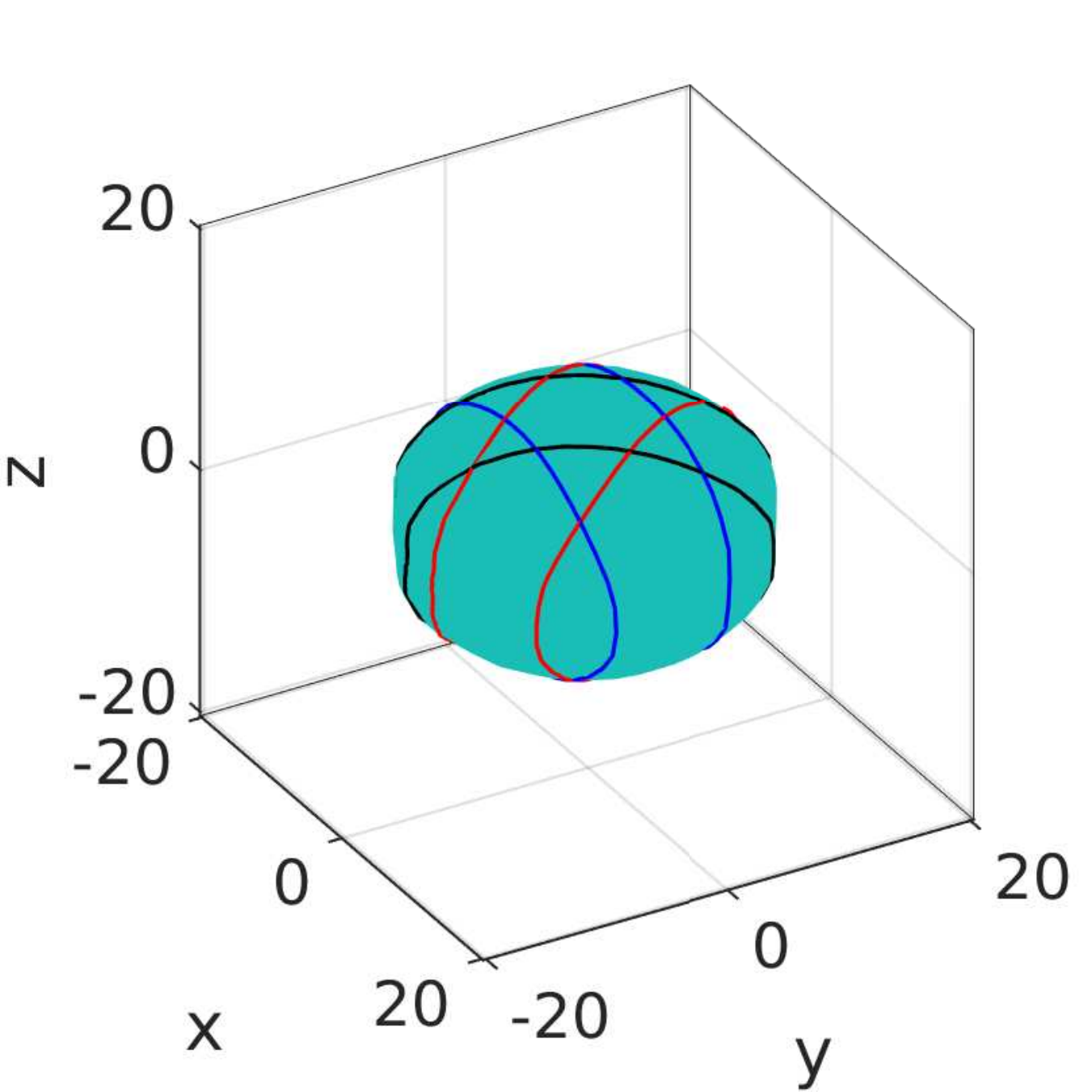}
			\includegraphics[width=\linewidth]{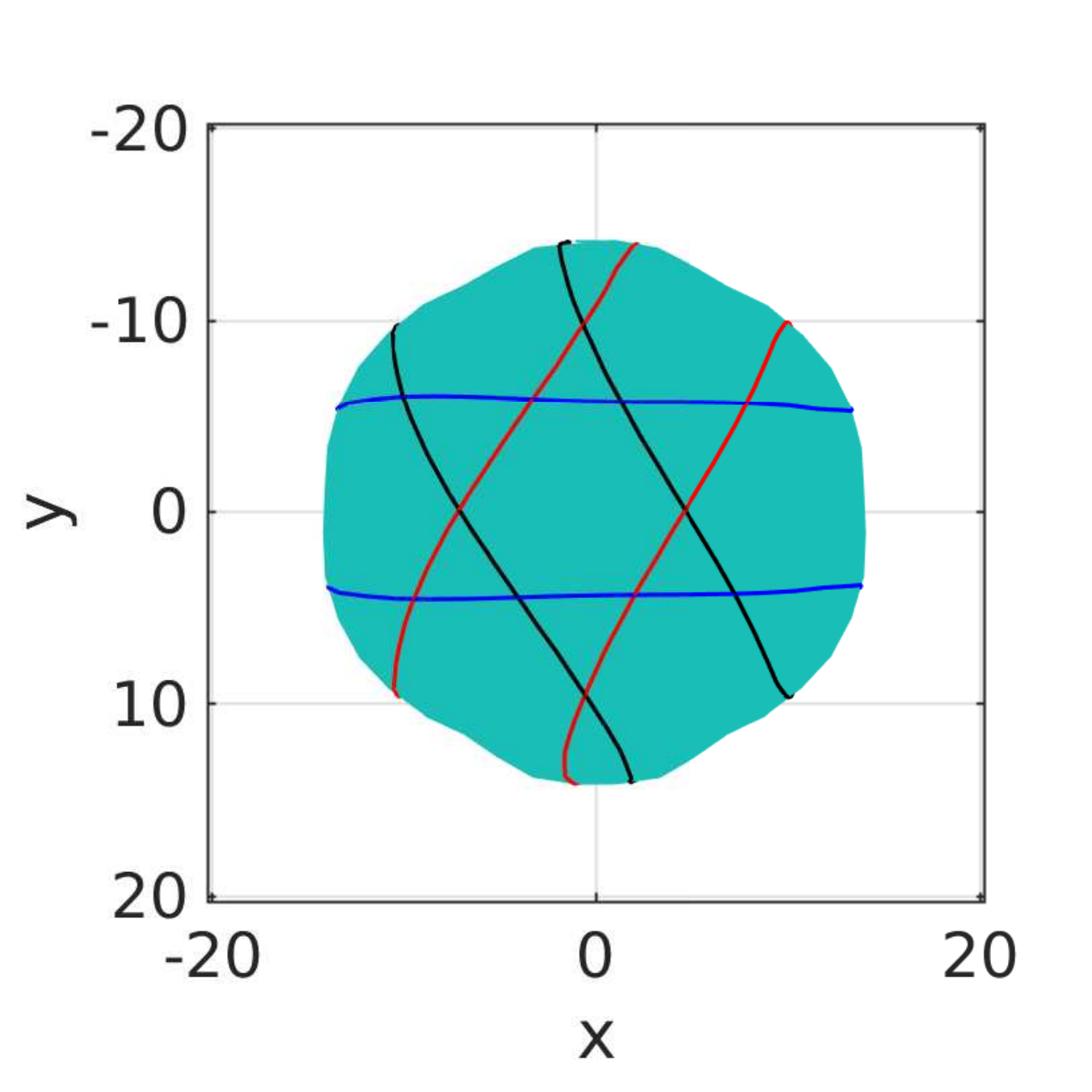}
			\includegraphics[width=\linewidth]{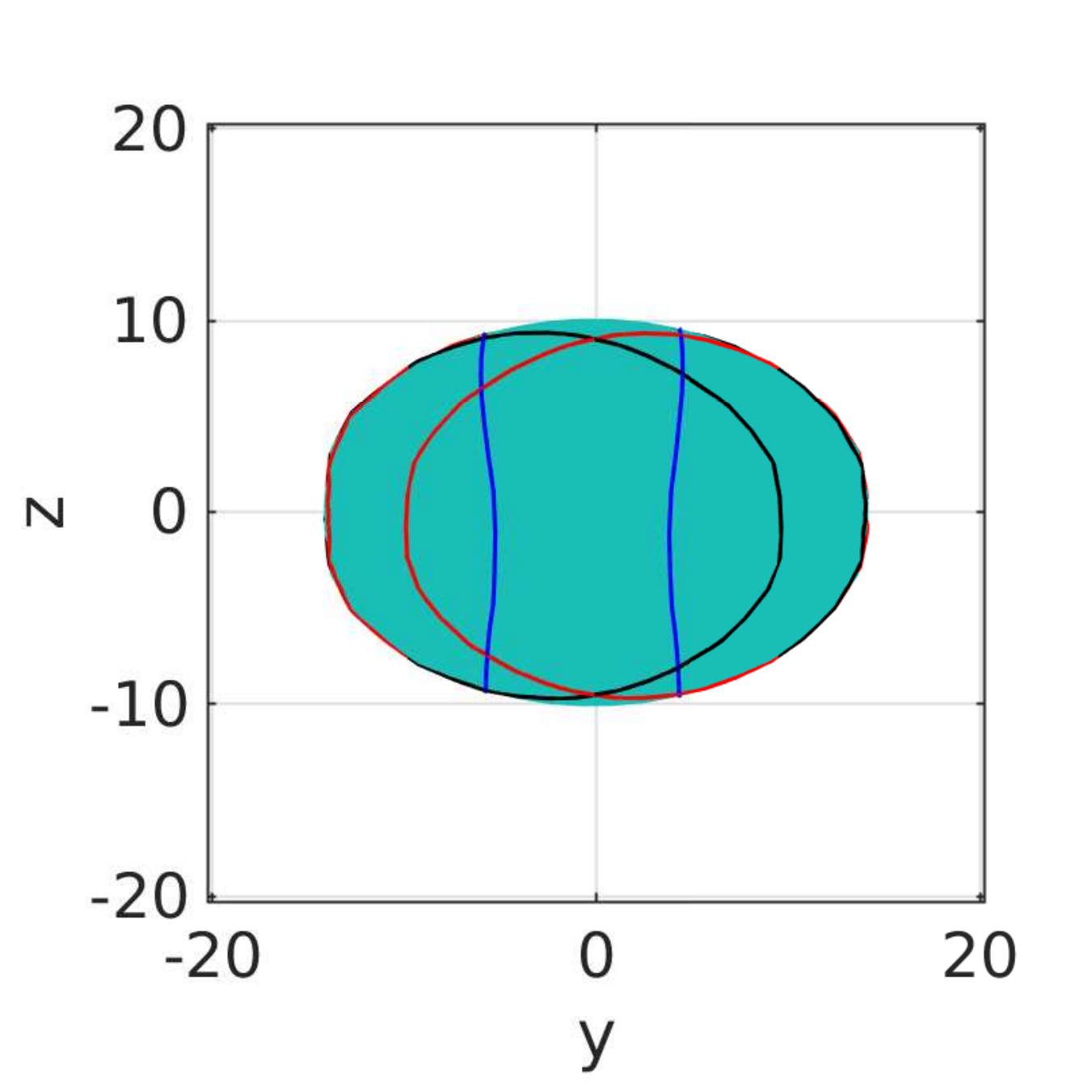}
		\end{minipage}
	}	
	\subfigure[]{
		\begin{minipage}{0.21\linewidth}			
			\includegraphics[width=\linewidth]{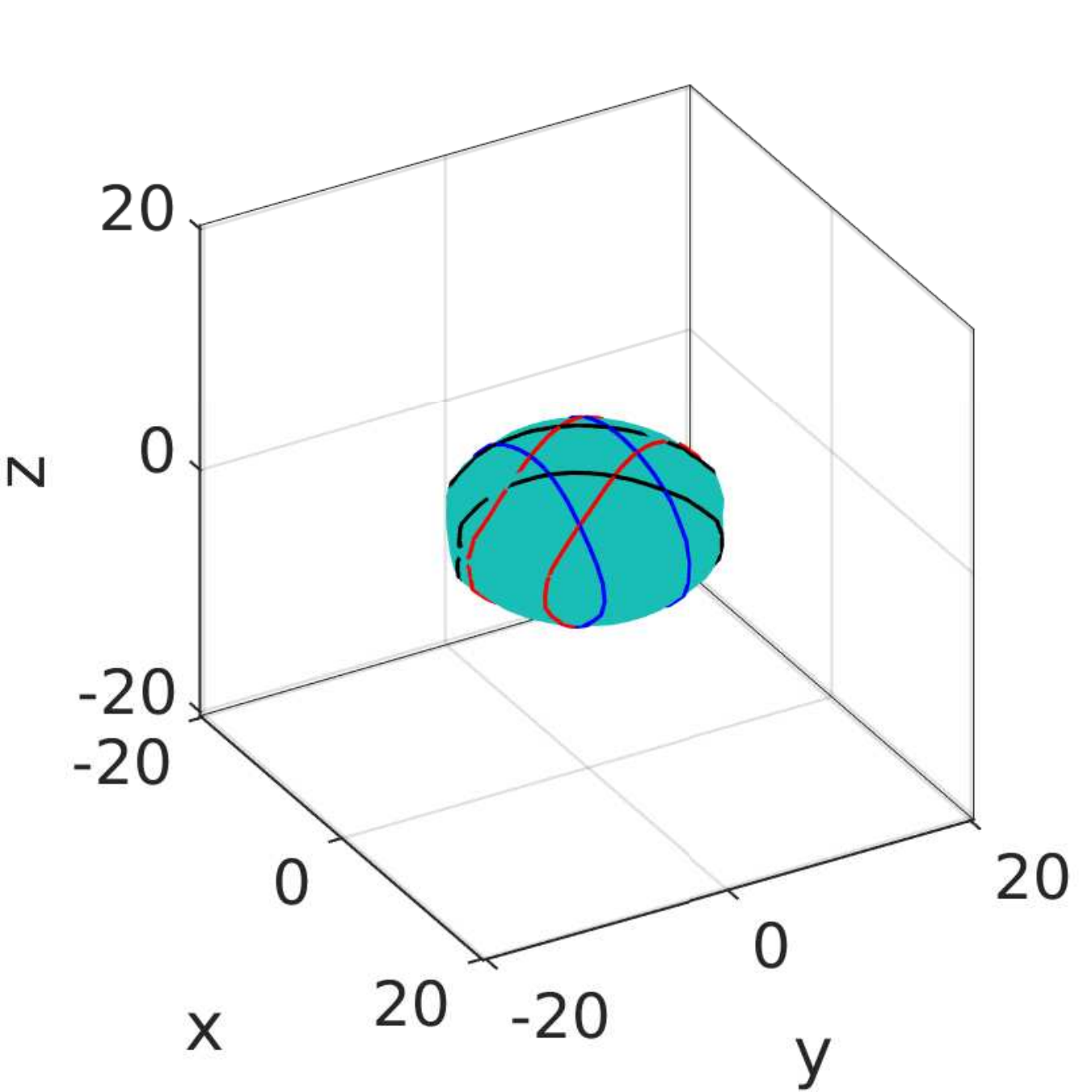}
			\includegraphics[width=\linewidth]{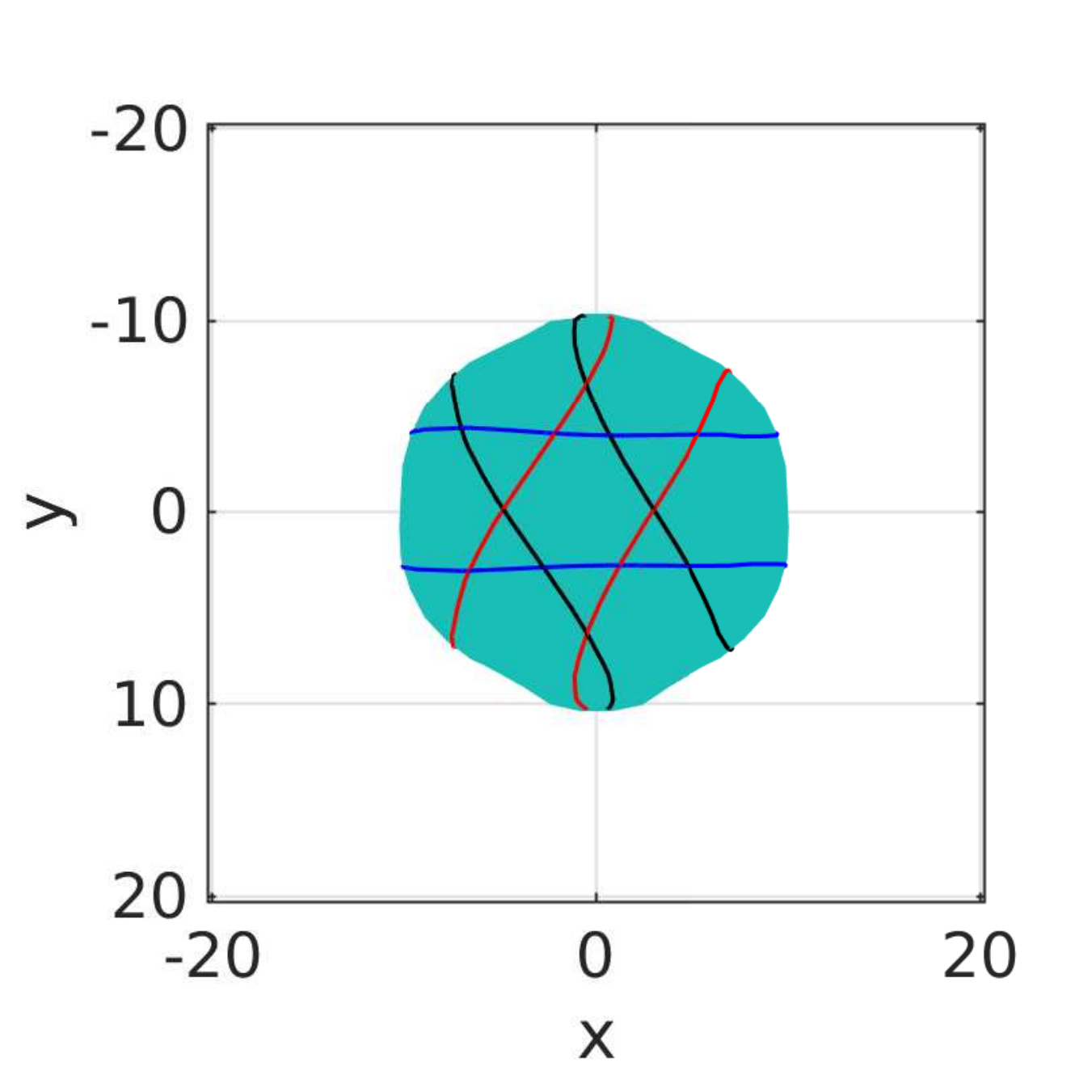}
			\includegraphics[width=\linewidth]{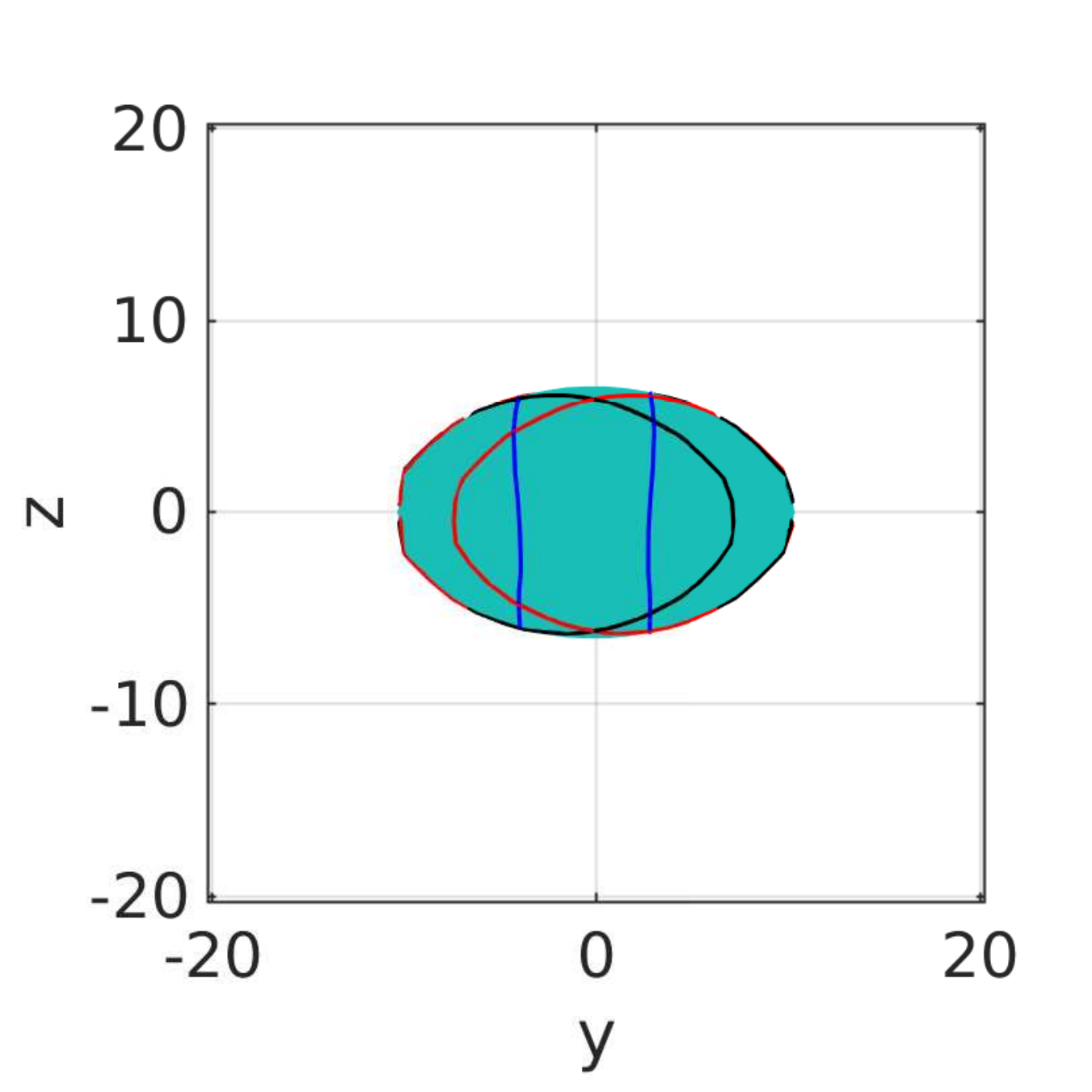}
		\end{minipage}
	}
	\caption{Shrinkage of an initially spherical grain boundary in fcc with dislocation reaction: $M_{\rm r}b^3/M_{\rm d}=1.83\times10^{-4}$. The rotation axis is the $z$ direction ($[111]$), and the initial misorientation angle $\theta=5^\circ$. The upper panel of images show the three-dimensional view of the grain boundary during evolution. The middle panel of images show the grain boundary during evolution viewed from the $+z$ direction ($[111]$), and the lower panel of images show the grain boundary during evolution viewed from the $+x$ direction ($[\bar{1}10]$). Dislocations with Burgers vectors $\mathbf b^{(1)}$, $\mathbf b^{(2)}$ and $\mathbf b^{(3)}$ are shown by blue, black and red lines, respectively. Length unit: $b$. (a) The initial spherical grain boundary. (b), (c), and (d) Configurations at time $t=10/M_{\rm d}\mu, 15/M_{\rm d}\mu, 20/M_{\rm d}\mu$, respectively.}\label{fig:fccfigure}
\end{figure}

Fig.~\ref{fig:fccfigure} shows the shrinkage of the initially spherical grain boundary with dislocation reaction, where the reaction mobility  $M_{\rm r}b^3/M_{\rm d}=1.83\times10^{-4}$.
We consider the cross-section of the grain boundary with the $z=0$ plane (i.e., cross-section normal to the $[111]$ rotation axis), which is the equator of the grain boundary in the three dimensional view in the upper panel in Fig.~\ref{fig:fccfigure} and the outer curve in the view from the $+z$ axis in
 the second panel in Fig.~\ref{fig:fccfigure}.
Along this curve, the grain boundary is pure tilt everywhere, and we have $ v^*_3=0$, i.e., the velocity is always in the $z=0$ plane during the evolution.  The evolution of this curve is similar to that of the two-dimensional grain boundary discussed in \cite{zhang2018motion,zhang2019new}.  The initial circular cross-section gradually changes to a hexagonal shape as it shrinks. Each edge in this hexagon is pure tilt that consists of dislocations of only one Burgers vector. This behavior is consistent with the fact that the energy density of the grain boundary is anisotropic and the pure tilt boundary has the minimum energy of all tilt boundaries, and is the same as the evolution of two dimensional grain boundary with dislocation reaction obtained in \cite{zhang2018motion,zhang2019new}.

The lower panel of Fig.~\ref{fig:fccfigure} shows the evolution of the grain boundary in the view from the $+x$ direction ($[\bar{1}10]$ direction). The cross-section of the grain boundary with the $x=0$ plane gradually changes to an ellipse as it shrinks.
%%meaning that the grain boundary moves faster in the rotation axis direction than in a direction normal to the rotation axis. This behavior and the underlying reason are the same
%similar to those in the simulation without dislocation reaction shown and discussed in Sec.~\ref{subsec:coup}.
These behaviors of the evolution of the initially spherical grain boundary with dislocation reaction in an fcc crystals are similar to the phase field crystal simulation results of an initially spherical  grain boundary in a bcc crystal~\cite{yamanaka2017phase}.

\begin{figure}[htbp]
	\centering
	\subfigure[]{\includegraphics[width=0.45\textwidth]{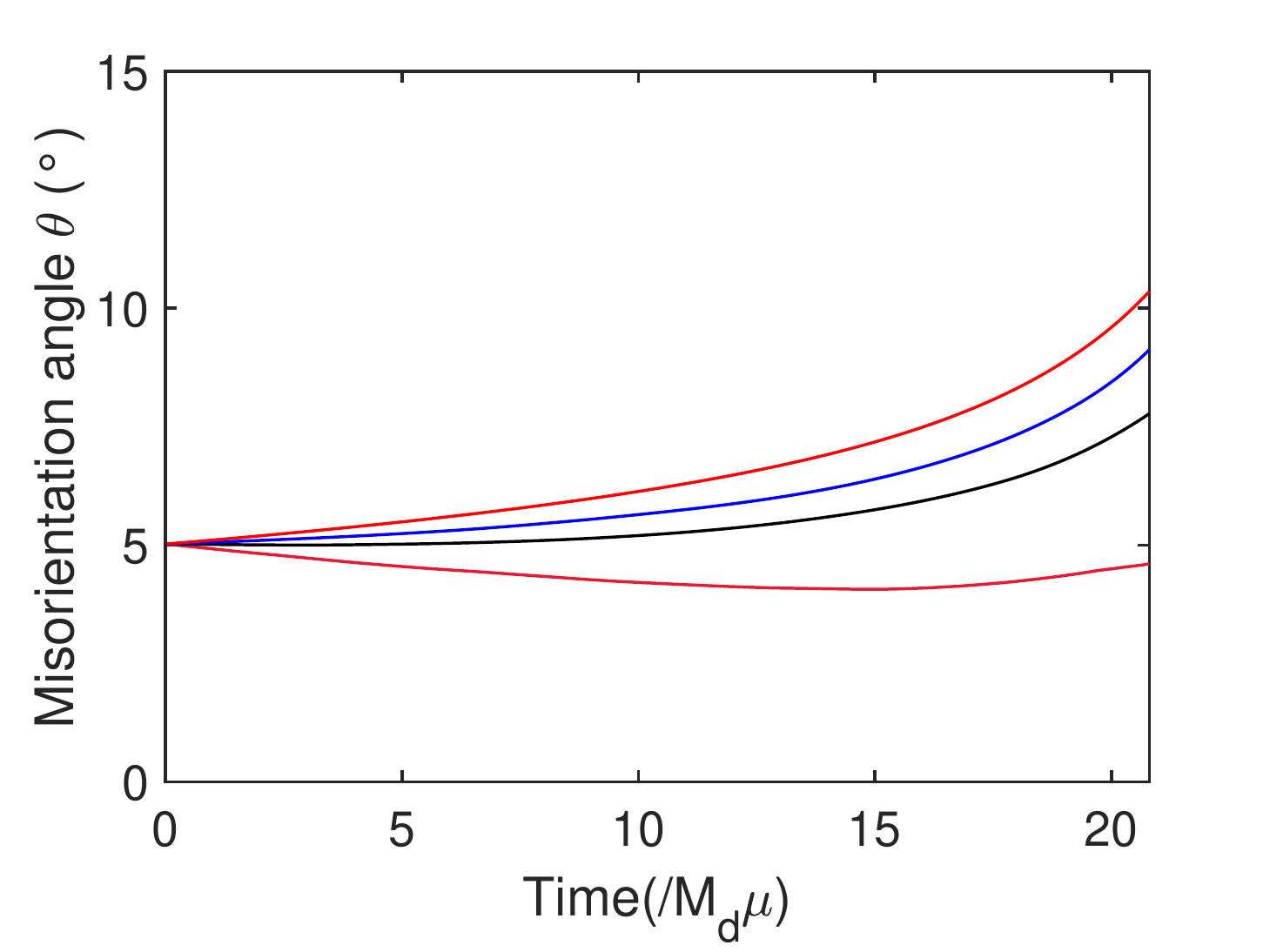}}
	\subfigure[]{\includegraphics[width=0.45\textwidth]{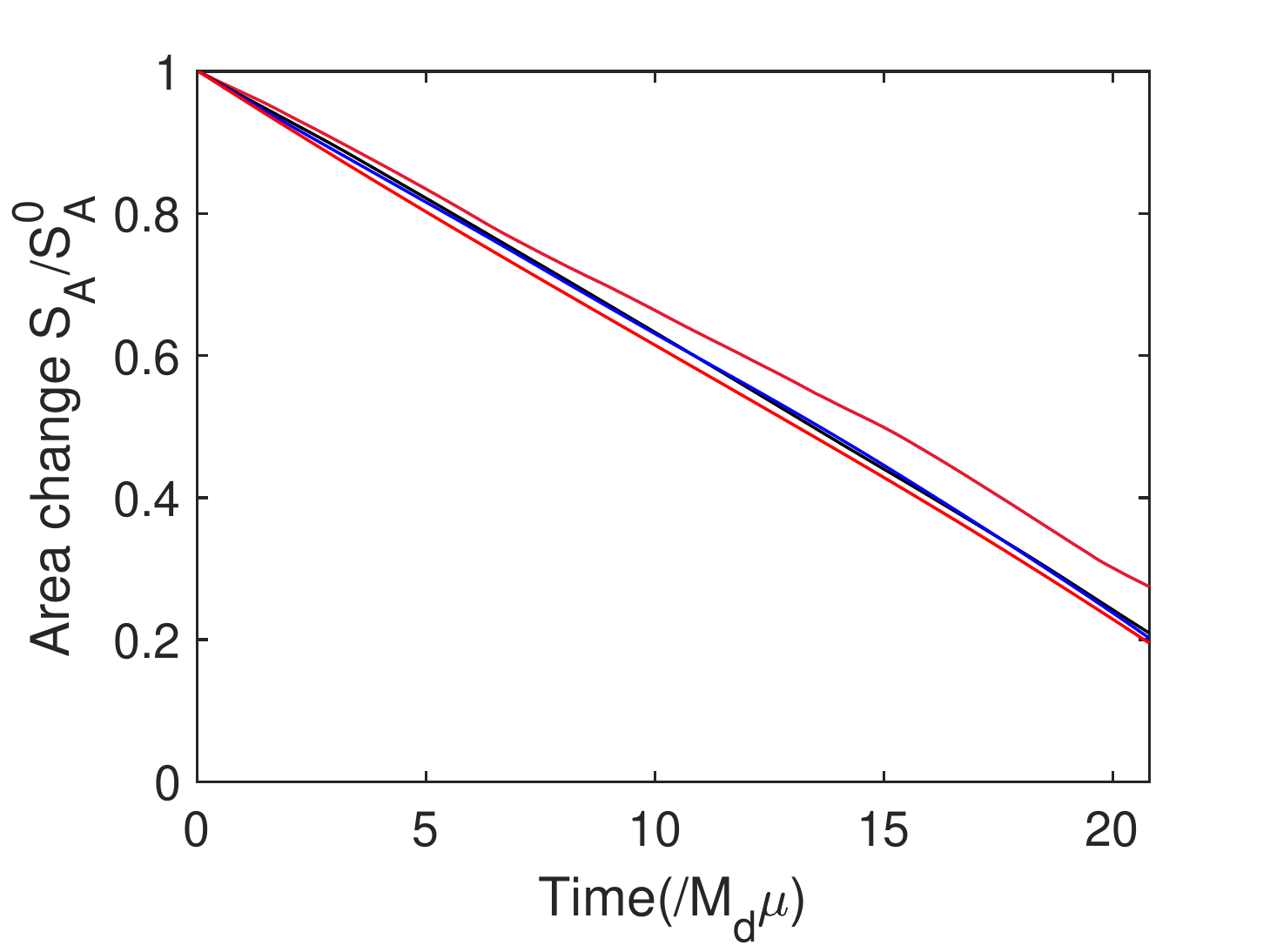}}
	\subfigure[]{\includegraphics[width=0.45\textwidth]{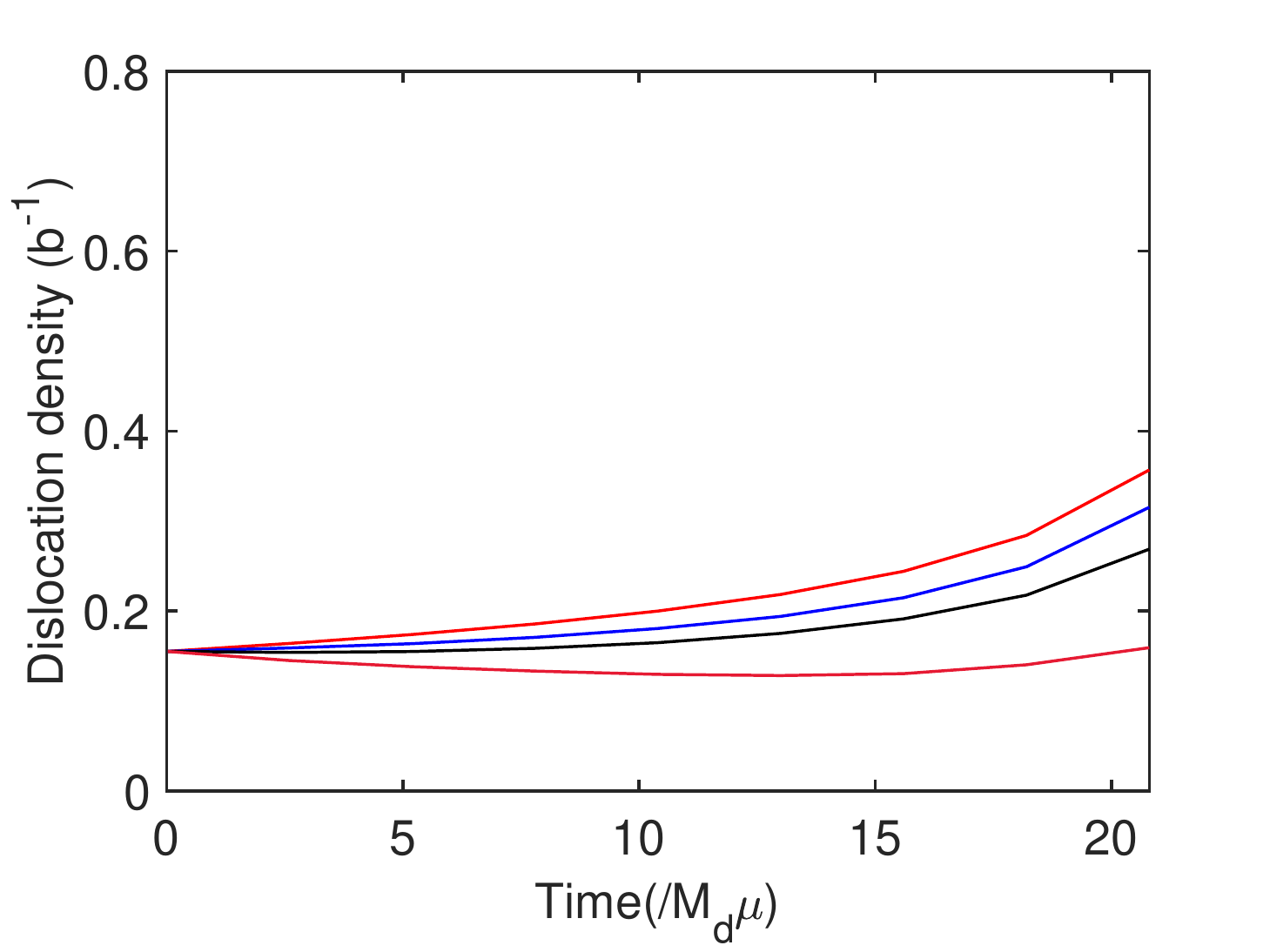}}
	\subfigure[]{\includegraphics[width=0.45\textwidth]{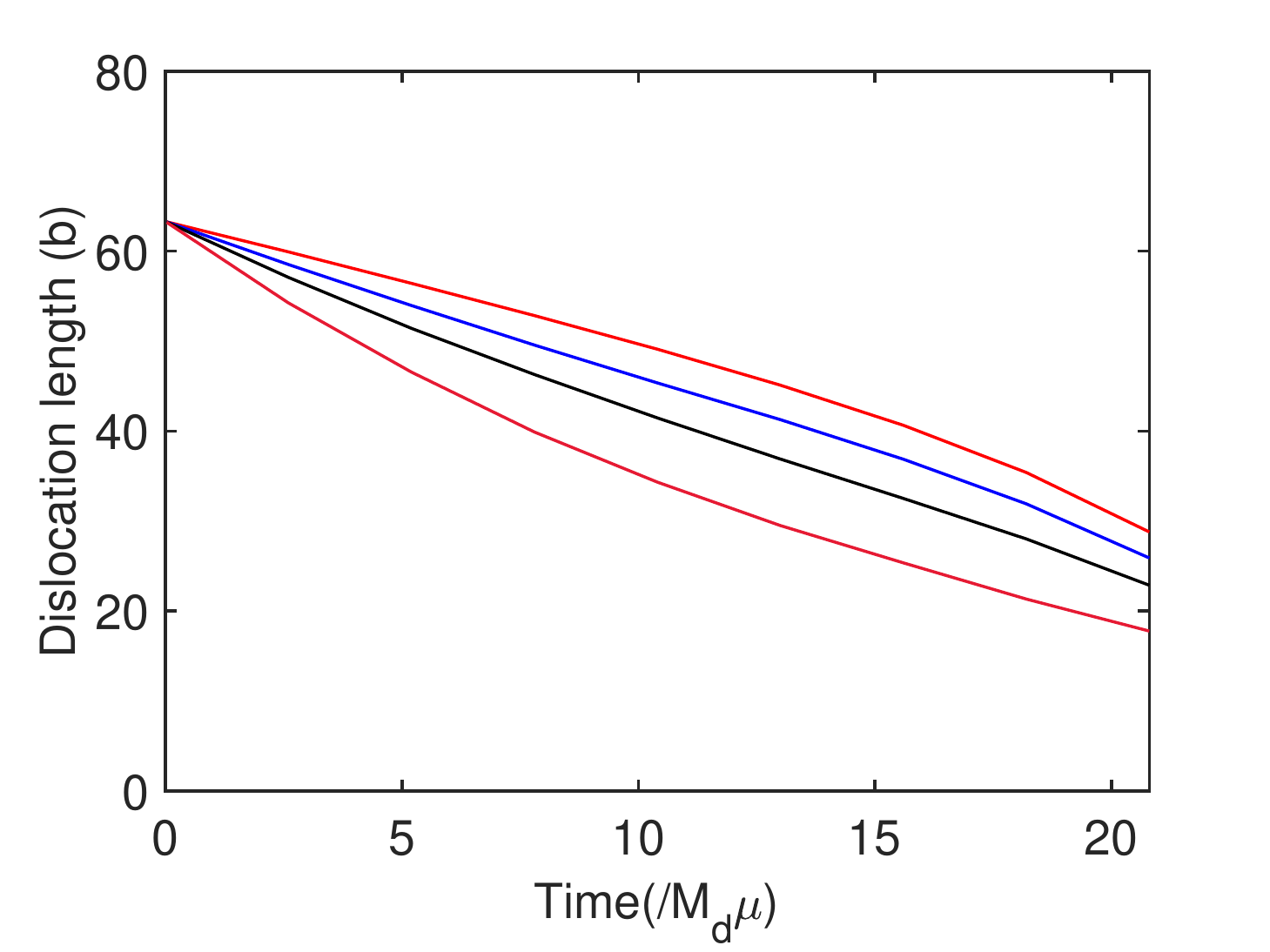}}
	\caption{Shrinkage of an initially spherical grain boundary in fcc with different values of reaction mobility $M_{\rm r}$. The rotation axis is the $z$ direction ($[111]$), and the initial misorientation angle $\theta=5^\circ$.   The reaction mobility $M_{\rm r}b^3/M_{\rm d}=0$, $9.16\times10^{-5}$, $1.83\times10^{-4}$, and $3.74\times10^{-4}$ from the top curve to the bottom one in (a), (c) and (d), and from the bottom to the top ones in (b).
(a) Evolution of  misorientation angle $\theta$. (b) Evolution of grain boundary area $S_A$, where $S_A^0$ is the area of the initial grain boundary. (c) Evolution of density of dislocations with Burgers vector $\mathbf b^{(1)}$/ $\mathbf b^{(2)}$/$\mathbf b^{(3)}$ on the grain boundary. (d) Evolution of the total length of dislocations with Burgers vector $\mathbf b^{(1)}$/ $\mathbf b^{(2)}$/$\mathbf b^{(3)}$ on the grain boundary. In (c) and (d), the densities and total lengths of dislocations with these three Burgers vectors are almost identical.  }\label{fig:fcctheta}
\end{figure}

Evolution of the misorietation angle $\theta$ with different values of reaction mobility $M_{\rm r}$ is shown in Fig.~\ref{fig:fcctheta}(a). When $M_{\rm r}\neq 0$, the evolution of misorientation angle is controlled by both the coupling effect and sliding effect. As can be seen from  Fig.~\ref{fig:fcctheta}(a), the misorientation angle $\theta$ is increasing during the evolution except for the case with very high dislocation reaction mobility;
  as the dislocation reaction mobility $M_{\rm r}$ increases, meaning the sliding effect due to dislocation reaction is becoming stronger, the increase rate of $\theta$ decreases, and when the sliding effect is strong enough, the misorientation angle $\theta$ is decreasing.  These properties are the same as those in the two-dimensional cases \cite{zhang2018motion,zhang2019new}: the coupling motion of grain boundary associated with the conservation of dislocations will increase the misorentation angle $\theta$ during the evolution, and the sliding motion generated by dislocation reaction will decrease $\theta$.  These results also suggest a way to tune the parameter $M_{\rm r}$ based on the evolution of misorientation angle measured by experiments or atomistic simulations.

Fig.~\ref{fig:fcctheta}(b) shows the evolution of grain boundary area with different values of dislocation reaction mobility $M_{\rm r}$. It can be seen that except for the case with very high dislocation reaction mobility, the decrease of grain boundary area still follows the linear law in Eq.~\eqref{eqn:area}, and is almost unchanged with different values of dislocation reaction mobility. In the case with very high dislocation reaction mobility $M_{\rm r}=  3.74\times10^{-4}M_{\rm d}/b^3$, the decrease of grain boundary area starts to deviate from the linear law with slower deceasing rate, which is due to the resulting significant decrease in the grain boundary energy density that slows down the shrinking of the grain boundary. Again, the linear decrease of grain boundary area is consistent with the available phase field crystal and amplitude expansion phase field crystal simulation results \cite{yamanaka2017phase,salvalaglio2018defects}.

Evolutions of dislocation densities on the grain boundary and total length of dislocations with different values of reaction mobility $M_{\rm r}$ are shown in Figs.~\ref{fig:fcctheta}(c) and (d). As can be seen from  Fig.~\ref{fig:fcctheta}(c),  the densities of the dislocations with all the three Burgers vectors are increasing during the evolution except for the case with very high dislocation reaction mobility;
as the dislocation reaction mobility $M_{\rm r}$ increases, the increase rate of dislocation densities decreases, and when the dislocation reaction mobility is high enough, the dislocation densities are decreasing.
 These behaviors are consistent with the increase of misorientation angle $\theta$ during the evolution shown in Fig.~\ref{fig:fcctheta}(a).
Fig.~\ref{fig:fcctheta}(d) shows that
the total length of dislocations is decreasing as the grain boundary shrinks, and the decrease rate is higher for higher dislocation reaction mobility  $M_{\rm r}$.
 The decrease of the total length of dislocations is in agreement with the phase field crystal simulation results in Ref.~\cite{yamanaka2017phase}.

%Note that detailed dislocation reaction mechanisms on a curved grain boundary in three dimensions have been analyzed in Ref.~\cite{yamanaka2017phase} based on their phase field crystal simulations.  There are also mechanisms observed/proposed  based two dimensional molecular dynamics simulations \cite{srinivasan2002challenging,trautt2012grain}. In general, the small dislocation loops and segments of dislocation networks on a curved grain boundary in three dimensions are easier to react than the infinite, straight dislocations in the two dimensional cases.
%

%\newpage
\section{Conclusions}\label{sec:con}
We have developed a continuum model for the dynamics of grain boundaries in three dimensions that incorporates the motion and reaction of the constituent dislocations. The continuum model includes evolution equations for both the motion of the grain boundary and the evolution of dislocation structure on the grain boundary. The evolution of orientation-dependent continuous distributions of dislocation lines on the grain boundary is based on the simple representation using dislocation density potential functions. This simple representation method also guarantees continuity of the dislocation lines on the grain boundaries during the evolution.

In order to overcome the illposedness in formulation that comes from the nonconvexity of the energy density, we use the components of the surface gradients of the dislocation density potential functions instead of these functions directly. Relationship between the components of these surface gradients (i.e. continuity of dislocation lines) is maintained by the projection method during the evolution. The critical but computationally expensive long-range elastic interaction of dislocations is replaced by a projection formulation that maintains the constraint of the Frank's formula describing the equilibrium of the strong long-range interaction. This continuum model is able to describe the grain boundary motion and grain rotation due to both coupling and sliding effects, to which the classical motion by mean curvature model does not apply.

Using the obtained continuum model, simulations are performed for the dynamics of initially spherical low angle grain boundaries in fcc Al, under the conditions without dislocation reaction (pure coupling motion) and with dislocation reaction (with sliding motion). The simulations have shown increase of the misorientation angle as the grain boundary shrinks under the effect of conservation of dislocations, anisotropic motion in the directions along and normal the rotation axis, anisotropic motion in the normal plane with respect to the rotation axis due to dislocation reaction, and linear decrease of grain boundary area.  These results
agree well with those of atomistic simulations (phase field crystal and amplitude expansion phase field crystal simulations) \cite{yamanaka2017phase,salvalaglio2018defects}. The simulation results are also consistent with previously obtained results using continuum model in two dimensions \cite{zhang2018motion,zhang2019new}. In particular, we explain the anisotropic motion in the directions along and normal the rotation axis by the fact that the constraint of Frank's formula only has effect in a direction normal to the rotation axis, and the motion is free in the direction of the rotation axis.

The continuum model presented in this paper provides a basis for continuum simulations of evolution of grain boundary networks at larger length scales \cite{ChenLQ2002,SrolovitzNature2007,DuQ2009}. This will be explored in the future work.
This continuum simulation framework for the distribution and dynamics of curves on  curved surfaces can also be applied more generally beyond  the dynamics of dislocations and grain boundaries.

\section*{Acknowledgement}
This work was supported by the Hong Kong Research Grants Council General Research
Fund 16301720 and 16302818.

\section*{Data availability}
The datasets generated in study are available upon reasonable
 request.

\appendix
\section{Derivation of the formula for misorientation angle $\theta$ in \eqref{eqn:mod1t}}

Substituting $\mathbf V_1= \mathbf r_u$ and $\mathbf V_2= \mathbf r_v$ into Frank's formula in Eq.~\eqref{eqn:mod2frank},
we have
\begin{flalign}
\theta(\mathbf r_u\times\mathbf{a})-\sum_{j=1}^{J}\mathbf{b}^{(j)}\eta_{ju}=&0,\label{eqn:F1} \\
\theta(\mathbf r_v\times\mathbf{a})-\sum_{j=1}^{J}\mathbf{b}^{(j)}\eta_{jv}=&0.\label{eqn:F2}
\end{flalign}
Here we have used $\nabla_S \eta_j\bm{\cdot}\mathbf r_u=\eta_{ju}$ and $\nabla_S \eta_j\bm{\cdot}\mathbf r_v=\eta_{jv}$. Adding the two equations \eqref{eqn:F1} and \eqref{eqn:F2}, multiplying both size of the summation by $( \mathbf r_u+ \mathbf r_v) \times\mathbf{a} $, we have
\begin{equation}
\theta \|(  \mathbf r_u+ \mathbf r_v) \times\mathbf{a}\|^2=
 \sum_{j=1}^J  ( \eta_{ju}+  \eta_{jv})( \mathbf r_u + \mathbf r_v )\times\mathbf{a}{\cdot}\mathbf{b}^{(j)}.
\end{equation}
Integrating over the entire grain boundary $S$, we obtain the formula of $\theta$ in Eq.~\eqref{eqn:mod1t}.
% which is exact if the Franks' formula holds and serves as an estimate formula otherwise.

\bibliography{ref}

\begin{thebibliography}{10}
\providecommand{\url}[1]{{#1}}
\providecommand{\urlprefix}{URL }
\expandafter\ifx\csname urlstyle\endcsname\relax
  \providecommand{\doi}[1]{DOI~\discretionary{}{}{}#1}\else
  \providecommand{\doi}{DOI~\discretionary{}{}{}\begingroup
  \urlstyle{rm}\Url}\fi

\bibitem{AdmalIJP2018}
Admal, N.C., Po, G., Marian, J.: A unified framework for polycrystal plasticity
  with grain boundary evolution.
\newblock Int. J. Plasticity \textbf{106}, 1--30 (2018)

\bibitem{AskJMPS2018}
Ask, A., Forest, S., Appolaire, B., Ammar, K., Salman, O.U.: A cosserat crystal
  plasticity and phase field theory for grain boundary migration.
\newblock J. Mech. Phys. Solids \textbf{115}, 167--194 (2018)

\bibitem{Gupta2014}
Basak, A., Gupta, A.: A two-dimensional study of coupled grain boundary motion
  using the level set method.
\newblock Modell. Simul. Mater. Sci. Eng. \textbf{22}, 055022 (2014)

\bibitem{Bilby1955}
Bilby, B.A.: Bristol conference report on defects in crystalline materials.
\newblock Phys. Soc., London p. 123 (1955)

\bibitem{cahn2006coupling}
Cahn, J.W., Mishin, Y., Suzuki, A.: Coupling grain boundary motion to shear
  deformation.
\newblock Acta Mater. \textbf{54}, 4953--4975 (2006)

\bibitem{cahn2004unified}
Cahn, J.W., Taylor, J.E.: A unified approach to motion of grain boundaries,
  relative tangential translation along grain boundaries, and grain rotation.
\newblock Acta Mater. \textbf{52}, 4887--4898 (2004)

\bibitem{chen1994computer}
Chen, L.Q., Yang, W.: Computer simulation of the domain dynamics of a quenched
  system with a large number of nonconserved order parameters: The grain-growth
  kinetics.
\newblock Phys. Rev. B \textbf{50}, 15752--15756 (1994)

\bibitem{Chorin1968}
Chorin, A.J.: Numerical solution of the navier-stokes equations.
\newblock Math. Comp. \textbf{22}, 745--762 (1968)

\bibitem{chorin1990mathematical}
Chorin, A.J., Marsden, J.E., Marsden, J.E.: A mathematical introduction to
  fluid mechanics, vol. 168.
\newblock Springer (1990)

\bibitem{ODE}
Coddingtong, E.A., Levinson, N.: Theory of Ordinary Differential Equations.
\newblock McGraw-Hill, New York (1955)

\bibitem{dai2018convergence}
Dai, S., Li, B., Lu, J.: Convergence of phase-field free energy and boundary
  force for molecular solvation.
\newblock Arch. Ration. Mech. Anal. \textbf{227}, 105--147 (2018)

\bibitem{Du-Feng2020}
Du, Q., Feng, X.B.: The phase field method for geometric moving interfaces and
  their numerical approximations.
\newblock Handbook of Numerical Analysis \textbf{21}, 425--508 (2020)

\bibitem{Selim2009}
Elsey, M., Esedoglu, S., Smereka, P.: Diffusion generated motion for grain
  growth in two and three dimensions.
\newblock J. Comput. Phys. \textbf{228}, 8015--8033 (2009)

\bibitem{epshteyn2019motion}
Epshteyn, Y., Liu, C., Mizuno, M.: Motion of grain boundaries with dynamic
  lattice misorientations and with triple junctions drag.
\newblock SIAM J. Math. Anal. \textbf{53}, 3072–3097 (2021)

\bibitem{esedoglu2016grain}
Esedoglu, S.: Grain size distribution under simultaneous grain boundary
  migration and grain rotation in two dimensions.
\newblock Comput. Mater. Sci. \textbf{121}, 209--216 (2016)

\bibitem{feng2003numerical}
Feng, X., Prohl, A.: Numerical analysis of the {Allen-Cahn} equation and
  approximation for mean curvature flows.
\newblock Numer. Math. \textbf{94}, 33--65 (2003)

\bibitem{Frank1950}
Frank, F.C.: The resultant content of dislocations in an arbitrary
  intercrystalline boundary.
\newblock pp. 150--154. Office of Naval Research, Pittsburgh (1950)

\bibitem{Molodov2009}
Gorkaya, T., Molodov, D.A., Gottstein, G.: Stress-driven migration of
  symmetrical $<100>$ tilt grain boundaries in al bicrystals.
\newblock Acta Mater. \textbf{57}, 5396--5405 (2009)

\bibitem{harris1998grain}
Harris, K., Singh, V., King, A.: Grain rotation in thin films of gold.
\newblock Acta Mater. \textbf{46}, 2623--2633 (1998)

\bibitem{Herring1951}
Herring, C.: Surface tension as a motivation for sintering.
\newblock In: W.E. Kingston (ed.) The Physics of Powder Metallurgy, pp.
  143--179. McGraw-Hill, New York (1951)

\bibitem{kazaryan2000generalized}
Kazaryan, A., Wang, Y., Dregia, S.A., Patton, B.R.: Generalized phase-field
  model for computer simulation of grain growth in anisotropic systems.
\newblock Phys. Rev. B \textbf{61}, 14275--14278 (2000)

\bibitem{liuchun2001}
Kinderlehrer, D., Liu, C.: Evolution of grain boundaries.
\newblock Math. Models Methods Appl. Sci. \textbf{4}, 713--729 (2001)

\bibitem{Kirch2006}
Kirch, D.M., Jannot, E., Barrales-Mora, L.A., Molodov, D.A., Gottstein, G.:
  Inclination dependence of grain boundary energy and its impact on the
  faceting and kinetics of tilt grain boundaries in aluminum.
\newblock Acta Mater. \textbf{56}, 4998--5011 (2006)

\bibitem{Kobayashi2000}
Kobayashi, R., Warren, J.A., Carter, W.C.: A continuum model of grain
  boundaries.
\newblock Phys. D \textbf{140}, 141--150 (2000)

\bibitem{ChenLQ2002}
{Krill III}, C., Chen, L.Q.: Computer simulation of 3-d grain growth using a
  phase-field model.
\newblock Acta Mater. \textbf{50}, 3059--3075 (2002)

\bibitem{lazar2010more}
Lazar, E.A., MacPherson, R.D., Srolovitz, D.J.: A more accurate two-dimensional
  grain growth algorithm.
\newblock Acta Mater. \textbf{58}, 364--372 (2010)

\bibitem{DuQ2009}
Le, T., Du, Q.: A generalization of the three-dimensional macpherson-srolovitz
  formula.
\newblock Commun. Math. Sci. \textbf{7}, 511--520 (2009)

\bibitem{li1953stress}
Li, C.H., Edwards, E.H., Washburn, J., Parker, E.R.: Stress-induced movement of
  crystal boundaries.
\newblock Acta Metall. \textbf{1}, 223--229 (1953)

\bibitem{li1962possibility}
Li, J.C.: Possibility of subgrain rotation during recrystallization.
\newblock J. Appl. Phys. \textbf{33}, 2958--2965 (1962)

\bibitem{SrolovitzNature2007}
MacPherson, R., Srolovitz, D.: The von neumann relation generalized to
  coarsening of three-dimensional microstructures.
\newblock Nature \textbf{446}, 1053--105 (2007)

\bibitem{mcreynolds2016grain}
McReynolds, K., Wu, K.A., Voorhees, P.: Grain growth and grain translation in
  crystals.
\newblock Acta Mater. \textbf{120}, 264--272 (2016)

\bibitem{molodov2007low}
Molodov, D.A., Ivanov, V.A., Gottstein, G.: Low angle tilt boundary migration
  coupled to shear deformation.
\newblock Acta Mater. \textbf{55}, 1843--1848 (2007)

\bibitem{mullins1956two}
Mullins, W.W.: Two-dimensional motion of idealized grain boundaries.
\newblock J. Appl. Phys. \textbf{27}, 900--904 (1956)

\bibitem{Qin2020}
Qin, X.X., Gu, Y.J., Zhang, L.C., Xiang, Y.: Continuum model and numerical
  method for dislocation structure and energy of grain boundaries.
\newblock arXiv p. arXiv:2101.02596 (2021)

\bibitem{Rath2007}
Rath, B.B., Winning, M., Li, J.C.M.: Coupling between grain growth and grain
  rotation.
\newblock Appl. Phys. Lett. \textbf{90}, 161915 (2007)

\bibitem{salvalaglio2018defects}
Salvalaglio, M., Backofen, R., Elder, K., Voigt, A.: Defects at grain
  boundaries: A coarse-grained, three-dimensional description by the amplitude
  expansion of the phase-field crystal model.
\newblock Phys. Rev. Mater. \textbf{2}, 053804 (2018)

\bibitem{shewmon1966energy}
Shewmon, P.G.: In: H.~Margolin (ed.) Recrystallization, grain growth and
  textures, pp. 165--199. American Society of Metals, Metals Park (1966)

\bibitem{srinivasan2002challenging}
Srinivasan, S.G., Cahn, J.W.: Challenging some free-energy reduction criteria
  for grain growth.
\newblock In: S.~Ankem, C.S. Pande, I.~Ovid'ko, S.~Ranganathan (eds.) Science
  and Technology of Interfaces, pp. 3--14. TMS, Seattle (2002)

\bibitem{Sutton1995}
Sutton, A., Balluffi, R.: Interfaces in Crystalline Materials.
\newblock Clarendon Press, Oxford (1995)

\bibitem{Taylor2007}
Taylor, J.E., Cahn, J.W.: Shape accommodation of a rotating embedded crystal
  via a new variational formulation.
\newblock Interfaces and Free Boundaries \textbf{9}, 493--512 (2007)

\bibitem{trautt2012grain}
Trautt, Z., Mishin, Y.: Grain boundary migration and grain rotation studied by
  molecular dynamics.
\newblock Acta Mater. \textbf{60}, 2407--2424 (2012)

\bibitem{upmanyu2002boundary}
Upmanyu, M., Hassold, G.N., Kazaryan, A., Holm, E.A., Wang, Y., Patton, B.,
  Srolovitz, D.J.: Boundary mobility and energy anisotropy effects on
  microstructural evolution during grain growth.
\newblock Interface Sci. \textbf{10}, 201--216 (2002)

\bibitem{upmanyu2006simultaneous}
Upmanyu, M., Srolovitz, D.J., Lobkovsky, A.E., Warren, J.A., Carter, W.C.:
  Simultaneous grain boundary migration and grain rotation.
\newblock Acta Mater. \textbf{54}, 1707--1719 (2006)

\bibitem{wu2012phase}
Wu, K.W., Voorhees, P.W.: Phase field crystal simulations of nanocrystalline
  grain growth in two dimensions.
\newblock Acta Mater. \textbf{60}, 407--419 (2012)

\bibitem{yamanaka2017phase}
Yamanaka, A., McReynolds, K., Voorhees, P.W.: Phase field crystal simulation of
  grain boundary motion, grain rotation and dislocation reactions in a bcc
  bicrystal.
\newblock Acta Mater. \textbf{133}, 160--171 (2017)

\bibitem{zhang2005curvature}
Zhang, H., Upmanyu, M., Srolovitz, D.J.: Curvature driven grain boundary
  migration in aluminum: molecular dynamics simulations.
\newblock Acta Mater. \textbf{53}, 79--86 (2005)

\bibitem{zhang2009numerical}
Zhang, J., Du, Q.: Numerical studies of discrete approximations to the
  {Allen-Cahn} equation in the sharp interface limit.
\newblock SIAM J. Sci. Comput. \textbf{31}, 3042--3063 (2009)

\bibitem{zhang2019new}
Zhang, L., Xiang, Y.: A new formulation of coupling and sliding motions of
  grian boundaries based on dislocation structure.
\newblock SIAM J. Appl. Math. \textbf{80}, 2365--2387 (2020)

\bibitem{zhang2017energy}
Zhang, L.C., Gu, Y.J., Xiang, Y.: Energy of low angle grain boundaries based on
  continuum dislocation structure.
\newblock Acta Mater. \textbf{126}, 11--24 (2017)

\bibitem{zhang2018motion}
Zhang, L.C., Xiang, Y.: Motion of grain boundaries incorporating dislocation
  structure.
\newblock J. Mech. Phys. Solids \textbf{117}, 157--178 (2018)

\bibitem{Zhu-Xiang2012}
Zhu, X.H., Xiang, Y.: A continuum model for the dynamics of dislocation arrays.
\newblock Commun. Math. Sci. \textbf{10}, 1081--1103 (2012)

\bibitem{zhu2014continuum}
Zhu, X.H., Xiang, Y.: Continuum framework for dislocation structure, energy and
  dynamics of dislocation arrays and low angle grain boundaries.
\newblock J. Mech. Phys. Solids \textbf{69}, 175--194 (2014)

\end{thebibliography}


\begin{thebibliography}{1}
\providecommand{\url}[1]{{#1}}
\providecommand{\urlprefix}{URL }
\expandafter\ifx\csname urlstyle\endcsname\relax
  \providecommand{\doi}[1]{DOI~\discretionary{}{}{}#1}\else
  \providecommand{\doi}{DOI~\discretionary{}{}{}\begingroup
  \urlstyle{rm}\Url}\fi

\bibitem{salvalaglio2018defects}
Salvalaglio, M., Backofen, R., Elder, K., Voigt, A.: Defects at grain
  boundaries: A coarse-grained, three-dimensional description by the amplitude
  expansion of the phase-field crystal model.
\newblock Phys. Rev. Mater. \textbf{2}, 053804 (2018)

\bibitem{srinivasan2002challenging}
Srinivasan, S.G., Cahn, J.W.: Challenging some free-energy reduction criteria
  for grain growth.
\newblock In: S.~Ankem, C.S. Pande, I.~Ovid'ko, S.~Ranganathan (eds.) Science
  and Technology of Interfaces, pp. 3--14. TMS, Seattle (2002)

\bibitem{Taylor2007}
Taylor, J.E., Cahn, J.W.: Shape accommodation of a rotating embedded crystal
  via a new variational formulation.
\newblock Interfaces and Free Boundaries \textbf{9}, 493--512 (2007)

\bibitem{wu2012phase}
Wu, K.W., Voorhees, P.W.: Phase field crystal simulations of nanocrystalline
  grain growth in two dimensions.
\newblock Acta Mater. \textbf{60}, 407--419 (2012)

\bibitem{yamanaka2017phase}
Yamanaka, A., McReynolds, K., Voorhees, P.W.: Phase field crystal simulation of
  grain boundary motion, grain rotation and dislocation reactions in a bcc
  bicrystal.
\newblock Acta Mater. \textbf{133}, 160--171 (2017)

\bibitem{zhang2019new}
Zhang, L., Xiang, Y.: A new formulation of coupling and sliding motions of
  grian boundaries based on dislocation structure.
\newblock SIAM J. Appl. Math. \textbf{80}, 2365--2387 (2020)

\bibitem{zhang2018motion}
Zhang, L.C., Xiang, Y.: Motion of grain boundaries incorporating dislocation
  structure.
\newblock J. Mech. Phys. Solids \textbf{117}, 157--178 (2018)

\end{thebibliography}

\end{document}

%\begin{frontmatter}

\title{A Three-Dimensional Continuum Simulation Method for Grain Boundary Motion Incorporating Dislocation Structure\\
{\bf Supplementary Materials}}%\tnoteref{mytitlenote}}
%\tnotetext[mytitlenote]{Fully documented templates are available in the elsarticle package on %\href{http://www.ctan.org/tex-archive/macros/latex/contrib/elsarticle}{CTAN}.}

%%% Group authors per affiliation:
%\author{Xiaoxue Qin}
%\author{Luchan Zhang\corref{mycorrespondingauthor}}
%\ead{malczhang@ust.hk}
%\author{Yang Xiang\corref{mycorrespondingauthor}}
%\ead{maxiang@ust.hk}
%\address{Department of Mathematics, Hong Kong University of Science and Technology, Clear Water Bay, Kowloon, Hong Kong}
%%\author[mymainaddress,mysecondaryaddress]{Department of Mathematics, Hong Kong University of Science and Technology}
%%\ead[url]{www.elsevier.com}
%
%%\author[mysecondaryaddress]{Global Customer Service\corref{mycorrespondingauthor}}
%%\cortext[mycorrespondingauthor]{Corresponding author}
%%\ead{xqinac@connect.ust.hk, lzhangas@connect.ust.hk, maxiang@ust.hk}
%%
%%\address[mymainaddress]{Department of Mathematics, Hong Kong University of Science and Technology}
%%\address[mysecondaryaddress]{Clearwater Bay, Kowloon, Hong Kong}
%%\ead{malczhang@ust.hk,maxiang@ust.hk}

\titlerunning{A Three-Dimensional Continuum Simulation Method for Grain Boundary Motion}

\author{Xiaoxue Qin \and Luchan Zhang \and Yang Xiang}

\authorrunning{X. X. Qin, L. C. Zhang, Y. Xiang}

\institute{X. X. Qin \at Department of Mathematics, Hong Kong University of Science and Technology, Clear Water Bay, Kowloon, Hong Kong. \email{maxqin@ust.hk}
\and L. C. Zhang \at College of Mathematics and Statistics, Shenzhen University, Shenzhen, 518060,  China. Corresponding author. \email{zhanglc@szu.edu.cn}
\and Y. Xiang \at Department of Mathematics, Hong Kong University of Science and Technology, Clear Water Bay, Kowloon, Hong Kong. Corresponding author. \email{maxiang@ust.hk} }

\date{}

\maketitle

\section{Numerical simulations of bcc spherical grain boundaries with rotation axis [011]}\label{sec:nr}
In this section, we perform numerical simulations of grain boundary dynamics using our numerical algorithm presented in the main text.

Now we consider grain boundaries in bcc Fe. We choose the directions  $[100]$, $[01\bar 1]$ and $[011]$ to be the $x$, $y$ and $z$ directions, respectively. There seven possible Burgers vectors, and in this coordinate system, they are $\mathbf{b}^{(1)}=\left(\frac{1}{2},0,\frac{\sqrt{2}}{2}\right)a$, $\mathbf{b}^{(2)}=\left(\frac{1}{2},\frac{\sqrt{2}}{2},0\right)a$,
$\mathbf{b}^{(3)}=\left(\frac{1}{2},-\frac{\sqrt{2}}{2},0\right)a$,
$\mathbf{b}^{(4)}=\left(-\frac{1}{2},0,\frac{\sqrt{2}}{2}\right)a$,
$\mathbf{b}^{(5)}=(1,0,0)a$, $\mathbf{b}^{(6)}=\left(0,\frac{\sqrt{2}}{2},\frac{\sqrt{2}}{2}\right)a$, and
$\mathbf{b}^{(7)}=\left(0,-\frac{\sqrt{2}}{2},\frac{\sqrt{2}}{2}\right)a$, where $a$ is the lattice constant.  For bcc Fe, a$=0.2856 \rm nm$  and the Poisson ratio is $\nu=0.29$.  The rotation axis is $\mathbf{a}=(0,0,1)$, i.e., in the $[011]$ direction.

We also study the evolution of an initially spherical grain boundary, whose radius is $R=20b$ and misorientation angle  is $\theta = 4^\circ $.  There are three sets of dislocations with Burgers vectors $\mathbf b^{(2)}$, $\mathbf{b}^{(3)}$, $\mathbf{b}^{(5)}$, respectively,  in the equilibrium dislocation structure on this initial, spherical grain boundary; see the top image in Fig.~\ref{fig:bccfigure0}(a).
Parametrization and discretization of the grain boundary are the same as those for the grain boundary in fcc in the main text.

%
%The grain boundary is parameterized using spherical coordinates $R=R(\alpha,\beta)$ for $0\leq \alpha < 2\pi$, and $0\leq \beta\leq \pi$. Initially, $R$ is constant. The $(\alpha,\beta)$ domain is discretized into $40\times20$ grids with uniform. Here $\alpha$ is the angle between the position vector of points on grain boundary and the $x$ axis and $\beta$ is the angle between the position vector of points on grain boundary and the $z$ axis. During the evolution steps, we update the points $R(\alpha,\beta)$ on the grain boundary with this uniform mesh of the $(\alpha, \beta)$ domain fixed.
%%
%\begin{figure}[htbp]
%	\centering
%	\includegraphics[width=3in]{figures/bcc3D.eps}
%	\caption{The equilibrium dislocation structure with three sets of dislocations.}\label{fig:bcc3D}
%\end{figure}

\subsection{Pure coupling motion} \label{subsec:coupbcc}

We first consider the motion of this grain boundary  without dislocation reaction, i.e. the reaction mobility $M_{\rm r}=0$ in Eq.~(54) (and Eq.~(43)). This is the pure coupling motion.

%We firstly focus on the grain boundary motion with the conservation of the number of grain boundary dislocations. This means that the reaction mobility $M_r=0$, and the grain boundary motion is pure coupling motion. The initial misorientation angle is $\theta = 4^\circ $, the initial radius of the spherical grain boundary is $R=20b$.

\begin{figure}[htbp]
\centering
\subfigure[]{
	\begin{minipage}{0.21\linewidth}
		\includegraphics[width=\linewidth]{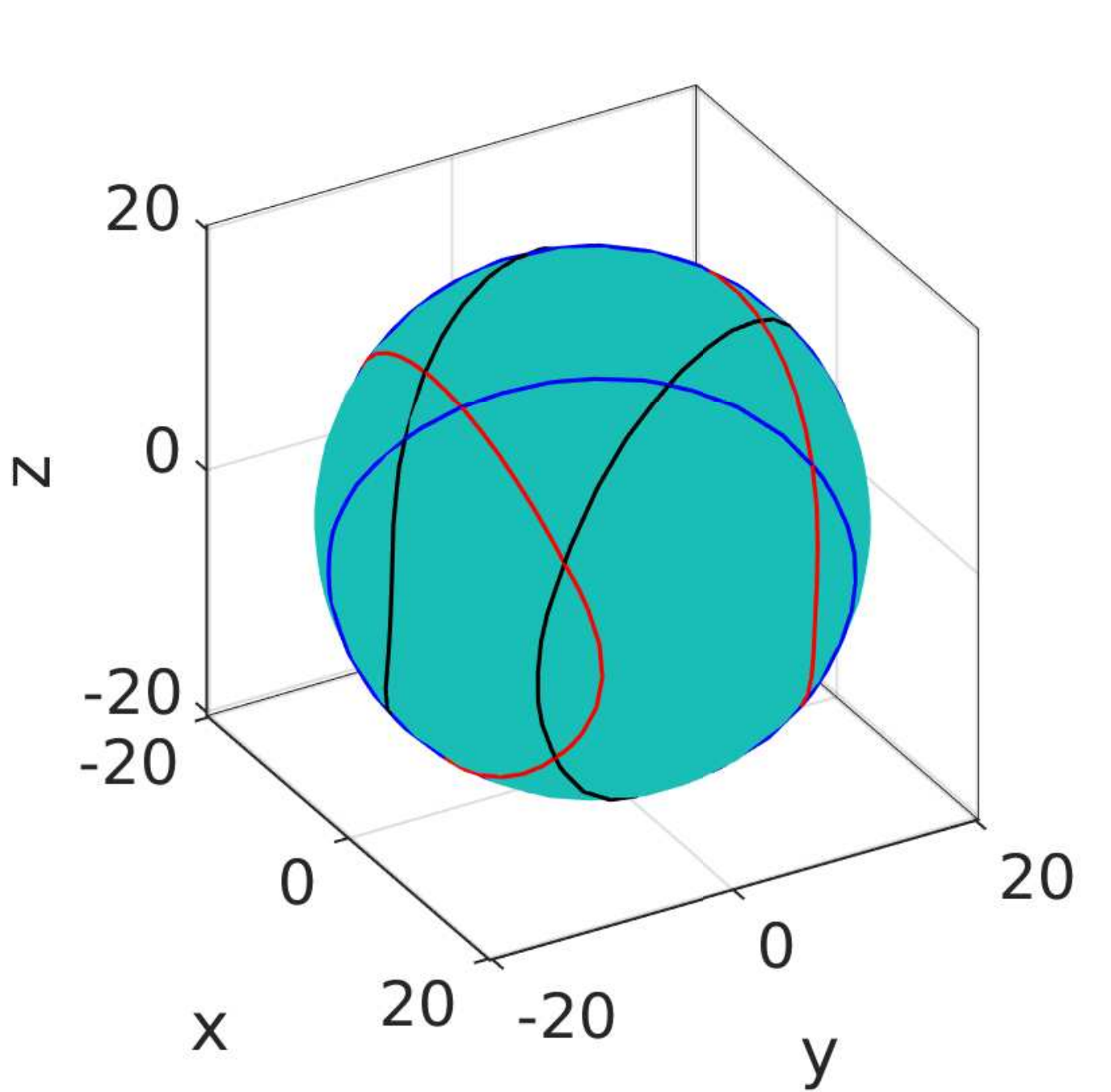}
		\includegraphics[width=\linewidth]{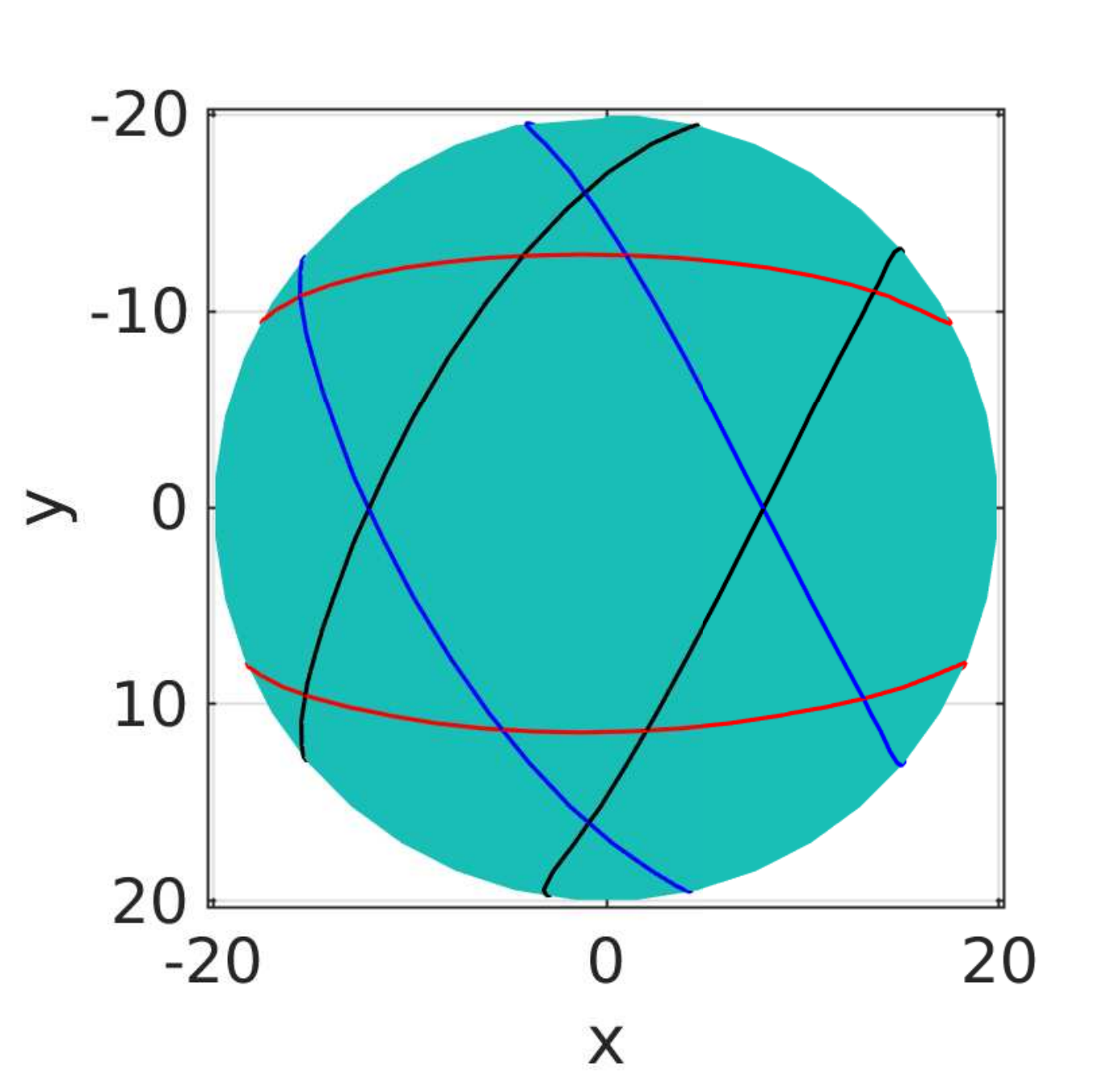}
		\includegraphics[width=\linewidth]{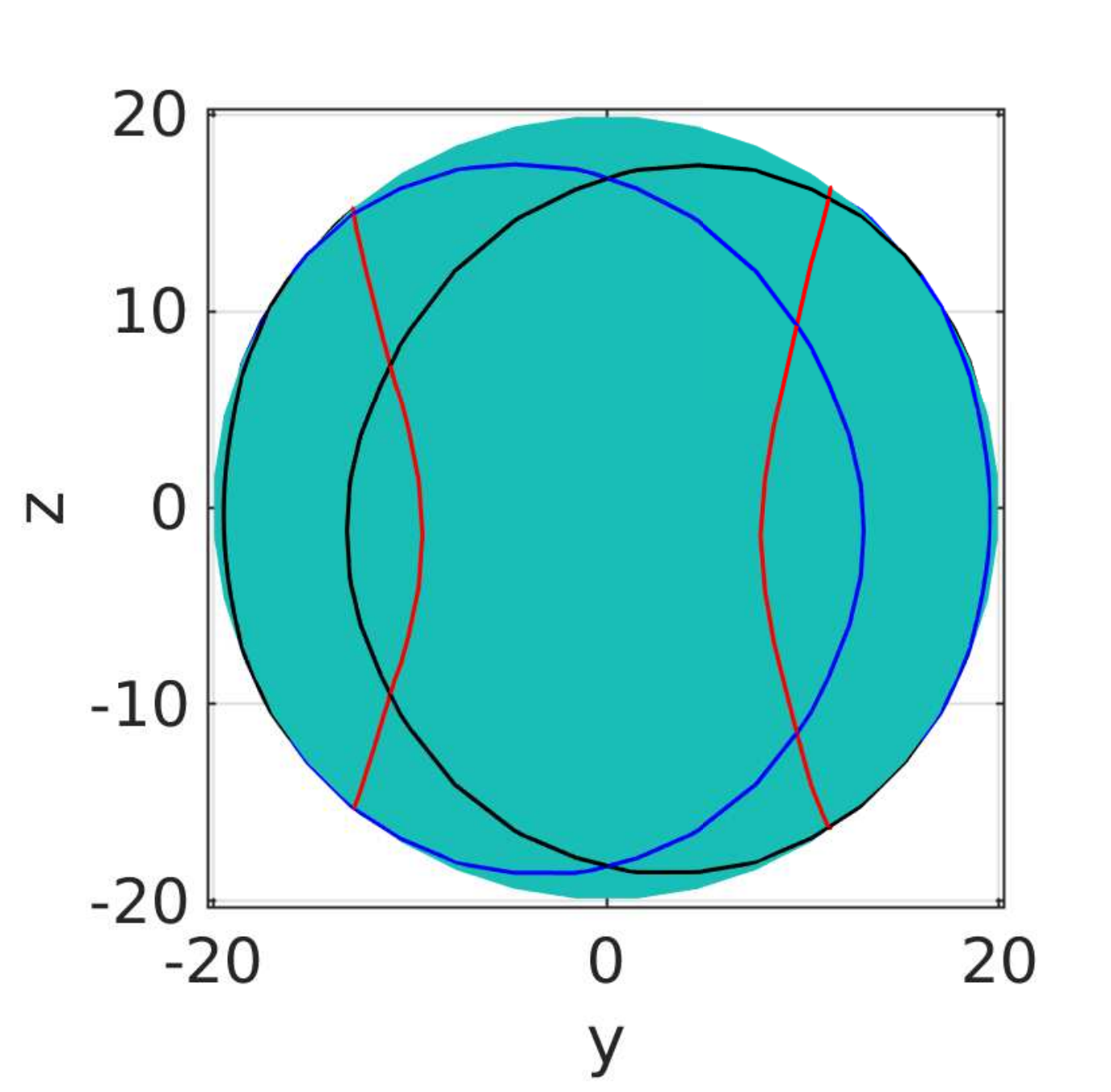}
	\end{minipage}
}
\subfigure[]{
	\begin{minipage}{0.21\linewidth}			
		\includegraphics[width=\linewidth]{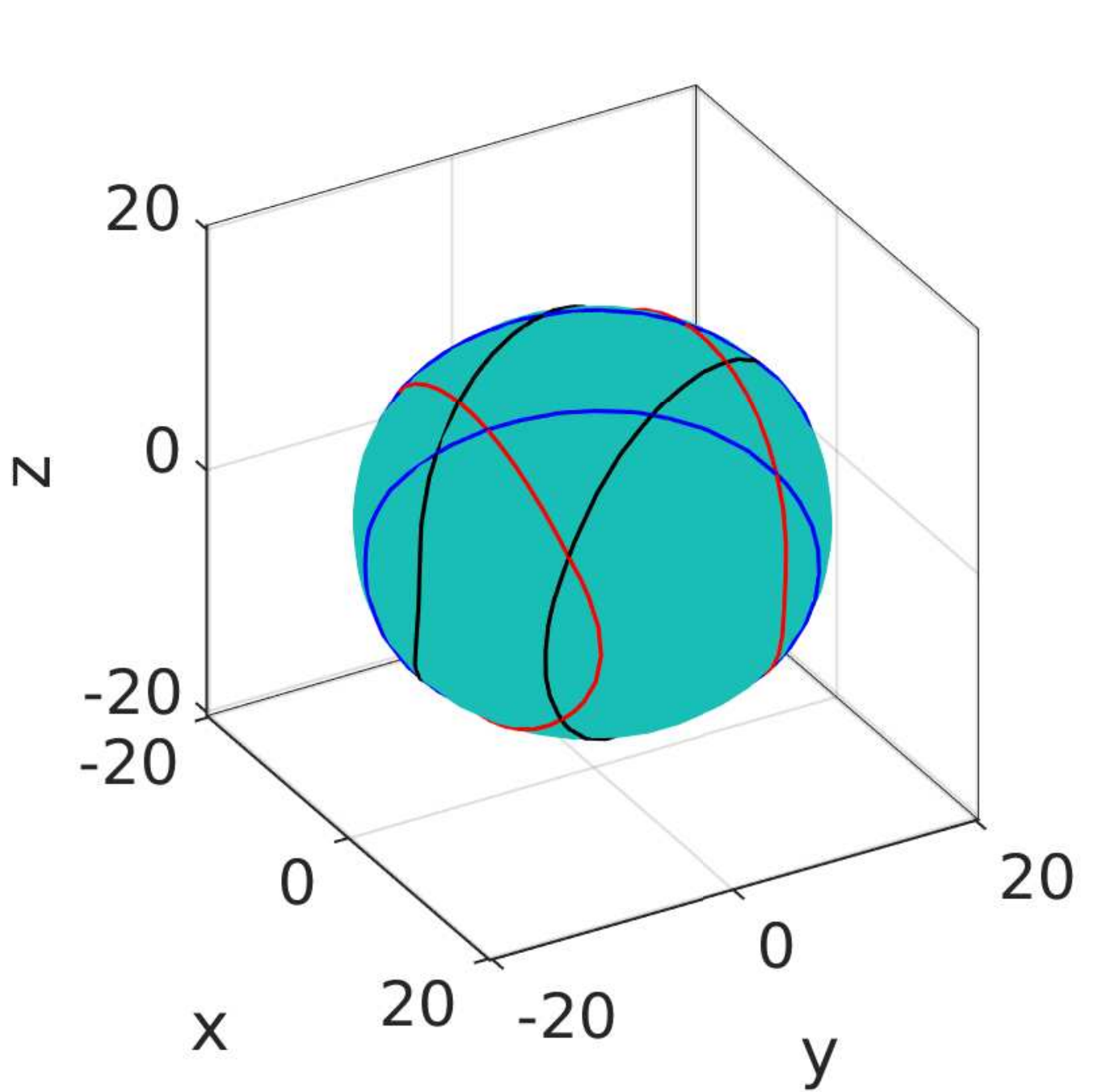}
		\includegraphics[width=\linewidth]{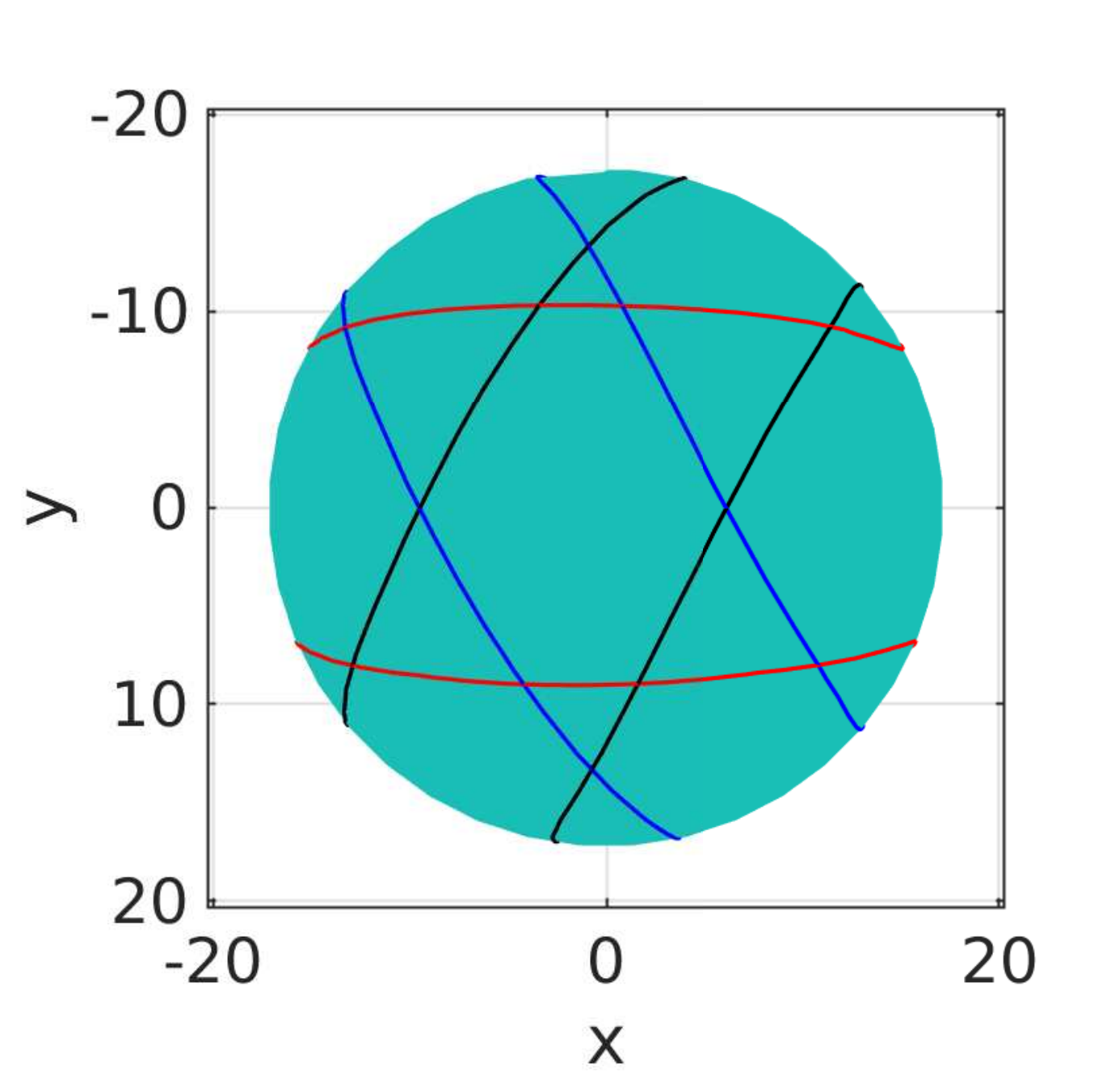}
		\includegraphics[width=\linewidth]{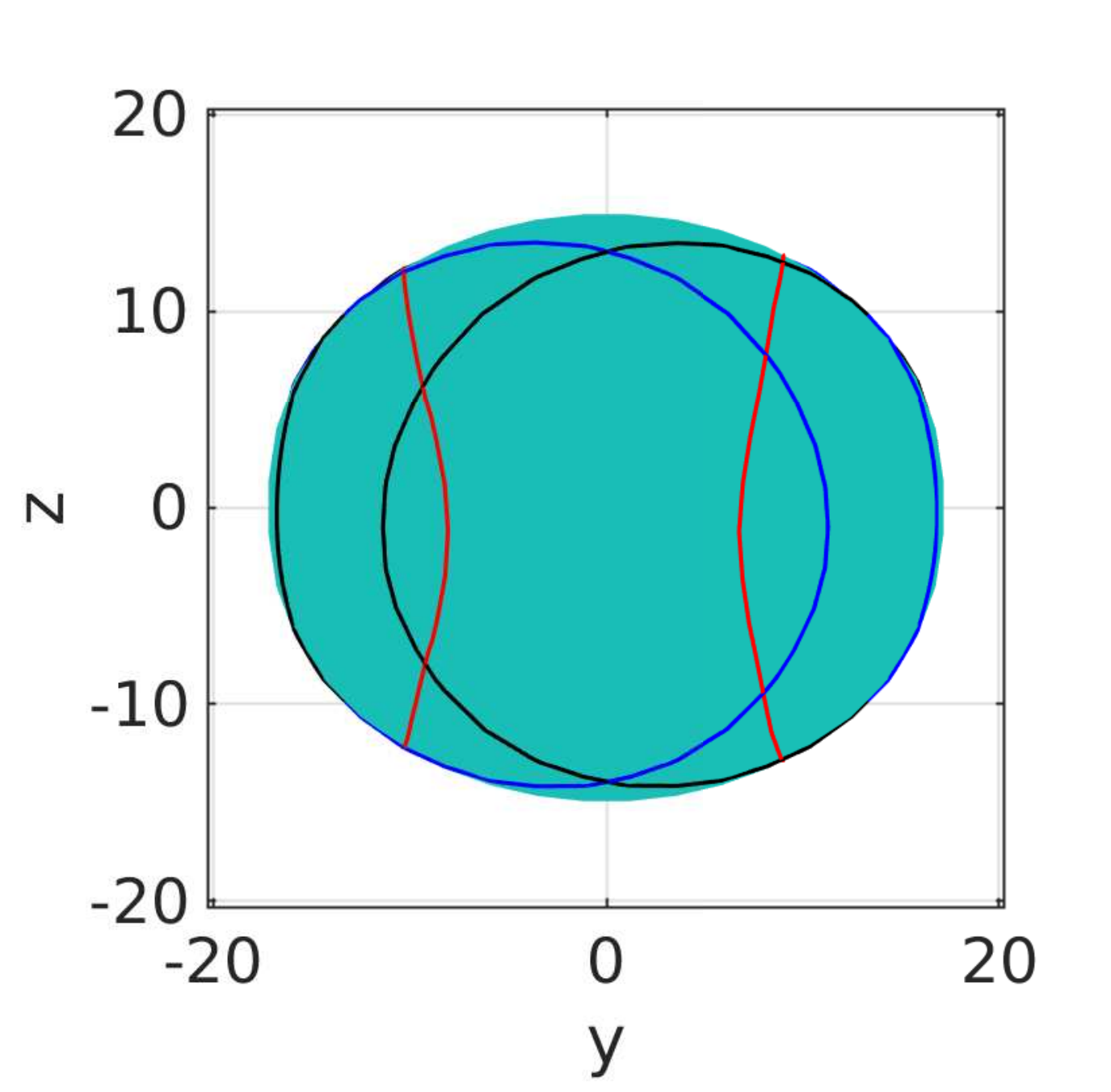}
	\end{minipage}
}	
\subfigure[]{
	\begin{minipage}{0.21\linewidth}			
		\includegraphics[width=\linewidth]{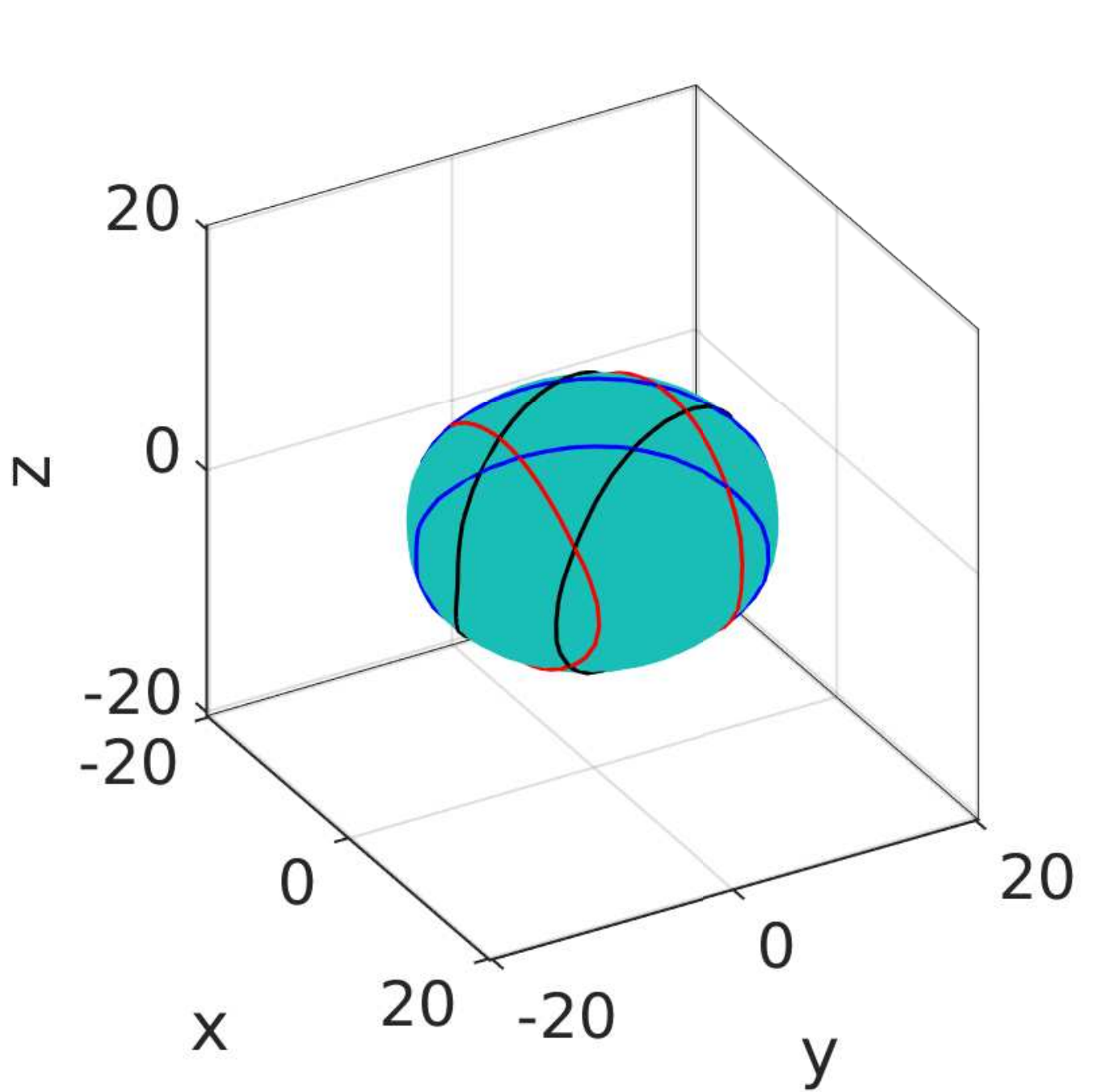}
		\includegraphics[width=\linewidth]{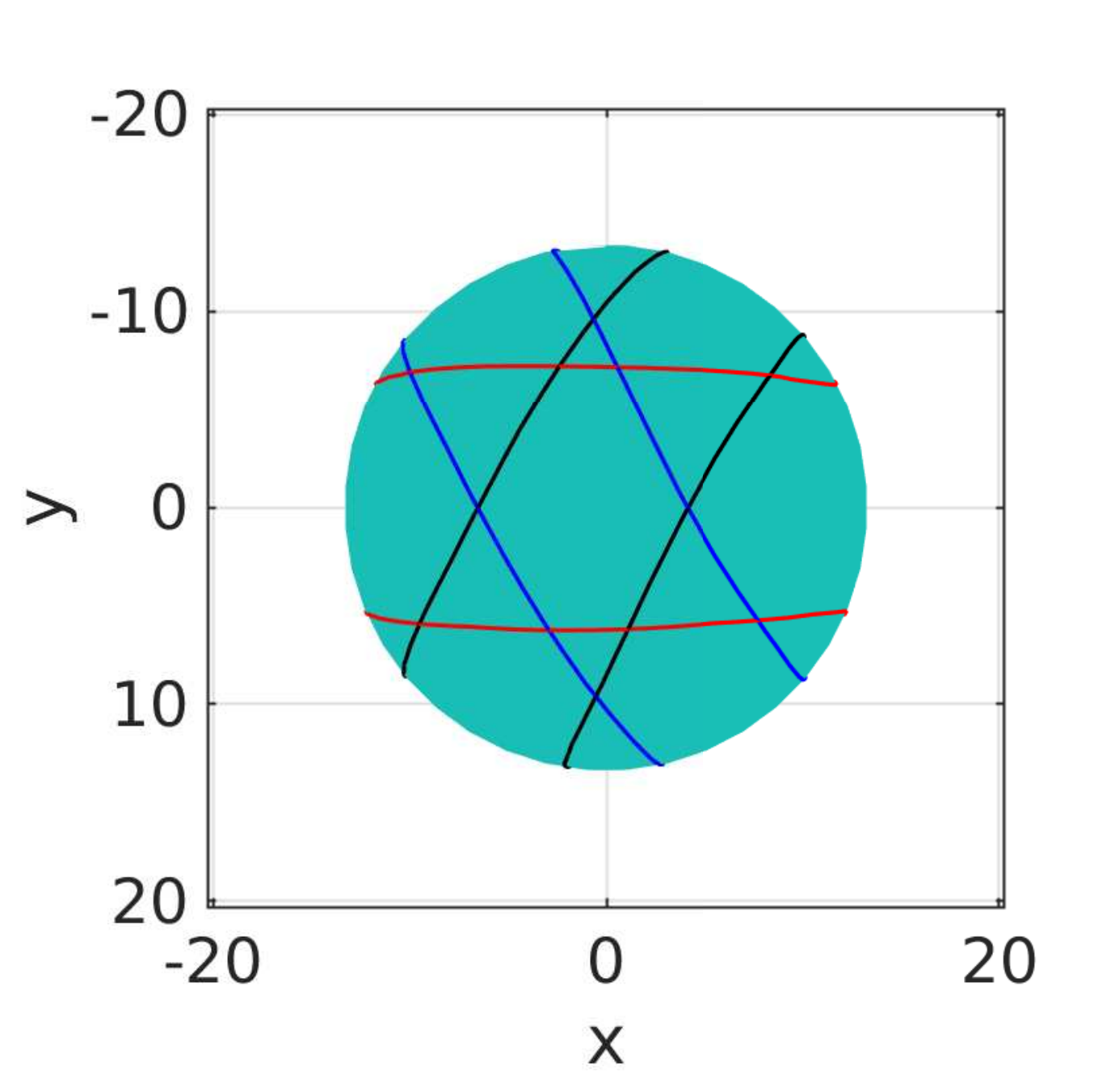}
		\includegraphics[width=\linewidth]{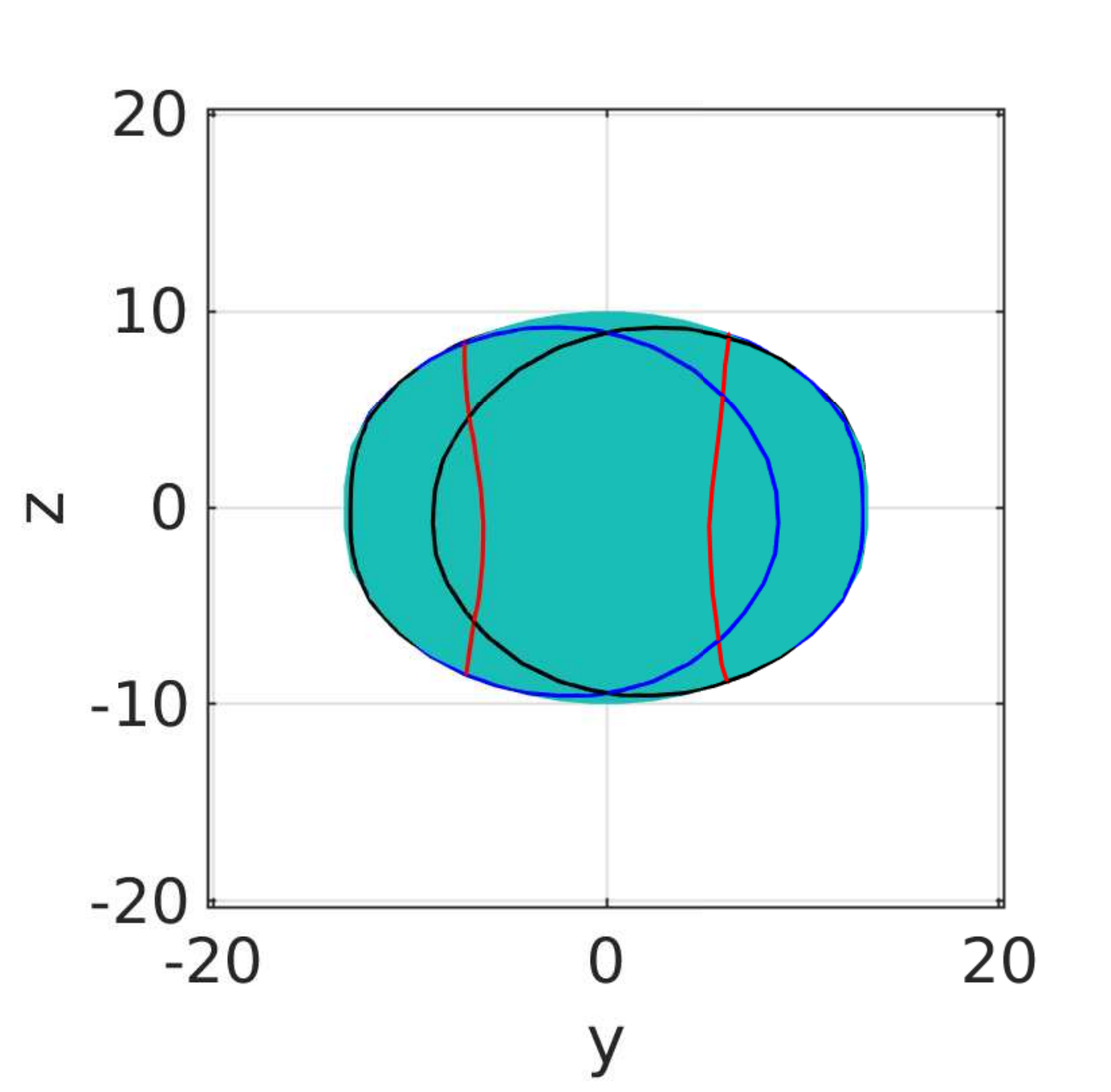}
	\end{minipage}
}	
\subfigure[]{
	\begin{minipage}{0.21\linewidth}			
		\includegraphics[width=\linewidth]{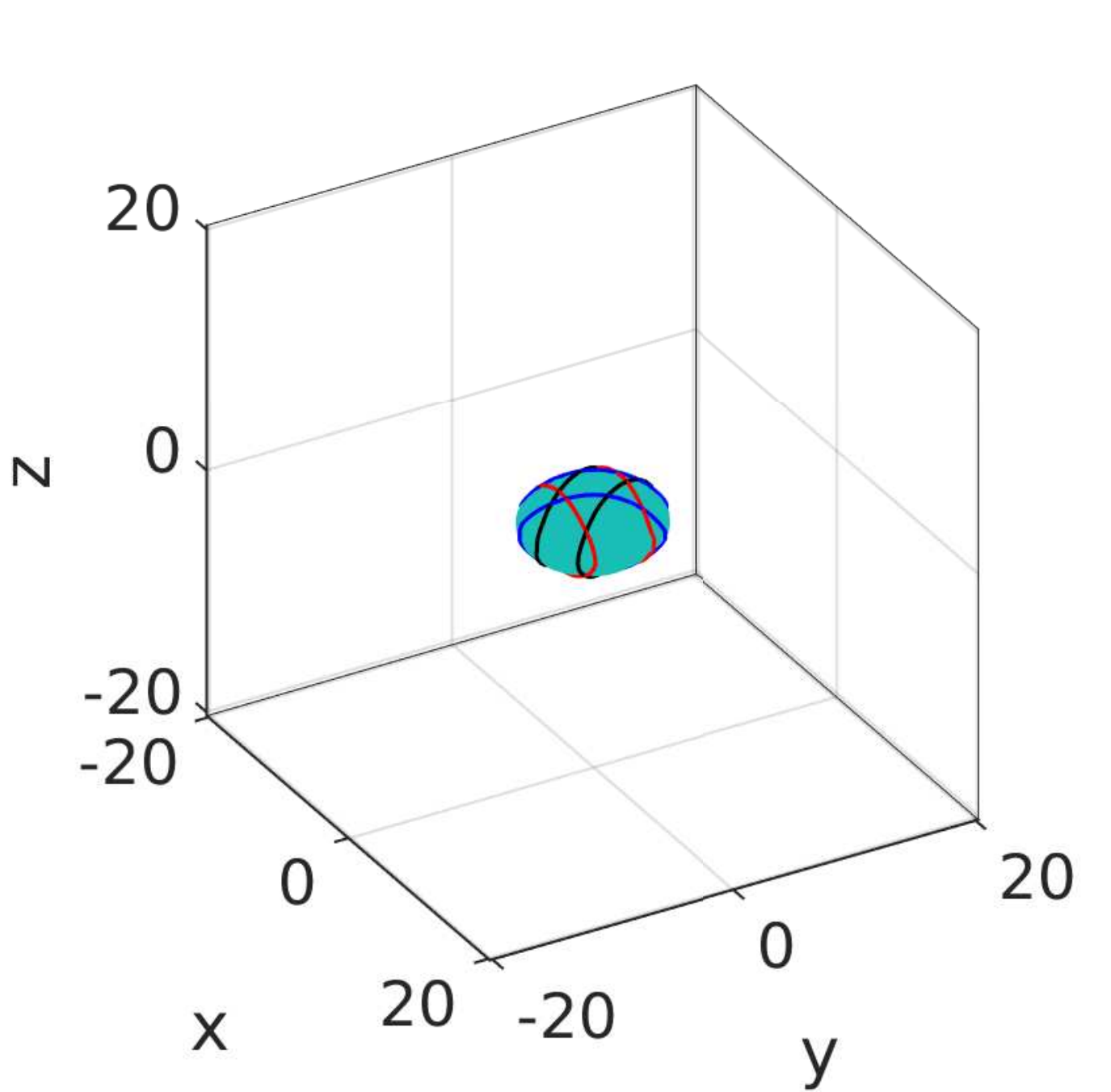}
		\includegraphics[width=\linewidth]{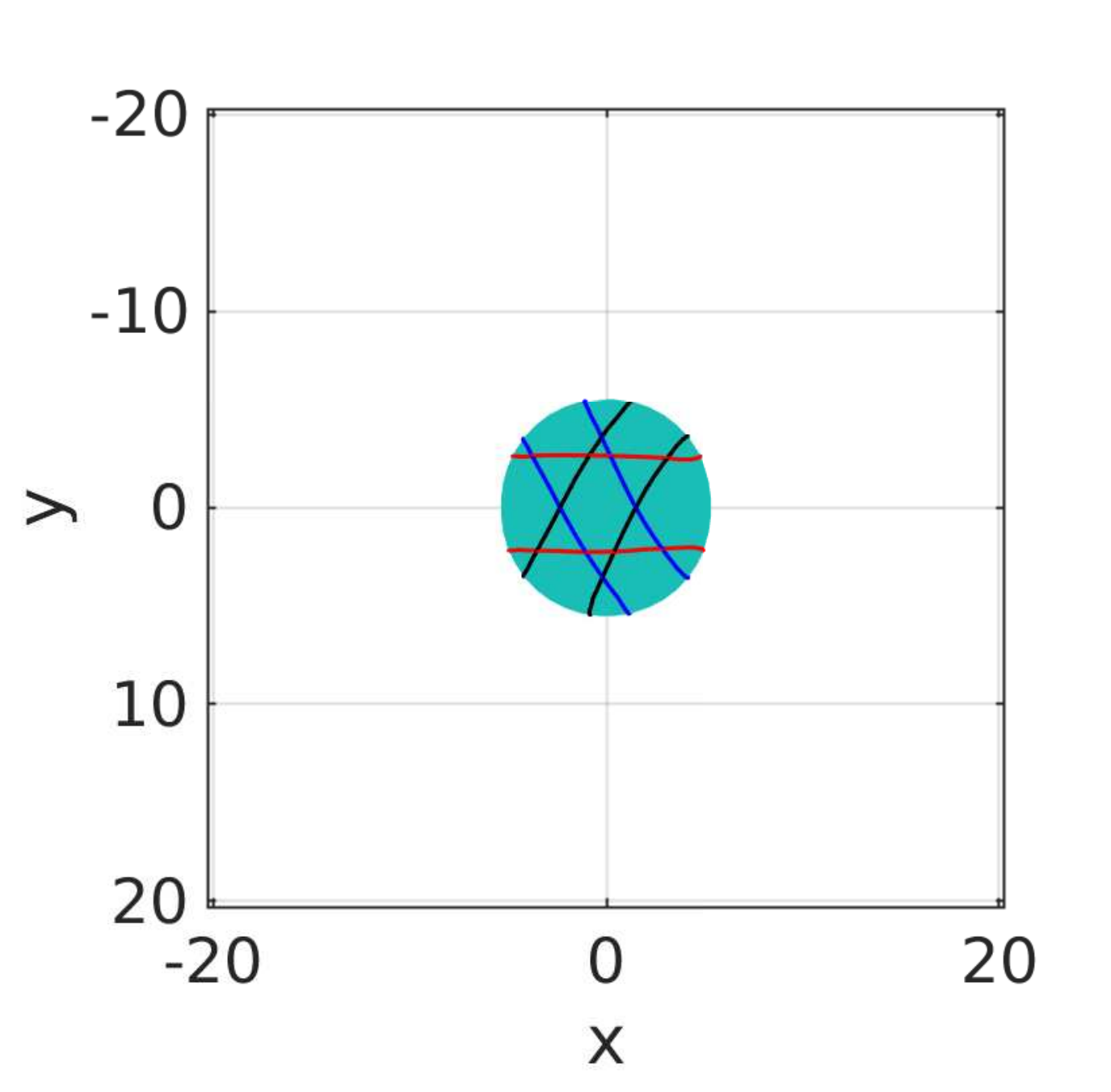}
		\includegraphics[width=\linewidth]{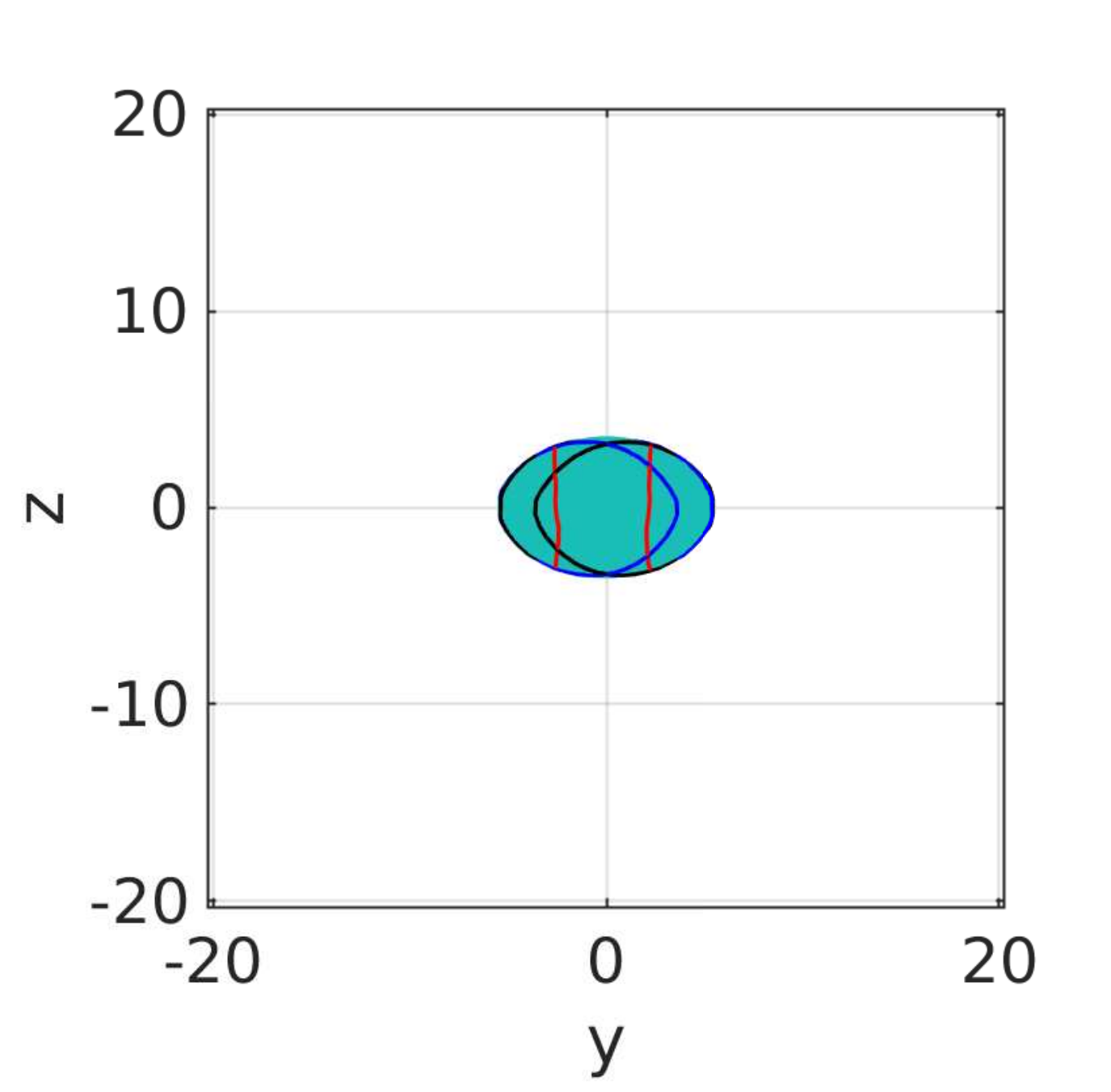}
	\end{minipage}
}
	\caption{Shrinkage of an initially spherical grain boundary in bcc under pure coupling motion, i.e., without dislocation reaction. The rotation axis is the $z$ direction ($[011]$), and the initial misorientation angle $\theta=4^\circ$. The upper panel of images show the three-dimensional view of the grain boundary during evolution. The middle panel of images show the grain boundary during evolution viewed from the $+z$ direction ($[011]$), and the lower panel of images show the grain boundary during evolution viewed from the $+x$ direction ($[100]$). Dislocations with Burgers vectors $\mathbf b^{(2)}$, $\mathbf b^{(3)}$ and $\mathbf b^{(5)}$ are shown by blue, black and red lines, respectively. Length unit: $b$. (a) The initial spherical grain boundary. (b), (c), and (d) Configurations at time $t=3/M_{\rm d}\mu, 6/M_{\rm d}\mu, 9/M_{\rm d}\mu$, respectively.
}\label{fig:bccfigure0}
\end{figure}

Fig.~\ref{fig:bccfigure0} shows the shrinkage of this spherical grain boundary in bcc under the pure coupling motion.
Evolutions of the grain boundary and its dislocation structure are similar to those of the grain boundary in fcc discussed in the main text. Especially, the shrinkage of the grain boundary is faster in the direction of the rotation axis than in other directions, which agrees with the results of atomistic simulations using phase field crystal models \cite{yamanaka2017phase,salvalaglio2018defects}. Again, this anisotropic motion can be explained based on our continuum model by the constraint of Frank's formula in any direction normal to the rotation axis. The shape-preserving evolution of the equator of the grain boundary (with respect to the rotation axis) agrees with the results of the two-dimensional grain boundary dynamics models \cite{srinivasan2002challenging,Taylor2007,wu2012phase,zhang2018motion,zhang2019new}.

%Fig. \ref{fig:bccfigure0} shows the shrinkage of a spherical grain boundary under these conditions. In this case, since $M_r=0$, the grain boundary velocity is $\mathbf v =-\frac{\delta \theta}{\theta\delta t}(x,y,0)+(0,0, v^*_3)$ from Eq. \eqref{eqn:fv2}. If we focus on the circular  at $z=0$, which is the outer circular seen on the top column in Fig. \ref{fig:bccfigure0}, the grain boundary is pure tilt. On the spherical grain boundary, at $z=0$, we have $\mathbf v^*_3=0$, and the velocity is $\mathbf v =-\frac{\delta \theta}{\theta\delta t}(x,y,0)$, which is in the inward radial direction.
%
%Now we focus on the circle at $x=0$ on the grain boundary, and the lower column in Fig. \ref{fig:bccfigure0}. The circle whose evolution is shown in gradually changes to an ellipse during the evolution. This is because the velocity in the rotation axis direction,  i.e. $+z$ direction is much higher than the other two direction. If we view from $+z$ direction, the grain boundary at $x=0,y=0$ is pure twist, and the dislocations are screw dislocations. If we view from $+x$ direction, the grain boundary at $y=0,z=0$ is pure tilt,  and the dislocations are edge dislocations. The results indicate that the velocity of screw dislocations tends to be higher than that of the edge dislocation. This results agrees with the high temperature case of bcc iron spherical grain boundary in \cite{yamanaka2017phase} and the fcc, bcc spherical grain boundaries in \cite{salvalaglio2018defects} obtained using phase-field crystal model.

\begin{figure}[htbp]
	\centering
	\subfigure[]{\includegraphics[width=2.3in]{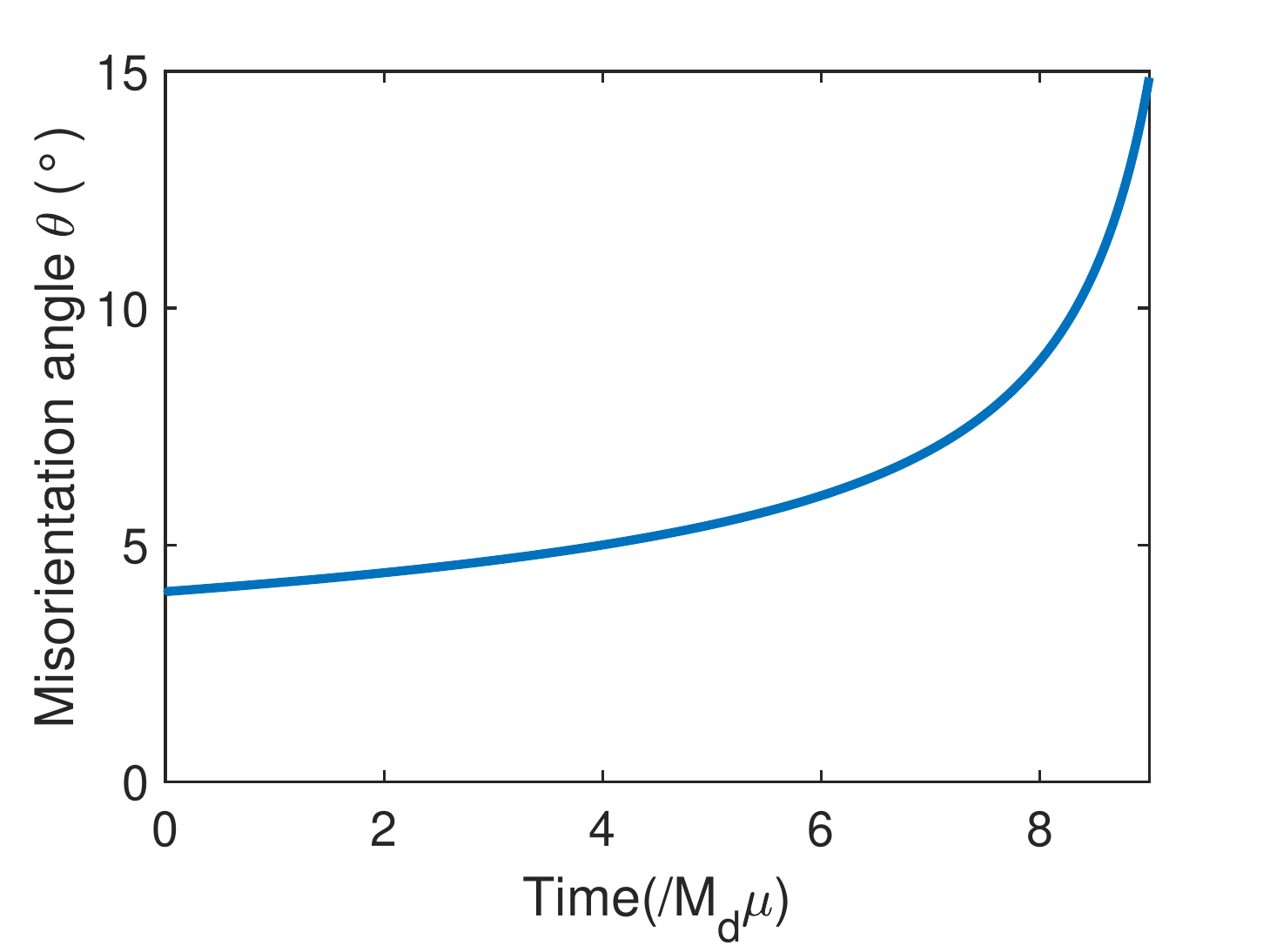}}
	\subfigure[]{\includegraphics[width=2.3in]{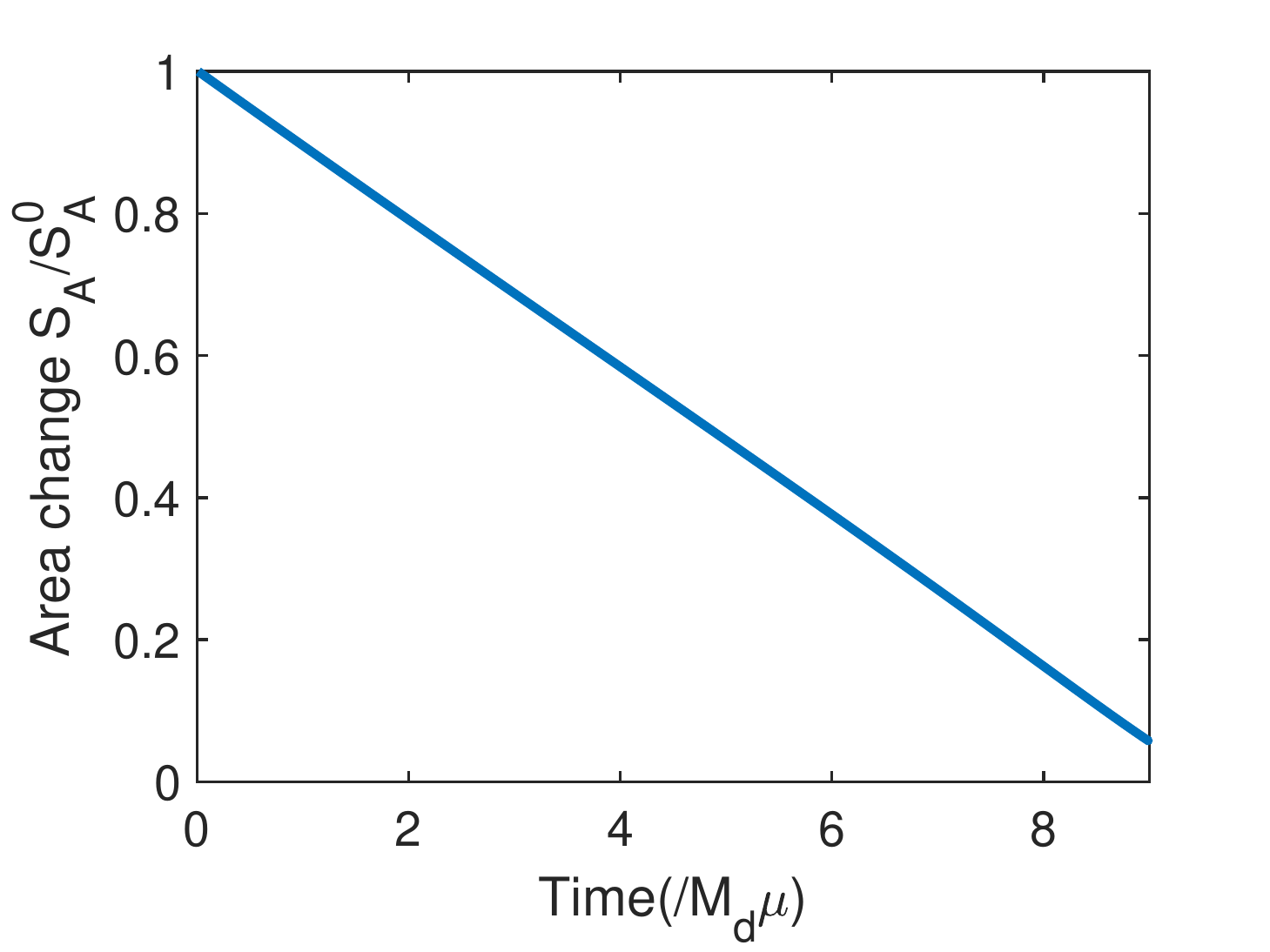}}
	\subfigure[]{\includegraphics[width=2.3in]{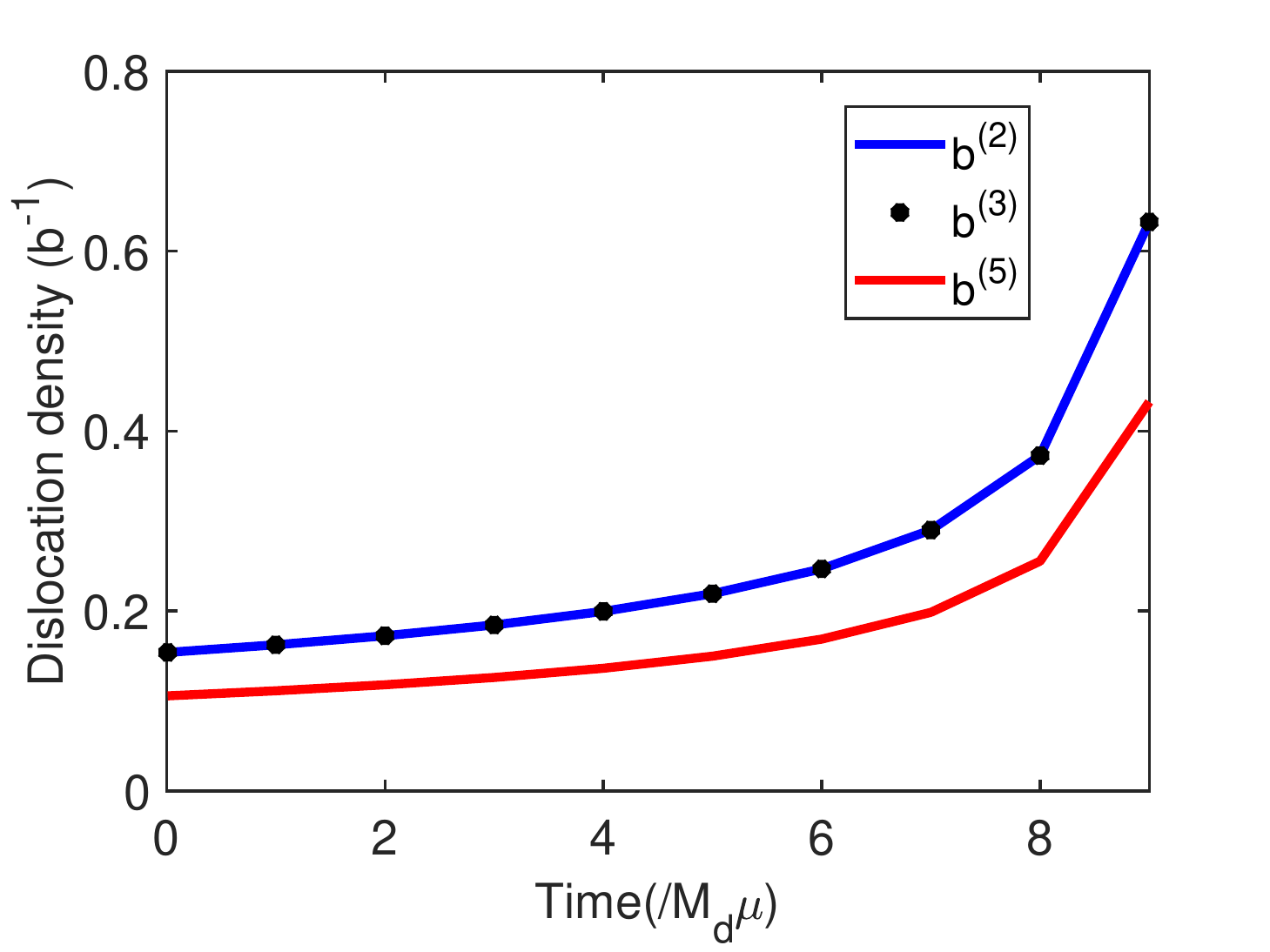}}
	\subfigure[]{\includegraphics[width=2.3in]{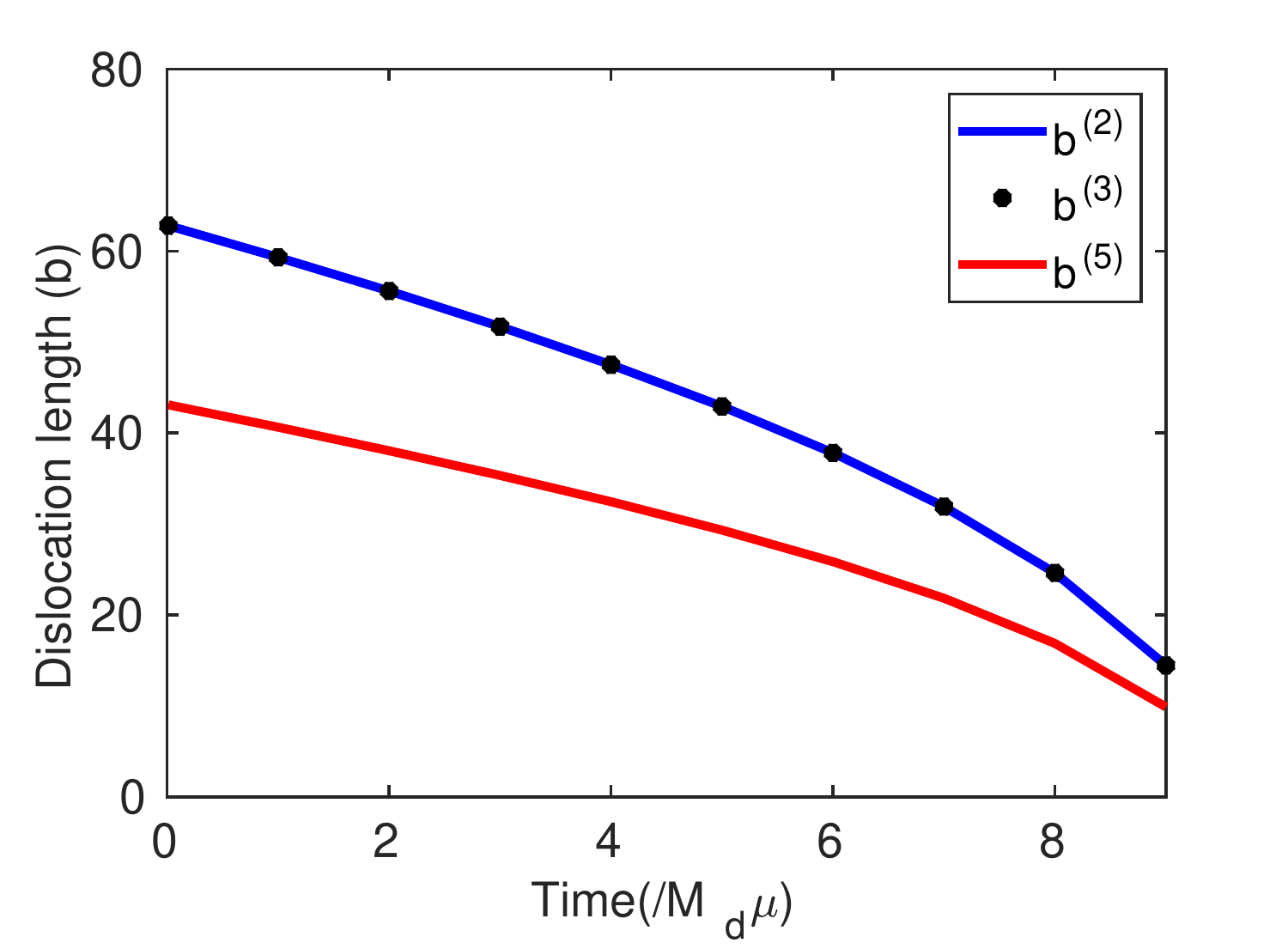}}
	\caption{Shrinkage of an initially spherical grain boundary in bcc under pure coupling motion. The rotation axis is the $z$ direction ($[011]$), and the initial misorientation angle $\theta=4^\circ$.  (a) Evolution of  misorientation angle $\theta$. (b) Evolution of grain boundary area $S_A$, where $S_A^0$ is the area of the initial grain boundary. (c) Evolution of densities of dislocations on the grain boundary. These dislocations have Burgers vectors
 $\mathbf b^{(2)}$, $\mathbf b^{(3)}$, and $\mathbf b^{(5)}$ on the grain boundary. (d) Evolution of the total lengths of dislocations with these Burgers vectors.
 The densities and total lengths of dislocations with the Burgers vectors $\mathbf b^{(2)}$, $\mathbf b^{(3)}$, and $\mathbf b^{(5)}$  are shown by blue lines, black dots, and red lines, respectively, in (c) and (d).
}\label{fig:bcctheta0}
\end{figure}

Fig.~\ref{fig:bcctheta0}  shows the increase of the misorientation angle $\theta$, linear decrease of the grain boundary area, increase of densities of dislocations, and decrease of the total lengths of dislocations during the shrinkage of the grain boundary. These results are similar to those of the grain boundary in fcc discussed in the main text and are consistent with available phase field crystal and two dimensional simulation results; see the discussion there. Note that for the dynamics of this grain boundary in bcc, the densities and total lengths of dislocations with Burgers vectors $\mathbf b^{(2)}$ and $\mathbf b^{(3)}$ are almost identical, and are greater than those of dislocations with Burgers vector $\mathbf b^{(5)}$. Recall that the lengths of Burgers vectors  $\mathbf b^{(2)}$ and $\mathbf b^{(3)}$ are equal, and are smaller than the length of Burgers vector $\mathbf b^{(5)}$.

\subsection{Motion with dislocation reaction}

Now we perform simulations using our continuum model considering dislocation reaction, i.e. $M_{\rm r}\neq 0$, for the motion of a grain boundary in bcc Fe.  We use the same initial spherical grain boundary as in Sec.~\ref{subsec:coupbcc} without dislocation reaction.

\begin{figure}[htbp]
\centering
\subfigure[]{
	\begin{minipage}{0.21\linewidth}
		\includegraphics[width=\linewidth]{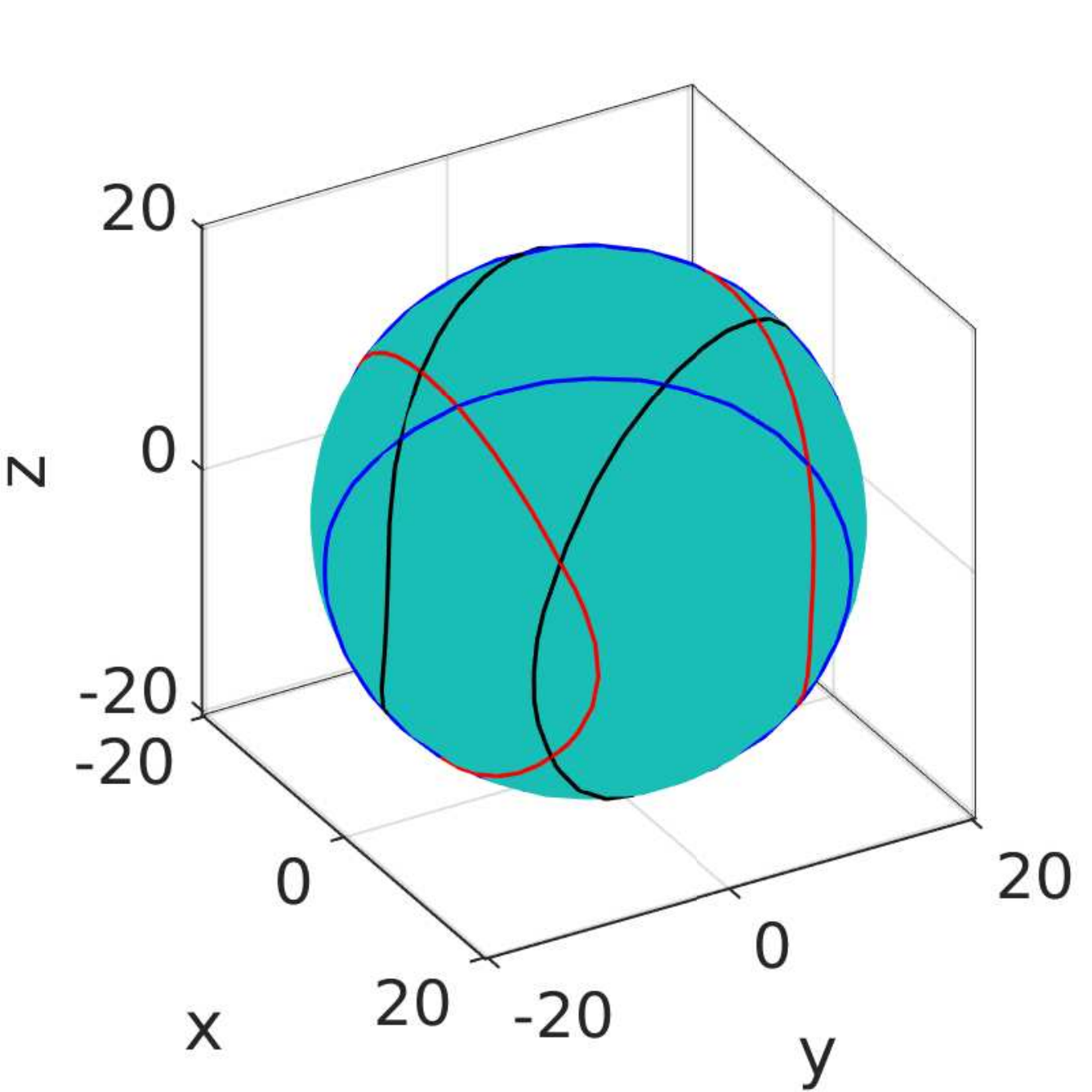}
		\includegraphics[width=\linewidth]{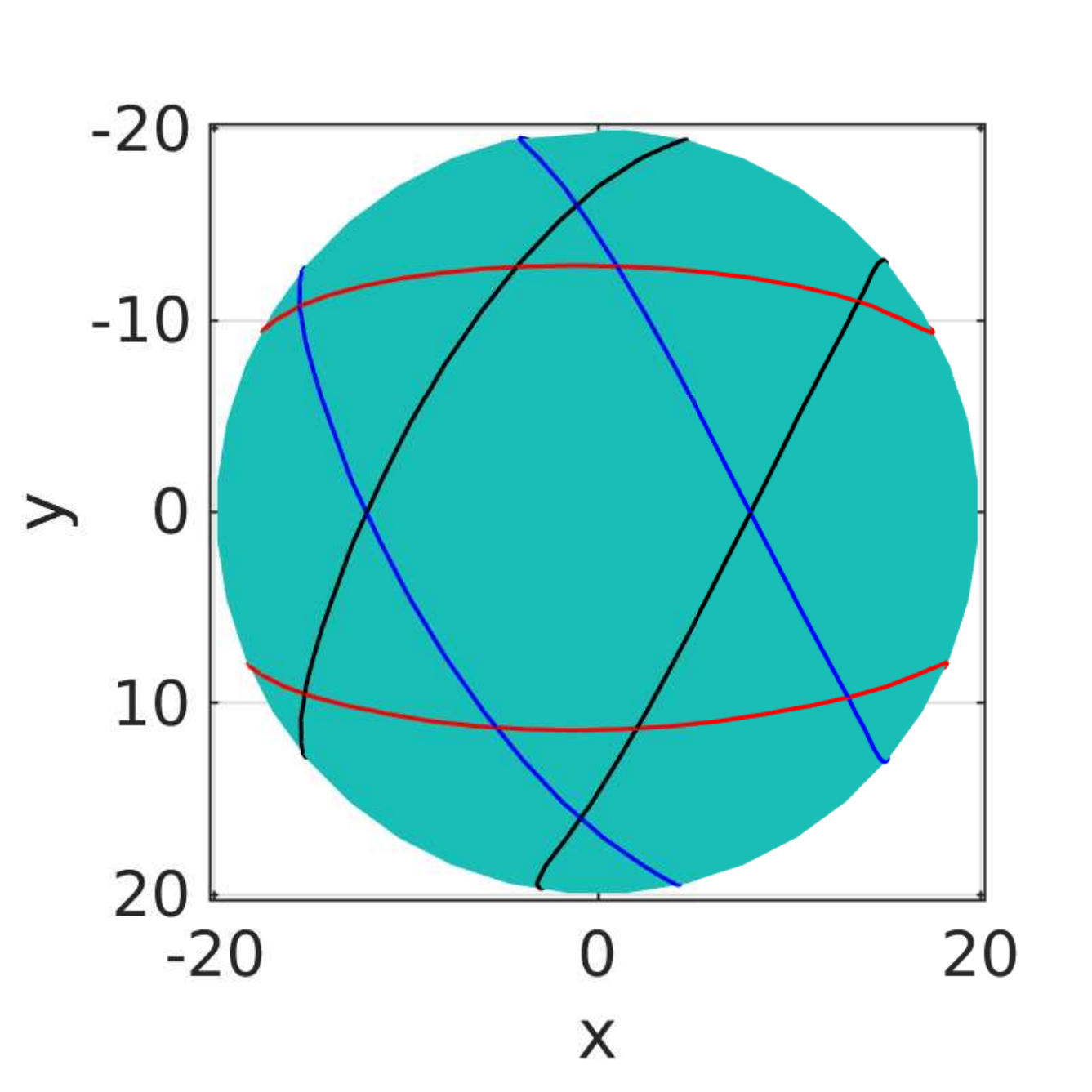}
		\includegraphics[width=\linewidth]{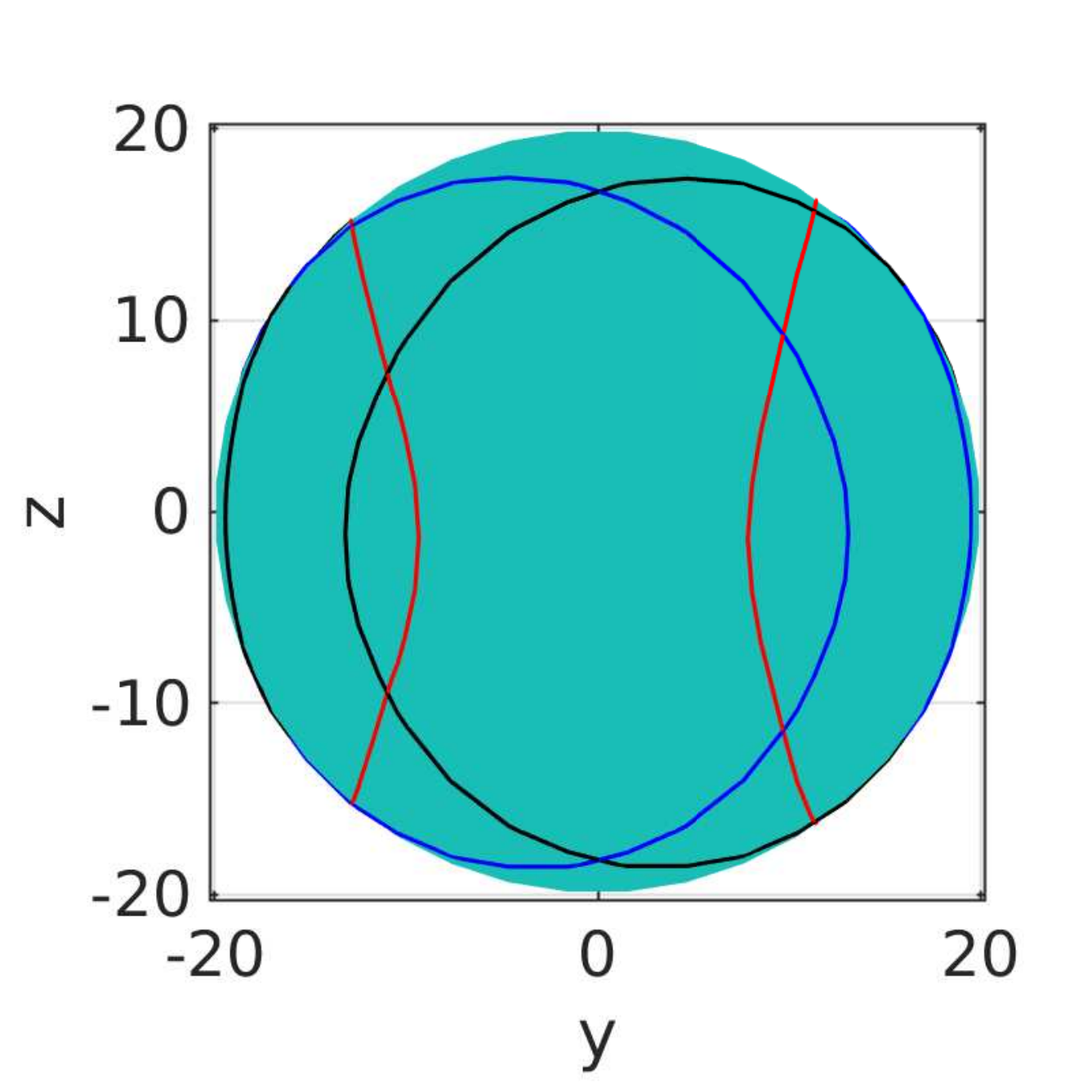}
	\end{minipage}
}
\subfigure[]{
	\begin{minipage}{0.21\linewidth}			
		\includegraphics[width=\linewidth]{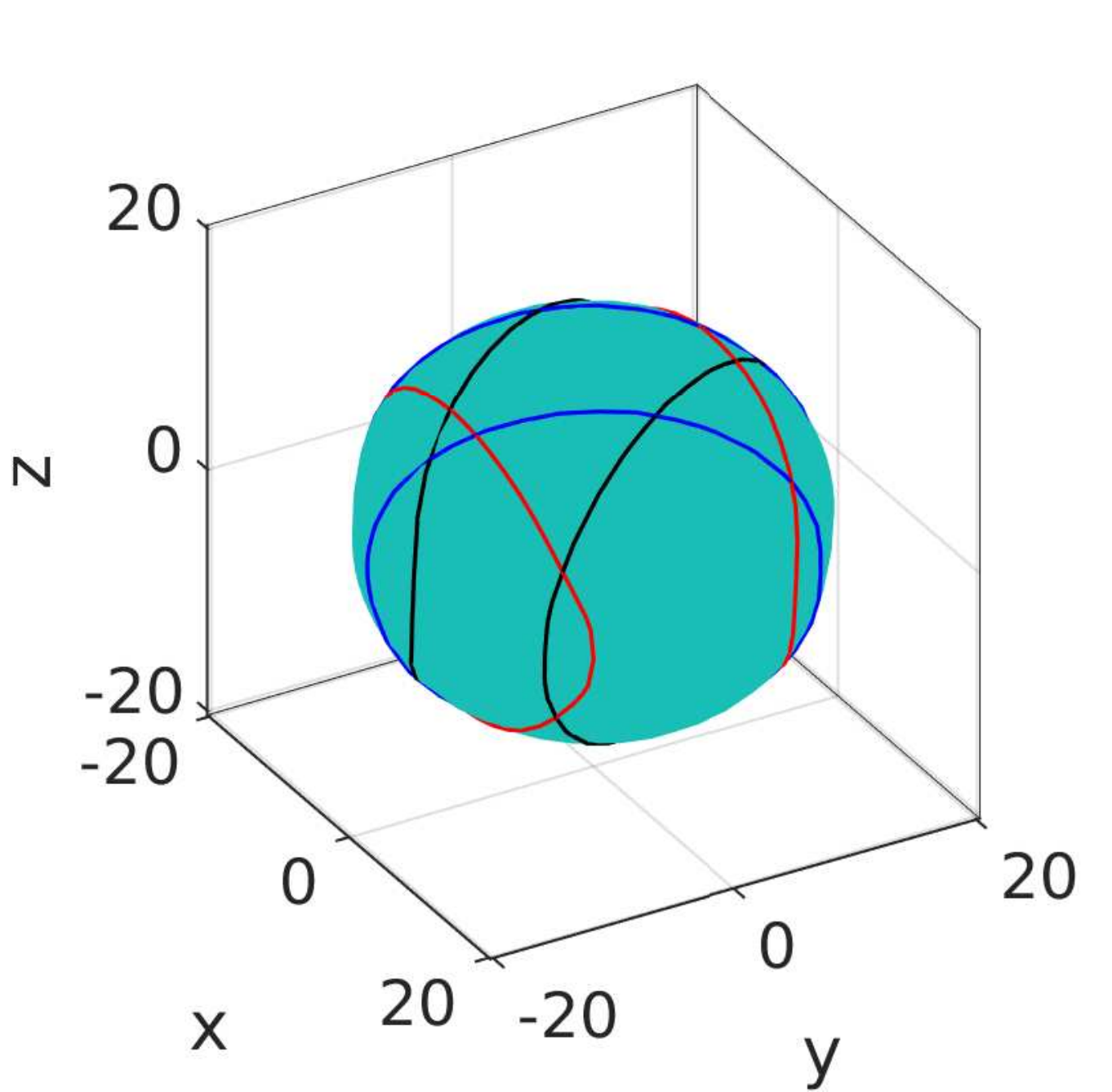}
		\includegraphics[width=\linewidth]{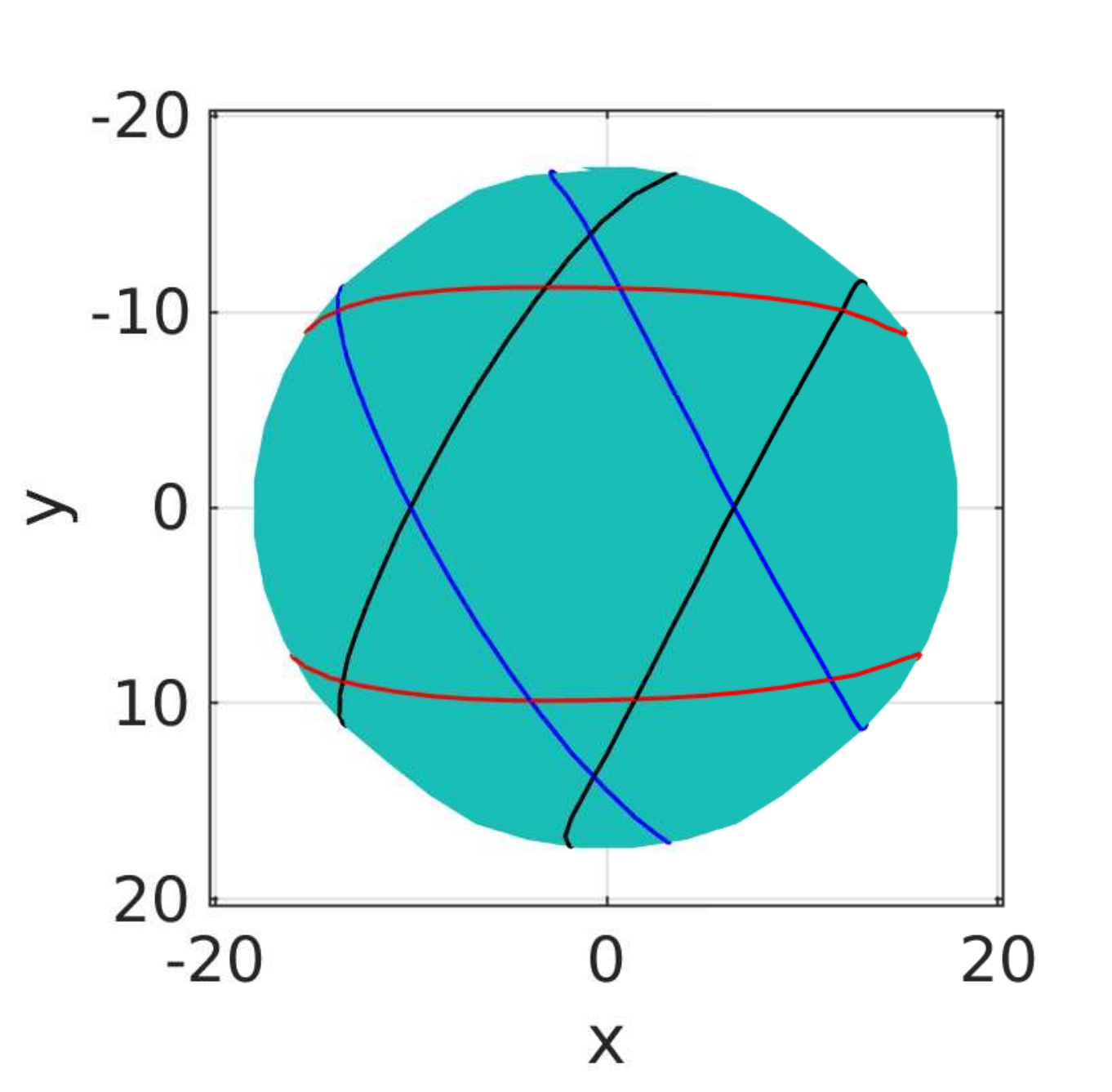}
		\includegraphics[width=\linewidth]{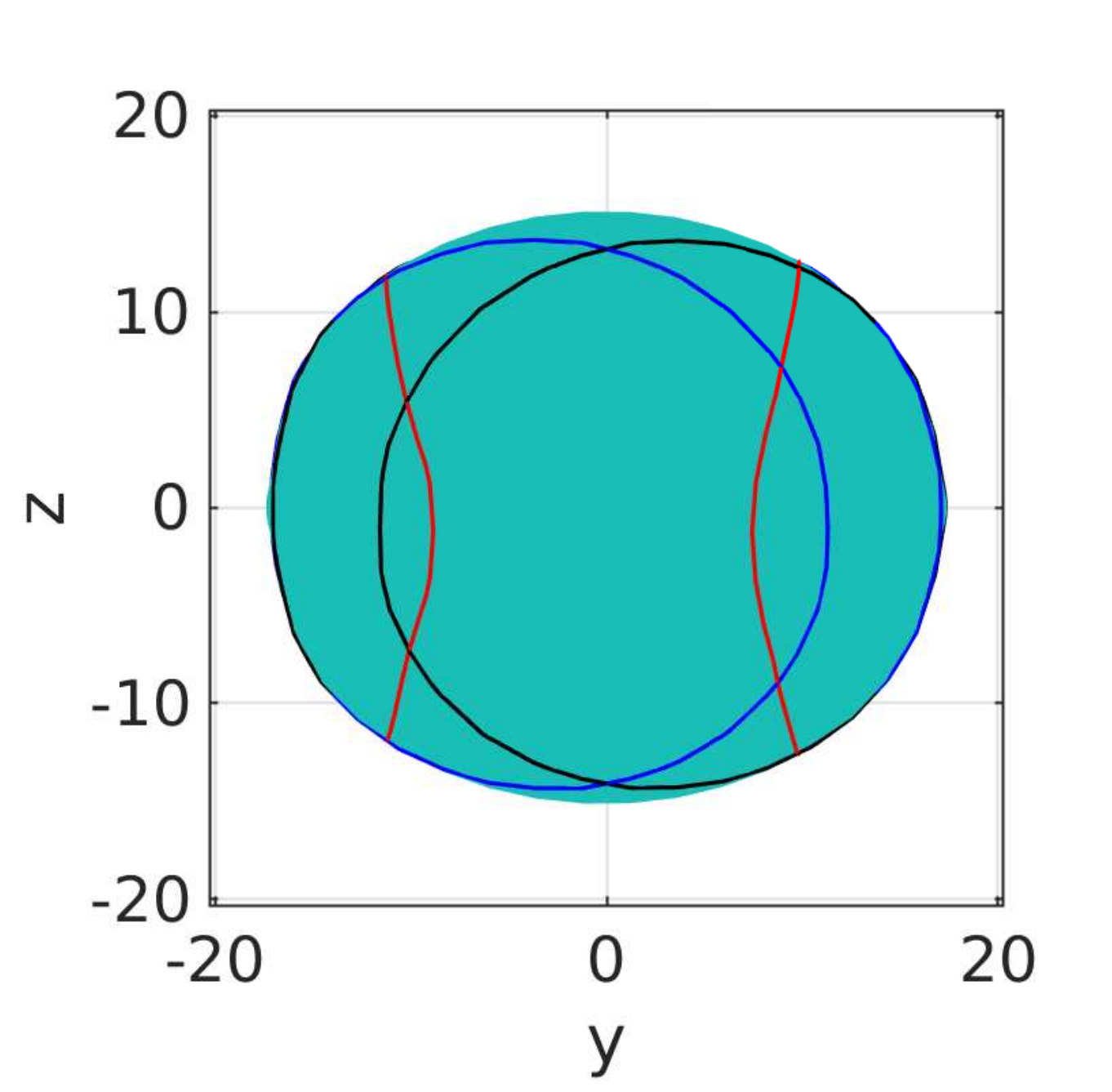}
	\end{minipage}
}	
\subfigure[]{
	\begin{minipage}{0.21\linewidth}			
		\includegraphics[width=\linewidth]{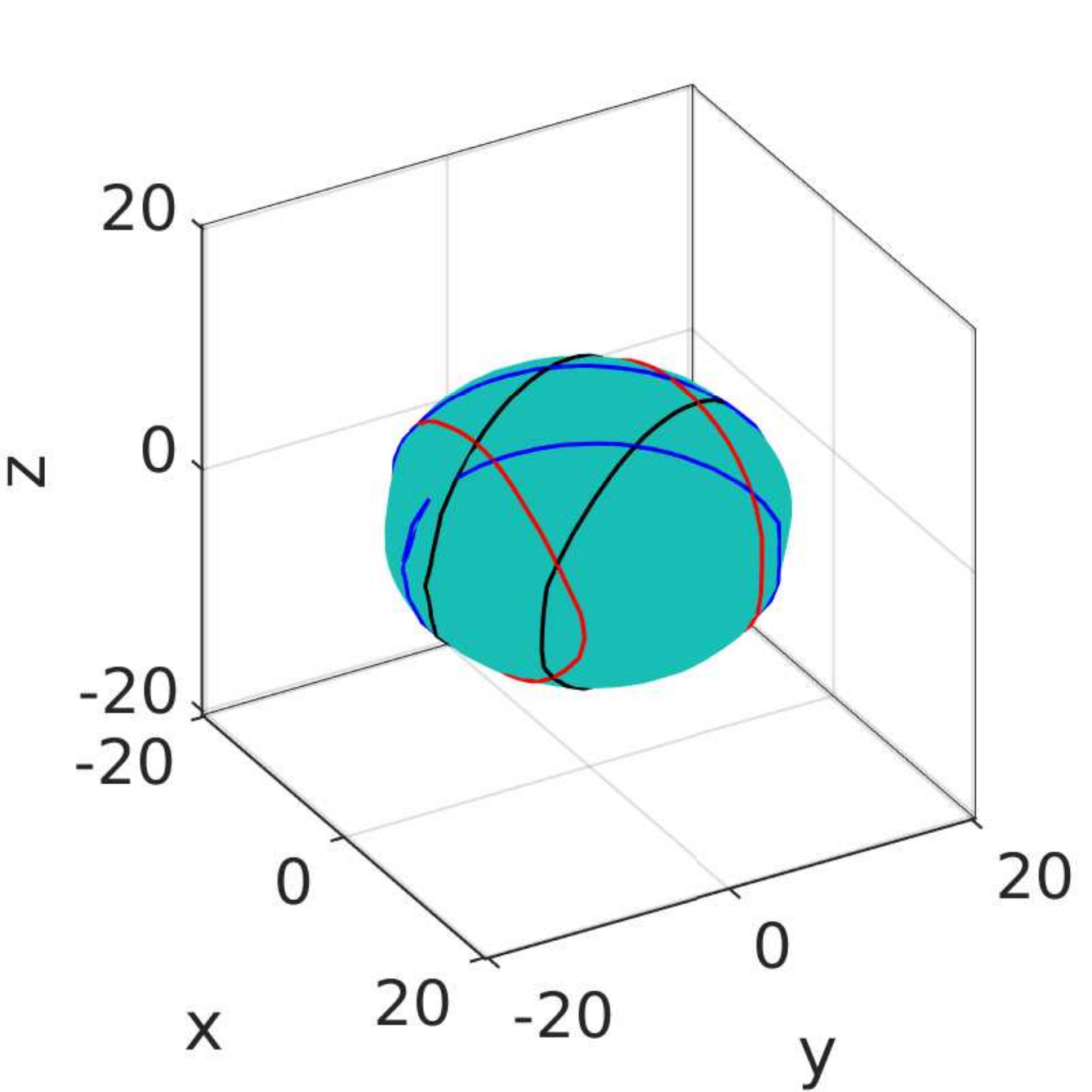}
		\includegraphics[width=\linewidth]{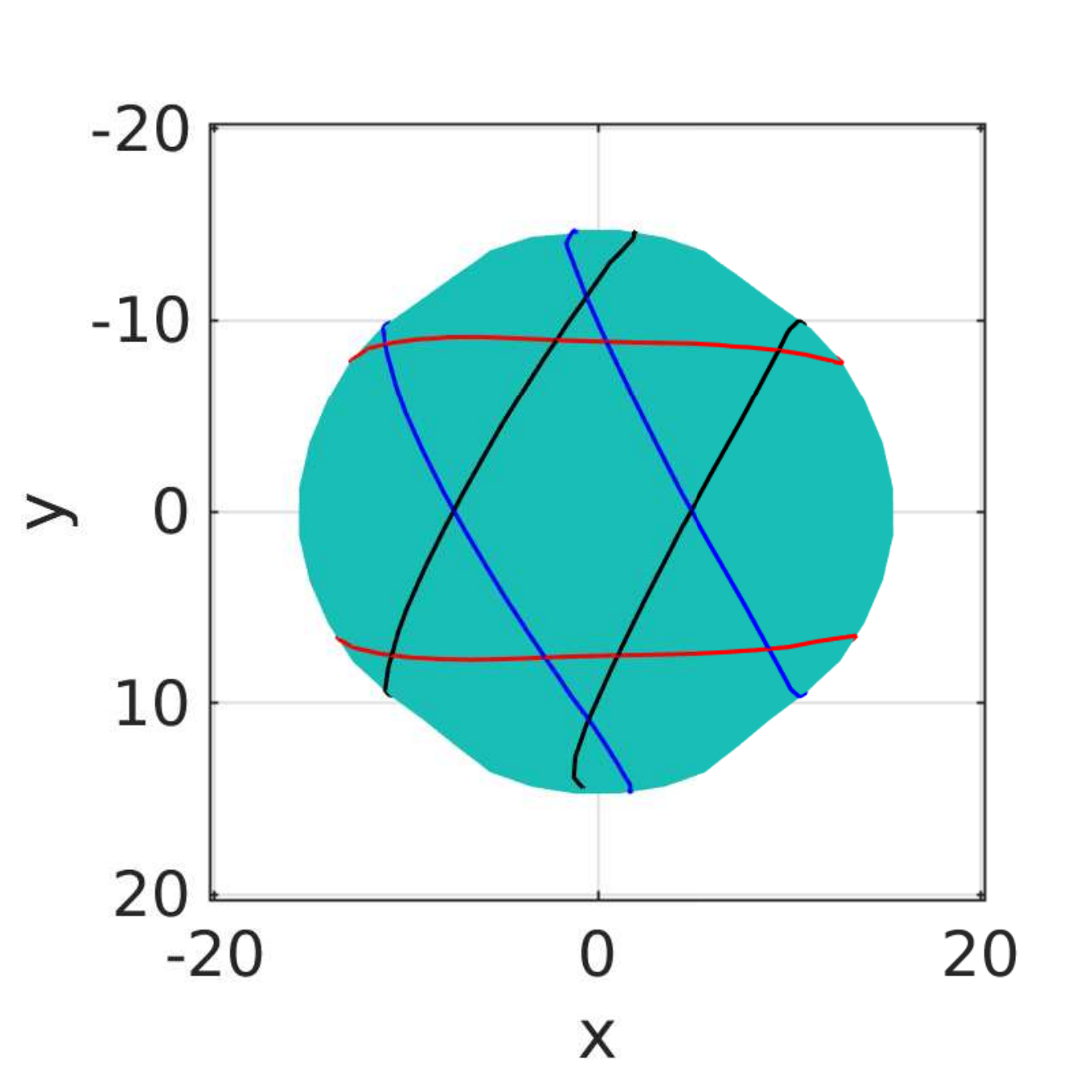}
		\includegraphics[width=\linewidth]{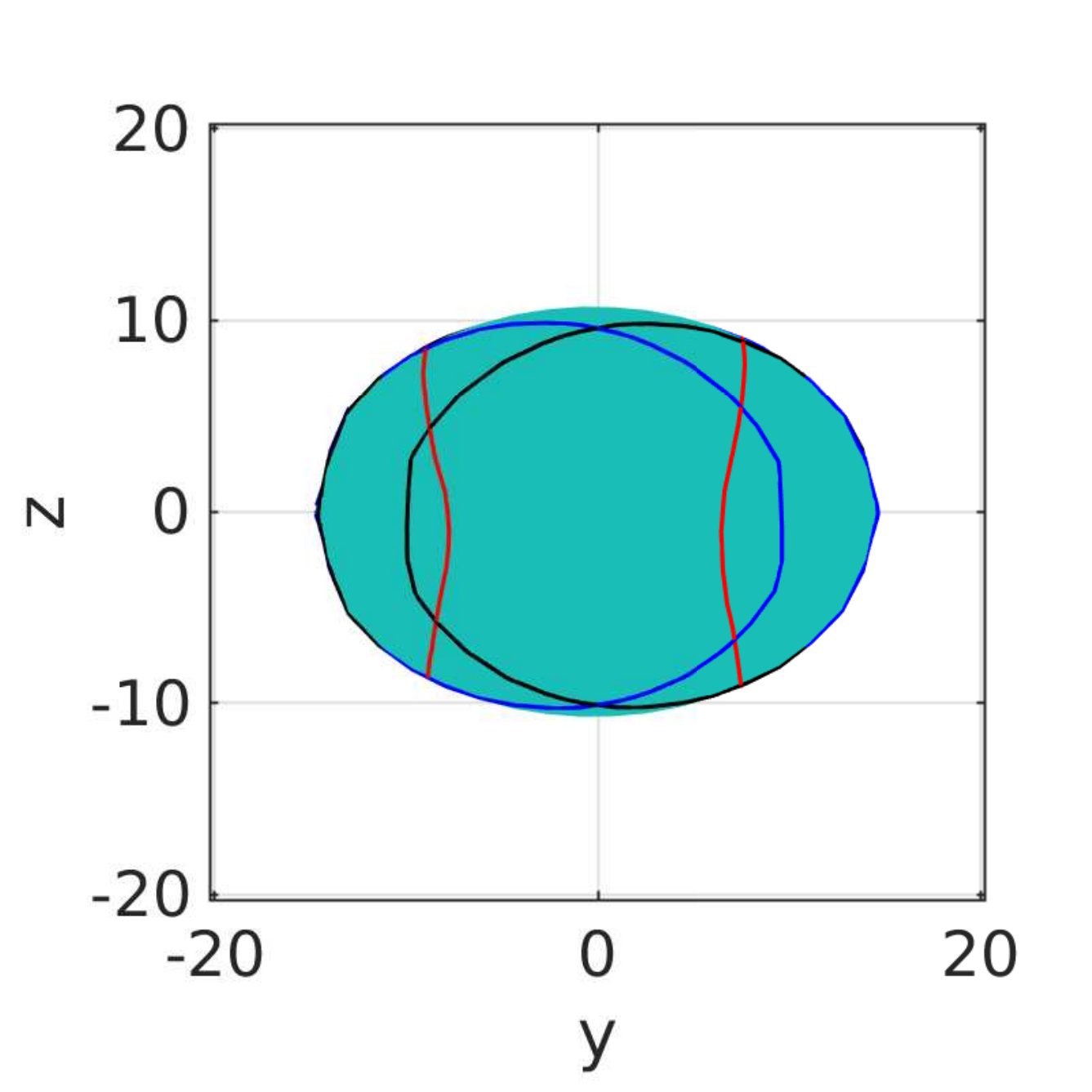}
	\end{minipage}
}	
\subfigure[]{
	\begin{minipage}{0.21\linewidth}			
		\includegraphics[width=\linewidth]{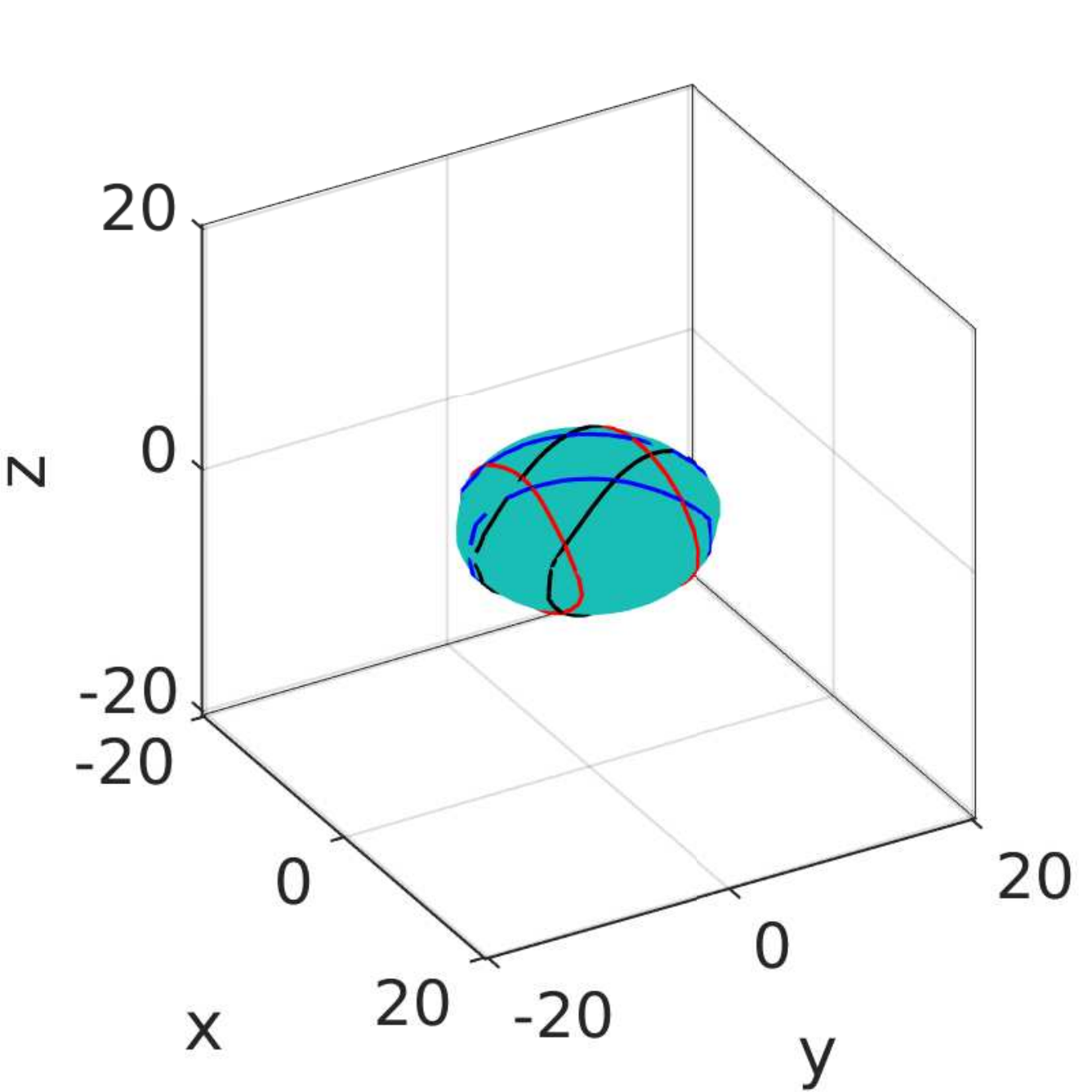}
		\includegraphics[width=\linewidth]{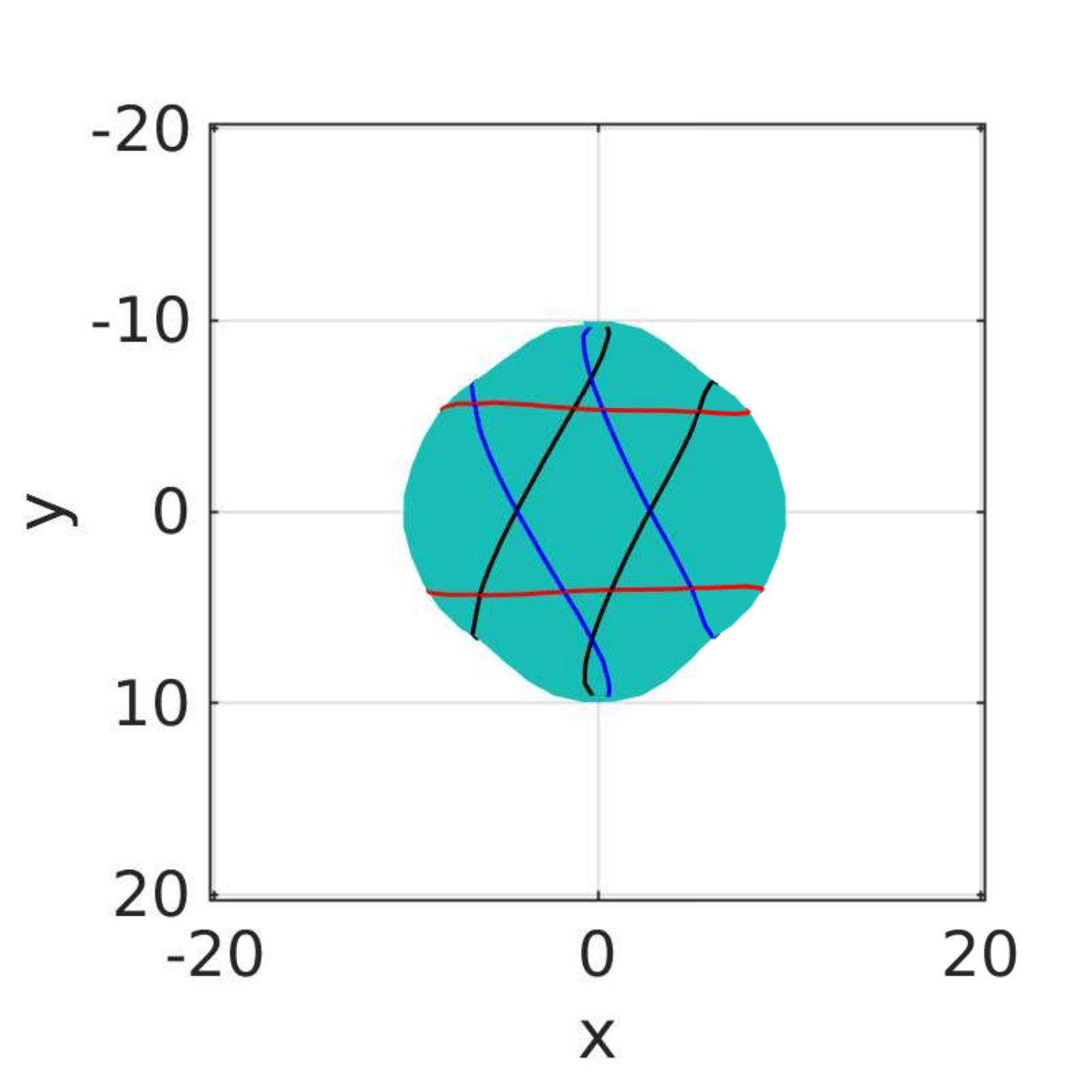}
		\includegraphics[width=\linewidth]{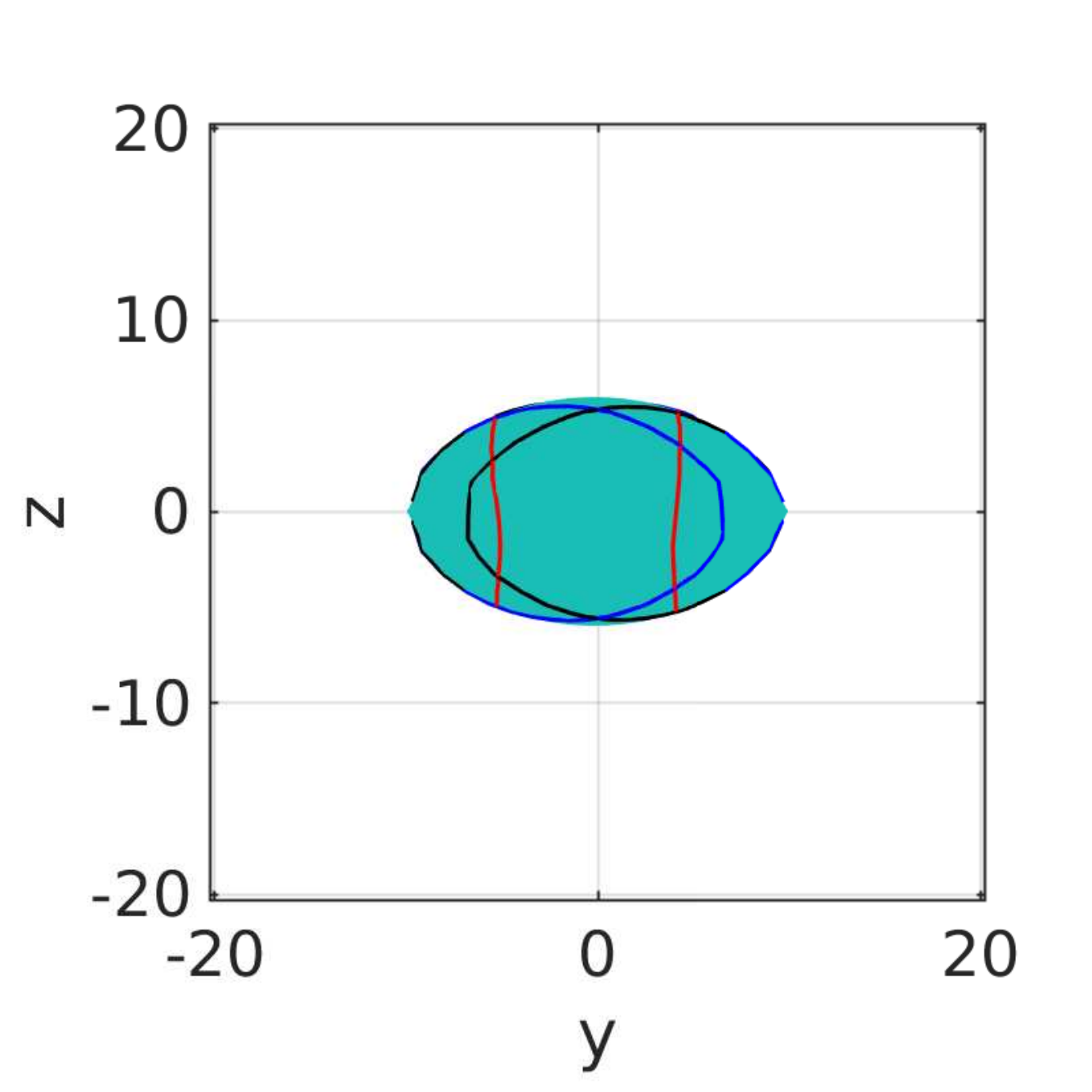}
	\end{minipage}
}
	\caption{Shrinkage of an initially spherical grain boundary in bcc with dislocation reaction: $M_{\rm r}b^3/M_{\rm d}=7.01\times10^{-5}$. The rotation axis is the $z$ direction ($[011]$), and the initial misorientation angle $\theta=4^\circ$. The upper panel of images show the three-dimensional view of the grain boundary during evolution. The middle panel of images show the grain boundary during evolution viewed from the $+z$ direction ($[011]$), and the lower panel of images show the grain boundary during evolution viewed from the $+x$ direction ($[100]$). Dislocations with Burgers vectors $\mathbf b^{(2)}$, $\mathbf b^{(3)}$ and $\mathbf b^{(5)}$ are shown by blue, black and red lines, respectively. Length unit: $b$. (a) The initial spherical grain boundary. (b), (c), and (d) Configurations at time $t=3/M_{\rm d}\mu, 6/M_{\rm d}\mu, 9/M_{\rm d}\mu$, respectively.}\label{fig:bccfigure}
\end{figure}

Fig.~\ref{fig:bccfigure} shows the shrinkage of the initially spherical grain boundary with dislocation reaction, where the reaction mobility  $M_{\rm r}b^3/M_{\rm d}=7.01\times10^{-5}$. As in the evolution of the grain boundary in fcc with dislocation reaction shown in the main text, the equator of the grain boundary (with respect to the rotation axis) gradually evolves from a circle into a hexagon. However, unlike  in fcc, here the grain boundary in bcc evolves into a non-regular hexagon. This is due to the fact that here the lengths of Burgers vectors  $\mathbf b^{(2)}$ and $\mathbf b^{(3)}$ are smaller than the length of Burgers vector $\mathbf b^{(5)}$, while in the grain boundary in fcc, the three Burgers vectors have the same length.
Again, we can see that the shrinkage of the grain boundary is faster in the direction of the rotation axis than in other directions. These results  agree with the those of phase field crystal simulations \cite{yamanaka2017phase}.

\begin{figure}[htbp]
	\centering
	\subfigure[]{\includegraphics[width=2.3in]{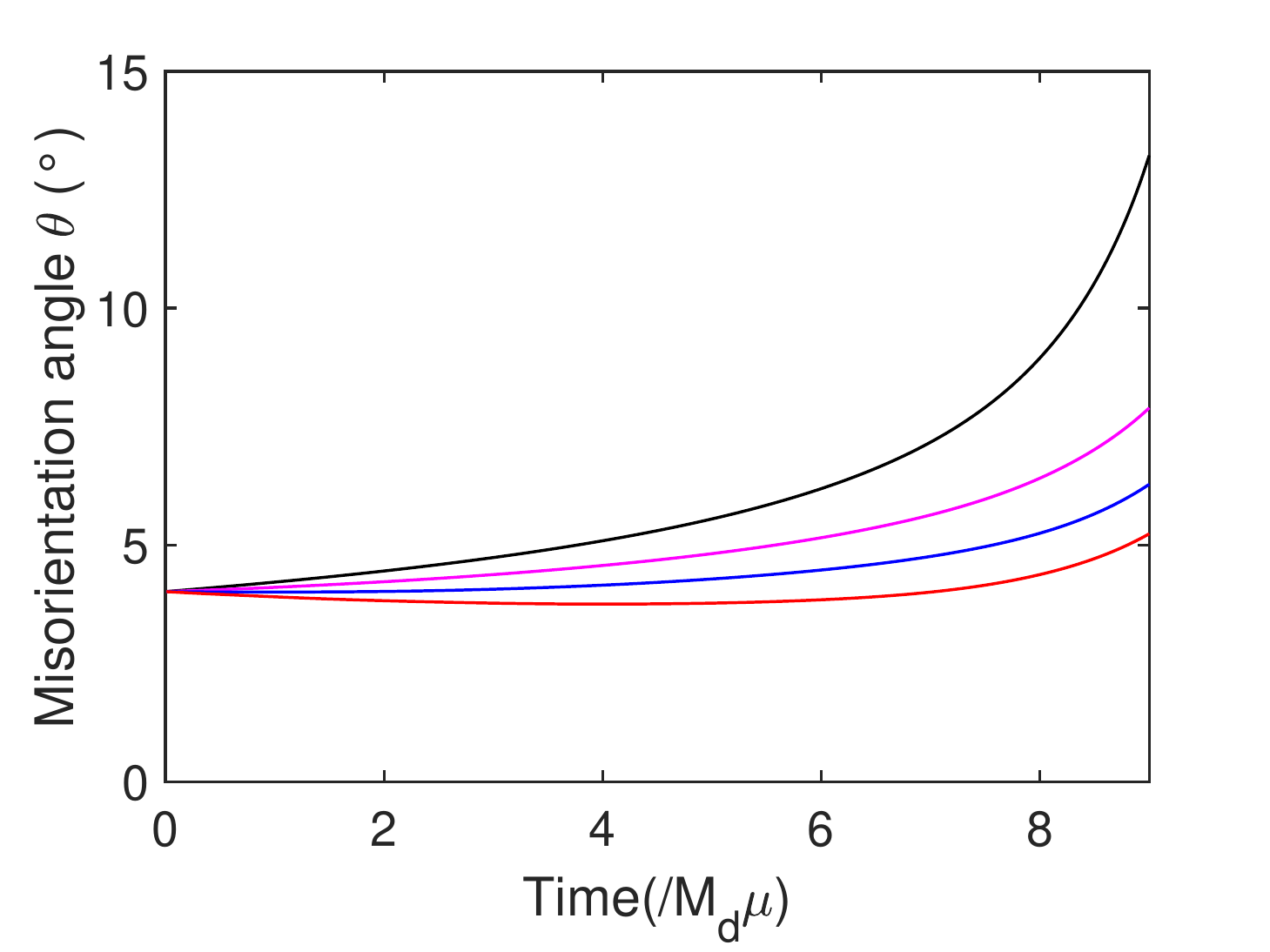}}
	\subfigure[]{\includegraphics[width=2.3in]{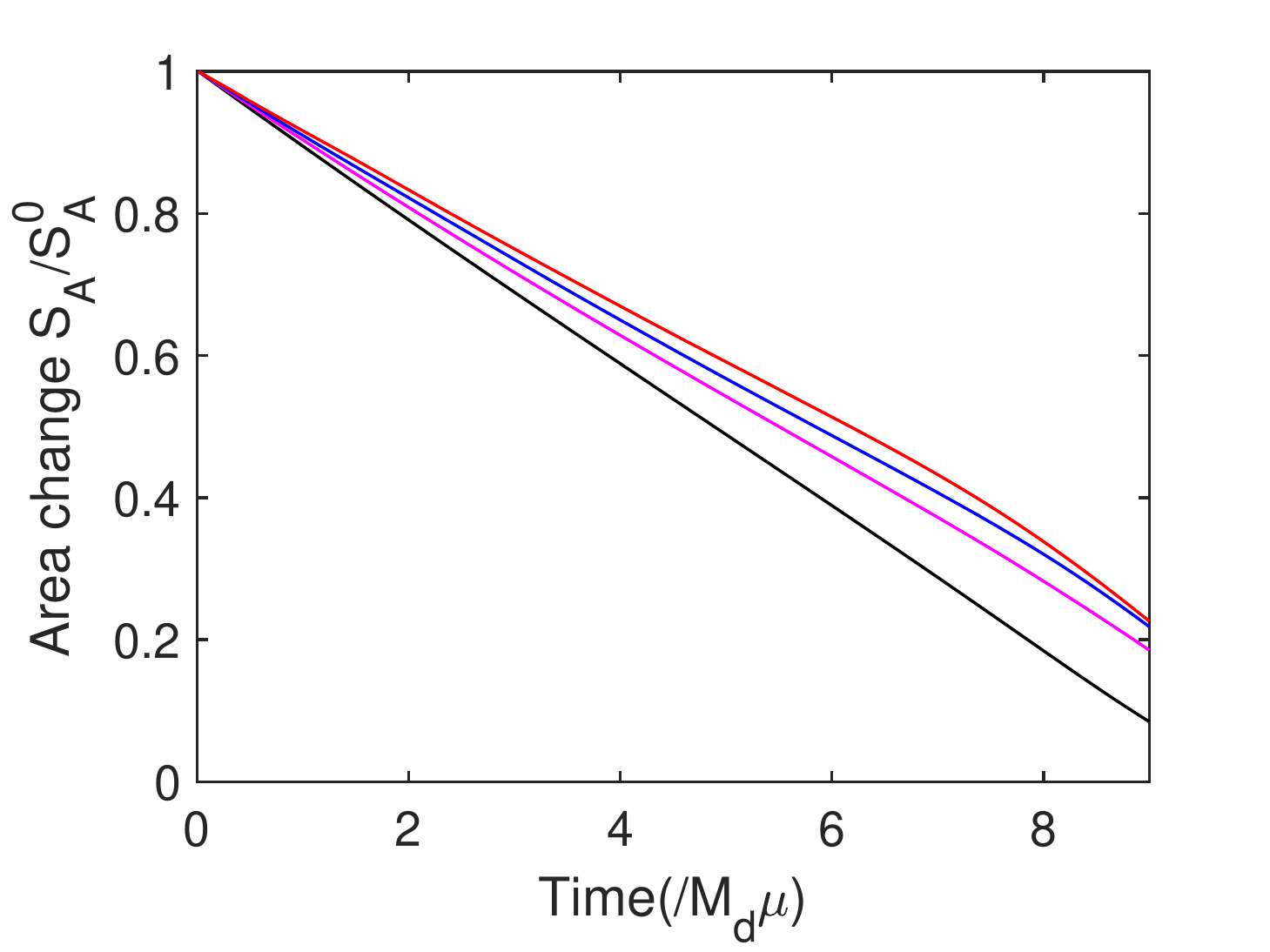}}
	\subfigure[]{\includegraphics[width=2.3in]{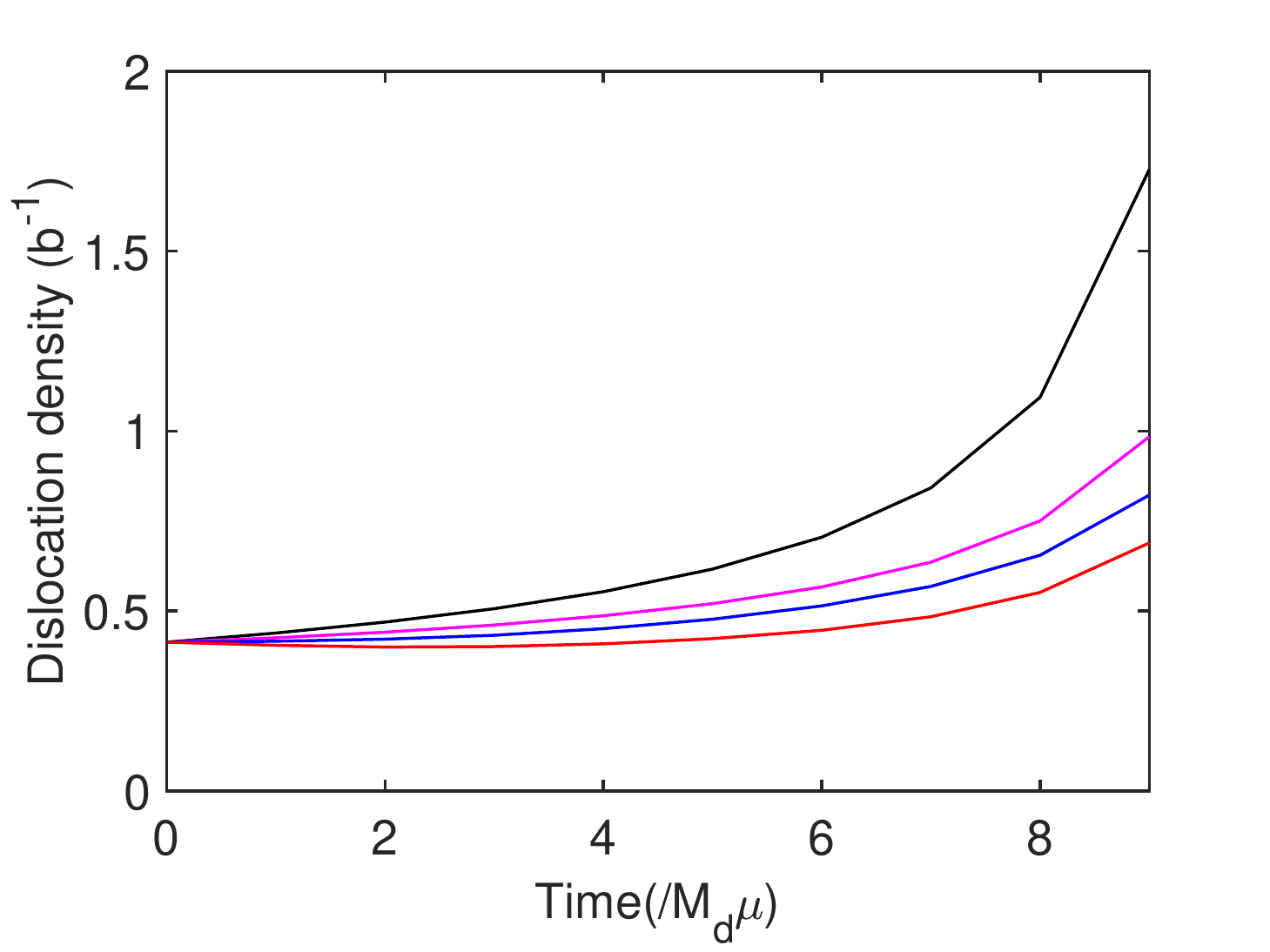}}
	\subfigure[]{\includegraphics[width=2.3in]{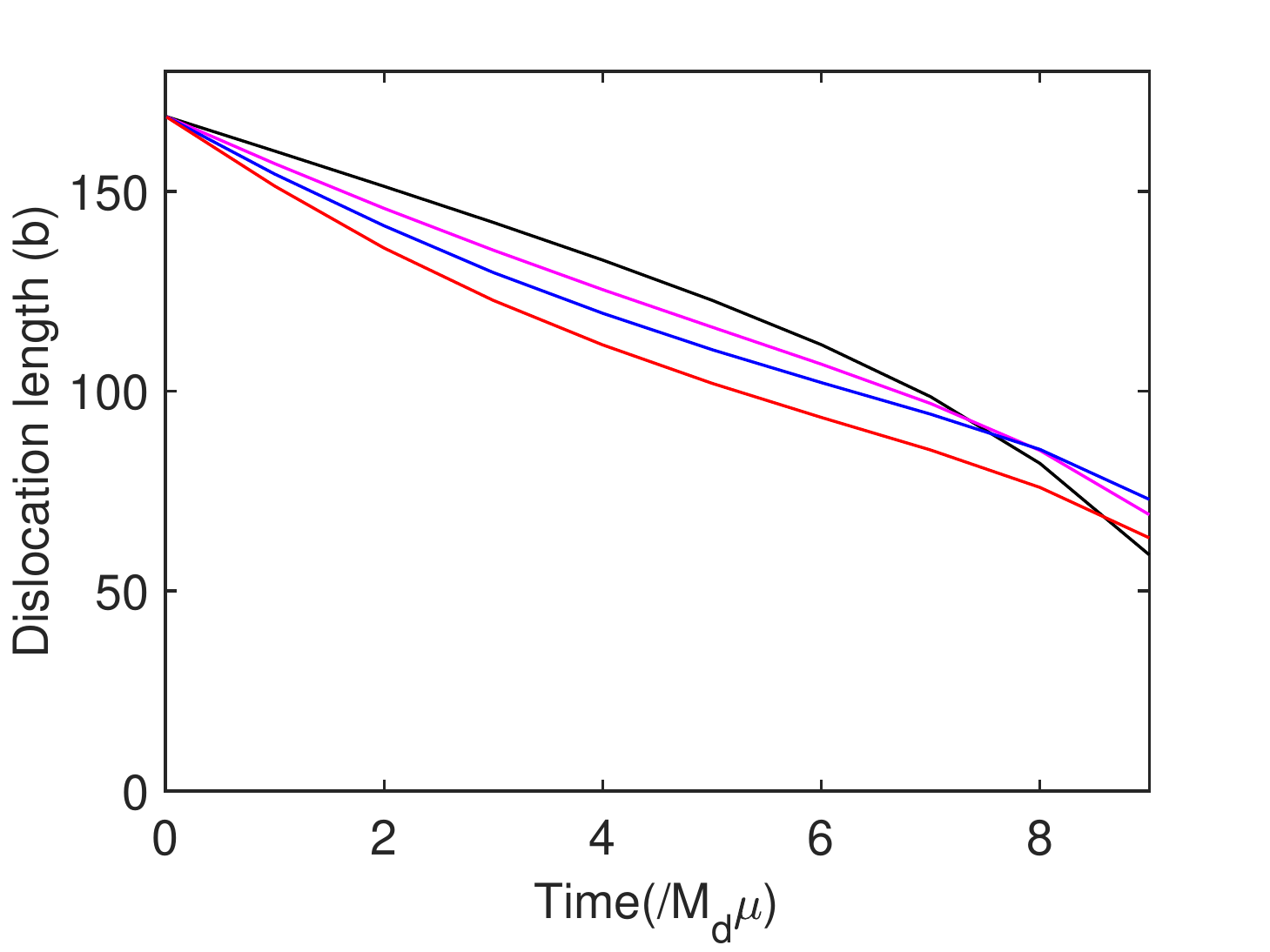}}
	\caption{ Shrinkage of an initially spherical grain boundary in bcc with different values of reaction mobility $M_{\rm r}$. The rotation axis is the $z$ direction ($[011]$), and the initial misorientation angle $\theta=4^\circ$.   The reaction mobility $M_{\rm r}b^3/M_{\rm d}=0$, $7.01\times10^{-5}$, $1.40\times10^{-4}$, and $2.11\times10^{-4}$ from the top curve to the bottom one in (a), (c) and (d), and from the bottom to the top ones in (b).
(a) Evolution of  misorientation angle $\theta$. (b) Evolution of grain boundary area $S_A$, where $S_A^0$ is the area of the initial grain boundary. (c) Evolution of the density of all the dislocations on the grain boundary. (d) Evolution of the total length of all the dislocations on the grain boundary. }\label{fig:bcctheta}
\end{figure}

Evolution of the misorietation angle $\theta$ with different values of reaction mobility $M_{\rm r}$ is shown in Fig.~\ref{fig:bcctheta}(a). When $M_{\rm r}\neq 0$, the evolution of misorientation angle is controlled by both the coupling effect (which is associated with the conservation of dislocations and increases $\theta$) and sliding effect (which is associated with dislocation reaction and decreases $\theta$). While the misorientation angle $\theta$ is increasing during the evolution, the increase rate of $\theta$ decreases as
  the dislocation reaction mobility $M_{\rm r}$ increases.
Fig.~\ref{fig:bcctheta}(b) shows the evolution of grain boundary area with different values of dislocation reaction mobility $M_{\rm r}$. The decrease of grain boundary area still follows the linear law in Eq.~(62) in the main text except in the later stage of the evolution with high dislocation reaction mobility,  and is slower for higher dislocation reaction mobility.
Figs.~\ref{fig:bcctheta}(c) and (d) show the evolutions of dislocation densities on the grain boundary and total length of dislocations with different values of reaction mobility $M_{\rm r}$. The density of the dislocations on the grain boundary is increasing whereas the total length of dislocations is decreasing for these values of dislocation reaction mobility $M_{\rm r}$.
As the dislocation reaction mobility $M_{\rm r}$ increases, the increase rate of dislocation density  decreases, and the decrease rate of the total length of dislocations increases except for the later stage of evolution without dislocation reaction. These behaviors are similar to those of the grain boundary in fcc shown in  Fig.~5 in the main text, and more discussion can be found there. These results also agree with the available phase field crystal simulation results \cite{yamanaka2017phase,salvalaglio2018defects}.

\bibliography{ref}